\documentclass[12pt]{article}

\usepackage{graphicx}
\usepackage{amsfonts}
\usepackage[usenames]{color}
\usepackage{amsmath}
\usepackage{feynmp}

\def\Z{\mathbb{Z}}

\begin{document}
\baselineskip 0.6cm
\newcommand{\gsim}{ \mathop{}_{\textstyle \sim}^{\textstyle >} }
\newcommand{\lsim}{ \mathop{}_{\textstyle \sim}^{\textstyle <} }
\newcommand{\vev}[1]{ \left\langle {#1} \right\rangle }
\newcommand{\bra}[1]{ \langle {#1} | }
\newcommand{\ket}[1]{ | {#1} \rangle }
\newcommand{\Dsl}{\mbox{\ooalign{\hfil/\hfil\crcr$D$}}}
\newcommand{\sslash}[1]{\mbox{\ooalign{\hfil/\hfil\crcr{${#1}$}}}}
\newcommand{\nequiv}{\mbox{\ooalign{\hfil/\hfil\crcr$\equiv$}}}
\newcommand{\nsupset}{\mbox{\ooalign{\hfil/\hfil\crcr$\supset$}}}
\newcommand{\nni}{\mbox{\ooalign{\hfil/\hfil\crcr$\ni$}}}
\newcommand{\EV}{ {\rm eV} }
\newcommand{\KEV}{ {\rm keV} }
\newcommand{\MEV}{ {\rm MeV} }
\newcommand{\GEV}{ {\rm GeV} }
\newcommand{\TEV}{ {\rm TeV} }

\renewcommand{\tilde}{\widetilde}
\renewcommand{\;}{\hspace{1pt}}

\def\diag{\mathop{\rm diag}\nolimits}
\def\tr{\mathop{\rm tr}}
\def\Tr{\mathop{\rm Tr}}

\def\Spin{\mathop{\rm Spin}}
\def\SO{\mathop{\rm SO}}
\def\O{\mathop{\rm O}}
\def\SU{\mathop{\rm SU}}
\def\U{\mathop{\rm U}}
\def\Sp{\mathop{\rm Sp}}
\def\SL{\mathop{\rm SL}}
\def\simgt{\mathrel{\lower2.5pt\vbox{\lineskip=0pt\baselineskip=0pt
           \hbox{$>$}\hbox{$\sim$}}}}
\def\simlt{\mathrel{\lower2.5pt\vbox{\lineskip=0pt\baselineskip=0pt
           \hbox{$<$}\hbox{$\sim$}}}}

\def\change#1#2{{\color{blue} #1}{\color{red} [#2]}\color{black}\hbox{}}

\makeatletter
 \renewcommand{\theequation}{%
   \thesection.\arabic{equation}}
  \@addtoreset{equation}{section}
 \makeatother

\begin{fmffile}{fmf}
\fmfstraight
\fmfset{wiggly_len}{5mm}
%\fmfset{wiggly_slope}{75}
\fmfset{arrow_len}{2.2mm}
\fmfset{arrow_ang}{30}
\unitlength = 1mm

%\def\change#1#2{#2}

%%%%%%%%%%
%%%%%%%%%%      title page
%%%%%%%%%%

\begin{titlepage}

\begin{flushright}
UT-08-01 \\
\end{flushright}
 
\vskip 1cm
\begin{center}
{\large \bf A Theoretical Framework for R-parity Violation} 
 
\vskip 1.2cm

Minoru~Kuriyama, Hiroto~Nakajima and Taizan Watari,
 
\vskip 0.4cm
{\it Department of Physics, the University of Tokyo, Tokyo, 
113-0033, Japan} \\
 
\vskip 1.5cm

\abstract{We propose a theoretical framework for R-parity violation. 
It is realized by a class of Calabi--Yau compactification 
of Heterotic string theory.
Trilinear R-parity violation in superpotential is either absent 
or negligibly small without an unbroken symmetry, 
due to a selection rule based on charge counting 
of a spontaneously broken U(1) symmetry. 
Although such a selection rule cannot be applied in general 
to non-renormalizable operators in the low-energy effective
superpotential, it is valid for terms trilinear in low-energy degrees 
of freedom, and hence can be used as a solution to the dimension-4 
proton decay problem in the minimal supersymmetric standard model. 
Bilinear R-parity violation is generated, but there are good reasons 
why they are small enough to satisfy its upper bounds from neutrino 
mass and washout of baryon/lepton asymmetry. All R-parity violating 
dimension-5 operators can be generated. In this theoretical framework, 
nucleons can decay through squark-exchange diagrams combining 
dimension-5 and bilinear R-parity violating operators. 
$B-L$ breaking neutron decay is predicted.} 

\end{center}
\end{titlepage}

%%%%%%%%%%
%%%%%%%%%%      main part 
%%%%%%%%%%

\tableofcontents

%%%%%%%%%%%%%%%%%%%%%%%%%%%%%%%%%%%%%%%%%%%%%%%%%%%%%
\section{Introduction}
%%%%%%%%%%%%%%%%%%%%%%%%%%%%%%%%%%%%%%%%%%%%%%%%%%%%

In supersymmetric extensions of the standard model, renormalizable 
operators 
\begin{equation}
 W \ni \lambda \; L \; \bar{E} \; L 
     + \lambda' \; L \; Q \; \bar{D} 
     + \lambda'' \; \bar{D} \; \bar{U} \; \bar{D} 
  \sim  \bar{\bf 5} \; {\bf 10} \; \bar{\bf 5}
\label{eq:dim4}
\end{equation}
break baryon and lepton number symmetries, and hence a proton decays 
too rapidly. Either the coefficient $\lambda''$ of the baryon number
violating operator $\bar{D} \; \bar{U} \; \bar{D}$ or 
$\lambda$ and $\lambda'$ of the lepton number violating 
$L \; \bar{E} \; L$ and $L \; Q \; \bar{D}$ have to be
highly suppressed. 
The most popular solution to this dimension-4 proton decay problem 
is to assume an unbroken R parity (or matter parity), which removes 
the operators in (\ref{eq:dim4}) altogether. 

Various alternative solutions have also been discussed in the
literature, some of which are found in a review article~\cite{Barbieri}. 
Some solutions assume discrete symmetries other than R-parity
(see also \cite{Allanach} and references therein), so that either 
the first two operators or the last one are(is) forbidden by the 
discrete symmetry. Although phenomenological consequences of these 
solutions are quite different from those of R-parity preserving ones, 
these two classes of solutions share one thing in common, 
an unbroken discrete symmetry. In Calabi--Yau compactification 
of Heterotic $E_8 \times E'_8$ string theory, however, discrete 
symmetries are found only at special points (or subsets) 
in moduli space. Reasons are not clear why such vacua have to 
be chosen. References~\cite{WeinbergDim5, FaraggiA}, for example, 
clearly expressed dissatisfaction to solutions relying on unbroken 
discrete symmetries.  

This article presents an alternative solution to the 
% dimension-4 proton decay 
problem without assuming an unbroken symmetry. 
% An alternative solution to be presented in this article  
% does not rely on an unbroken symmetry. In Calabi--Yau compactification 
% of Heterotic $E_8 \times E_8'$ string theory, discrete symmetries 
% such as an R parity are preserved only in measure-zero subset of 
% moduli space. We find, however, that the dimension-4 proton decay 
% problem is solved in a class of compactification without restricting 
% to a moduli space with an enhanced symmetry. 
The essence of the solution is an extra U(1) gauge symmetry 
with a Fayet--Iliopoulos parameter. The U(1) symmetry is broken 
spontaneously at high energy,\footnote{There are also solutions in the 
literature where such a U(1) symmetry is broken by a hierarchically 
small expectation value. Such solutions, however, share the same
unsatisfactory aspect as those with discrete unbroken symmetries. 
We need to understand why our vacuum is very close to a U(1)-symmetry
enhanced point in the moduli space.} allowing for large Majorana masses 
of right-handed neutrinos. 
No unbroken symmetry is left at low energy, but its legacy 
still remains.
There is a selection rule \cite{SUSY-0} (also known as SUSY-zero 
mechanism) in how the U(1)-breaking vacuum expectation value (vev) 
can appear in superpotential of low-energy effective theory, and 
this rule may be used to make sure that the trilinear R-parity 
violating couplings (\ref{eq:dim4}) are absent. 
In this sense, the solution in this article is certainly along 
the line of models in \cite{GR, TW1, TW2, DESY-Rparity}.
It should be reminded, however, that there are many subtleties 
in how to use the selection rule in low-energy effective
superpotential. It is one of the purposes of section~\ref{sec:frame} 
of this article to clarify when the selection rule can be used 
and when it cannot.
This solution fits very well with Calabi--Yau compactification 
of the Heterotic $E_8 \times E'_8$ string theory (and its dual 
descriptions) \cite{TW1,TW2}. 
In such string compactification, moduli fields are not required 
to be at special points, and U(1)-symmetry breaking vev is not 
assumed to be hierarchically small in order to make sure that 
the trilinear couplings (\ref{eq:dim4}) are sufficiently small.
Thus, this solution does not share the unsatisfactory aspect of 
the solutions with discrete symmetries. 

In section~\ref{ssec:4132}, a class of compactification of the Heterotic 
string theory is discussed;\footnote{Partial results of 
sections~\ref{ssec:4132} and \ref{ssec:dim5} have been obtained 
in \cite{TW1, TW2}.} technically, it is to assume that a vector 
bundle has an extension structure, and various low-energy degrees of 
freedom are identified with cohomologies of appropriate sub-bundles. 
% Minimum pre-requisite knowledge in the Heterotic string theory is 
% summarized in section~\ref{ssec:Het}.
We will see in this framework that holomorphicity controls mixings 
between {\em massless} states with different U(1) charges.
Thus, the selection rule based on U(1)-charge counting is applied 
for terms trilinear in massless states, and the absence of R-parity 
violating terms (\ref{eq:dim4}) can be guaranteed from the selection
rule.
1-loop amplitudes generate a bilinear R-parity
violating mass term $W \ni \mu_i \; L_i \; H_u$ with $\mu_i$ 
proportional to supersymmetry breaking (SUSY-breaking), and 
the tree-level contribution can be even smaller.
Thus, at the renormalizable level, this framework 
predicts an R parity violation only in the bilinear terms, which 
is known not to be terribly bad in phenomenology \cite{Barbieri, HS}.
We also find that all the dimension-5 operators that violate R parity 
are generated; the selection rule does not have a predictive power 
at non-renormalizable level.
With a theoretical framework controlling all aspects of R parity 
violation, we can discuss how key parameters of short-distance 
description control the coefficients of various R parity violating
operators. 

Section \ref{sec:phen} is devoted to phenomenology that is expected when 
both bilinear and dimension-5 R parity violating operators exist. 
Although small trilinear R-parity violating couplings can be 
generated in the framework of section~\ref{sec:frame}, it turns out 
that they are so small that they are rarely relevant to phenomenology. 
Because of negligibly small trilinear R-parity violation, most of 
phenomenological constraints discussed so far are easily satisfied,
considerably simplifying phenomenological study.
Remaining constraints from low-energy neutrino masses and  
washout of baryon/lepton asymmetry are briefly discussed 
in sections~\ref{ssec:nu} and \ref{ssec:washout}, respectively.
Constraints on R-parity violating decay of the lightest supersymmetry
particle (LSP) are reanalyzed in section~\ref{ssec:BBN}, where 
we exploit the latest understanding of impact of new physics 
on the Big--Bang Nucleosynthesis (BBN). 
Section~\ref{ssec:proton} is devoted to limits on R-parity violating 
couplings from nucleon decay amplitudes. Although trilinear R-parity 
violating couplings do not induce too rapid a proton decay, 
squark-exchange diagrams combining bilinear and dimension-5 R-parity 
violating operators still induce nucleon decay. 
We will obtain a big picture of allowed region of parameter space of 
bilinear--dimension-5 R-parity violation, and find that natural 
expectation of these parameters that follows from the framework in 
section~\ref{sec:frame} fits well within the allowed region. 

{\bf If time is limited}, it is best to read only
summary sections~\ref{ssec:summary-frame} and \ref{ssec:summary-phen}, 
skipping all the rest. 
It is also possible to read sections~\ref{sec:frame} and \ref{sec:phen} 
separately, as the materials in these sections do not require perfect 
understanding of the contents of the other.

%%%%%%%%%%%%%%%%%%%%%%%%%%%%%%%%%%%%%%%%%%%%%%%%%%%%%%%%%
\section{Theoretical Framework}
\label{sec:frame}
%%%%%%%%%%%%%%%%%%%%%%%%%%%%%%%%%%%%%%%%%%%%%%%%%%%%%%%%%

Section~\ref{ssec:Het} provides basic knowledge in Calabi--Yau 
compactification of the Heterotic $E_8 \times E_8'$ string theory 
that is necessary later in section~\ref{sec:frame}.
We include this mini-review just to make this paper self-contained. 
Thus, there should be no problem in skipping section~\ref{ssec:Het} 
and proceeding directly to \ref{ssec:4132}.
%  with those who have minimum knowledge in this field. 

%%%%%%%%%%%%%%%%%%%%%%%%%%%%%%%%%%%%%%%%%%%%%%%%%%%%%%%%%
\subsection{Mini-review on Heterotic String Compactification}
\label{ssec:Het} 
%%%%%%%%%%%%%%%%%%%%%%%%%%%%%%%%%%%%%%%%%%%%%%%%%%%%%%%%%

{\bf Origin of Gauge Fields and Matter Multiplets}

Effective theories on 3+1 dimensions can be obtained by compactifying 
the Heterotic $E_8 \times E'_8$ string theory on a six-dimensional 
manifold $X$. ${\cal N} = 1$ supersymmetry is preserved in a low-energy 
effective theory when $X$ is a Calabi--Yau 3-fold (and hence is 
a complex manifold). Local complex coordinates are denoted 
by $z^\alpha$ ($\alpha = 1,2,3$).

The gauge group $E_8 \times E'_8$ can be reduced virtually to 
any subgroups in low-energy effective theories 
by turning on non-trivial gauge-field background on $X$.
We will refer to the gauge field background\footnote{To be more precise,
$V$ means a vector bundle in the fundamental representation that 
a gauge field background in an $\SU(n)$ subgroup of $E_8$ defines.} as $V$.
The gauge group of the effective theory is $\SU(5)_{\rm GUT}$
for unified theories, if the background gauge field configuration 
is contained within $\SU(5)' \times E'_8 \subset E_8 \times E'_8$, 
where $\SU(5)'$ commutes with $\SU(5)_{\rm GUT}$ within $E_8$.

A super Yang--Mills multiplet on 9+1 dimensions
consists of a vector field $A_M$ and a gauge fermion $\Psi$. 
A vector field on 9+1 dimensions,
$A_M(x,y)$ ($M=0,\cdots,9$), is decomposed into 
$A_\mu(x, y)$ ($\mu=0,\cdots,3$) and  
$A_{\bar{\alpha}}(x,y)$ ($\alpha=1,2,3$); 
$A_{\alpha}(x,y)$ is a complex conjugate of
$A_{\bar{\alpha}}$ and is not an independent degree of freedom.
Here, $x^\mu$ ($\mu=0,1,2,3$) denote four Minkowski coordinates, 
and $y^m$ ($m=1, \cdots, 6$) (or simply $y$) six real local coordinates 
on $X$, equivalent of three complex coordinates 
$(z^\alpha, \bar{z}^{\bar{\alpha}})$  ($\alpha = 1,2,3$).

Those fields on 9+1 dimensions are decomposed into infinite 
degrees of freedom on 3+1 dimensions:
their Kaluza--Klein decompositions are given by
\begin{eqnarray}
 A_\mu(x,y) & = & \sum_I A_{I ; \mu}(x) \; \varphi_I(y), \\
 A_{\bar{\alpha}}(x,y) & = & g_{\rm YM}
   \sum_J \phi_J(x) \; \varphi_{J ; \bar{\alpha}}(y),
\label{eq:gYM-normalize}
\end{eqnarray}
using mode functions $\varphi_I(y)$ and $\varphi_{J ; \bar{\alpha}}(y)$:
$A_{I ; \mu}(x)$ and $\phi_J(x)$ 
are vector and complex scalar fields on 3+1 dimensions, respectively.
We factored out $g_{\rm YM}$ in (\ref{eq:gYM-normalize}) in setting 
the normalization of the mode functions $\varphi_{J; \bar{\alpha}}$;    
$g_{\rm YM}$ is the gauge coupling constant of the $\SU(5)_{\rm GUT}$ 
effective theory on 3+1 dimensions.
A gauge fermion on 9+1 dimensions, 
\begin{equation}
 \Psi(x,y) = 
(\Psi_\alpha^a(x,y), \Psi^{\dot{\alpha}}_a(x,y)),
\end{equation}
splits into $a=1,2,3$ part and $a=4$ part on a Calabi--Yau 3-fold $X$, 
where $\alpha$ and $\dot{\alpha}$ denote doublet indices
of left- and right-handed spinors of $\SO(3,1)$, and 
%superscripts and subscripts
$a$ label four different entries of spinor representations 
${\bf 4} + \bar{\bf 4}$ of local Lorentz symmetry $\SO(6)$.
The two parts have Kaluza--Klein mode decompositions separately:
\begin{eqnarray}
 (\Psi_\alpha^{a=4}, \; \Psi^{\dot{\alpha}}_{a=4})(x,y) & = & 
    \sum_I (\lambda_{I ; \alpha}(x) \; \chi_{I}(y), \;
            \bar{\lambda}^{\dot{\alpha}}(x) \; \tilde{\chi}_I(y)), \\
 (\Psi_\alpha^{a}, \; \Psi^{\dot{\alpha}}_{a})(x,y) & = & g_{\rm YM}
    \sum_J (\psi_{J ; \alpha}(x) \; \chi_{J}^{\; a}(y), \;
            \bar{\psi}^{\dot{\alpha}}_J(x) \; \chi_{J ; a}(y))  
 \qquad (a=1,2,3).
\end{eqnarray}
%
% Here, $y$ stands for six real coordinates on $X$, 
% the same thing as $(z,\bar{z})$.

Mode functions $e_{\bar{\alpha} a} \; \chi_J^{a}(y)$ (being multiplied
by a sechsbein\footnote{Hermitian metric $h_{\alpha \bar{\alpha}}$ of a 
K\"{a}hler manifold $X$ is given by 
$h_{\alpha \bar{\alpha}} = \sum_{a=1}^3 e_{\alpha}^a e_{\bar{\alpha} a}$.} 
$e_{\bar{\alpha} a}$) are proportional to those of 
the vector fields $\varphi_{J ; \bar{\alpha}}(y)$ in a supersymmetric 
compactification, and $A_{\bar{\alpha}}(x,y)$ and $\Psi_\alpha^a(x,y)$
are grouped into a Kaluza--Klein tower of chiral multiplets 
$\Phi_I(x,\theta, \bar{\theta})$:  
\begin{equation}
 A_{\bar{\alpha}}(x,z,\bar{z},\theta,\bar{\theta}) d\bar{z}^{\bar{\alpha}} \equiv 
  \sum_J \left(\phi_J + \theta \psi_J + \cdots \right)(x) \; 
         \varphi_{J ; \bar{\alpha}}(y)d\bar{z}^{\bar{\alpha}} 
  \equiv \sum_J \Phi_J \; \varphi_{J ; \bar{\alpha}}
  d\bar{z}^{\bar{\alpha}}.
\label{eq:KK-chiral}
\end{equation}
The remaining $A_\mu$ ($\mu=0,\cdots, 3$) part and $\Psi_\alpha^{a=4}$ 
part are also grouped into a tower of vector multiplets 
$V_I(x,\theta,\bar{\theta})$:
\begin{equation}
 V(x,z,\bar{z},\theta,\bar{\theta}) \equiv 
  \sum_I \left( \theta \sigma^\mu \bar{\theta} A_{I ; \mu} 
                + \bar{\theta}^2 \theta \lambda_I + \cdots \right)(x) \;
         \varphi_I(y) 
  \equiv \sum_I V_I \; \varphi_I.
\label{eq:KK-vector}
\end{equation}

As we assume that gauge-field background is non-vanishing, mode functions 
$\varphi_{J; \bar{\alpha}}(y)$ and $\varphi_I(y)$ are not the same  
everywhere in $E_8$. The vector field and gauge fermion of the $E_8$ 
super Yang--Mills theory are in the adjoint representation of $E_8$. 
They split into irreducible components
\begin{equation}
{\bf 248} \rightarrow ({\bf adj.},{\bf 1}) + ({\bf 1},{\bf adj.}) 
+ \left[ ({\bf 5},{\bf 10}) + ({\bf 10},\bar{\bf 5}) \right] + {\rm
h.c.}
\label{eq:55decomp}
\end{equation}
of $\SU(5)' \times \SU(5)_{\rm GUT} \subset E_8$.
Each irreducible component has its own Kaluza--Klein decomposition
(\ref{eq:KK-chiral}, \ref{eq:KK-vector}):
Mode functions in the $(R', R)$-irreducible component are determined by mode equations
on the gauge-field background in the $R'$ representation,
and hence the spectrum and decomposition of one irreducible component 
are different from those of another.
Therefore we use such notations as
$(R', R)_J$ or $R_J$, instead of $\Phi_J$ for chiral multiplets.

On the $\SU(5)'$ gauge-field background,
massless vector multiplets are found only in the $({\bf 1}, {\bf adj.})$ component.
Massless chiral multiplets in the $\SU(5)_{\rm GUT}$-$R$ representation
% that appear at energy scale far below the Kaluza--Klein scale
are in one to one correspondence with the zero modes
$\varphi_{J; \; \bar{\alpha}}(y)$ in the $(R',R)$ irreducible component.
Difference between the number of massless chiral 
multiplets---net chirality---in a Hermitian conjugate pair of 
irreducible components, $(R',R)$--$(\overline{R'}, \overline{R})$,
is determined only by topology of the background gauge-field configuration $V$.
Phenomenological request is to find a topology of $(X,V)$
so that the net chirality in the
$({\bf 5}, {\bf 10})$--$(\bar{\bf 5},\overline{\bf 10})$
sector is three.
It then follows that the net chirality in the
$({\bf 10}, \bar{\bf 5})$--$(\overline{\bf 10},{\bf 5})$
sector also becomes three;
low-enrgy effective theories cannot be anomalous if they are obtained 
by compactifying an anomaly free theory in higher dimensions.

{\bf Vector-like Massless Pair}

Topology of $(X,V)$ determines the net chirality, but the number of 
massless chiral multiplets of each irreducible component can vary 
for a continuous deformation of gauge field background: 
% The numbers of massless chiral multiplets of 
% $\SU(5)_{\rm GUT}$-$\bar{\bf 5}$ 
% and $\SU(5)_{\rm GUT}$-${\bf 5}$ can still change under continuous 
% deformation of gauge field background; 
There can be $3+m$ massless chiral multiplets
in the $\bar{\bf 5}$ representation and $m$ in ${\bf 5}$;
$m$ can change while keeping the net chirality $(3+m)-m=3$.
% $m$ can change along the moduli space of gauge-field background.
For a given topology of $(X,V)$, however, there is still a minimum 
number of $m$ for the $\bar{\bf 5}$--${\bf 5}$ sector,  
% for a given topology of $(X, V)$, 
and it is not necessarily zero.\footnote{The same thing can happen 
for the multiplets in the $\SU(5)_{\rm GUT}$-${\bf 10}$ and 
$\overline{\bf 10}$ representations, in principle.}
See \cite{Penn5-2} for an explicit model, where $m \geq 34$ for the 
$\bar{\bf 5}$--${\bf 5}$ sector, while there can be no extra massless 
vector-like pair in the ${\bf 10}$--$\overline{\bf 10}$ sector. 
Similar examples can be found in the literature studying spectra on 
orbifold compactifications.
So far, top down principles have been unable to determine the topology 
of $(X, V)$, and hence the minimum number of $m$'s. 

There are a few bottom-up constraints on the number of vector-like 
massless pairs. First of all, there should be at least one vector-like 
pair in the doublet part of the $\bar{\bf 5}$--${\bf 5}$ 
sector.\footnote{
$\SU(5)_{\rm GUT}$ symmetry can be broken either by a
Wilson line \cite{WittenWilson} or by a line bundle \cite{Munich}. 
Spectra and mode functions in a given $\SU(5)_{\rm GUT}$
representation can be different for different irreducible components 
of the MSSM gauge group. Hence $m$ can be 1 for the doublet part, 
while $m=0$ for triplets.
We maintain $\SU(5)_{\rm GUT}$ notations in many places in this paper, 
mainly to avoid cluttered equations, at the cost of sacrificing 
rigorousness and unambiguousness. \label{fn:doublet-triplet}}
They are identified with the two Higgs doublets of the minimal 
supersymmetric standard model (MSSM).
If there were too many $\SU(5)_{\rm GUT}$-charged vector-like pairs, 
however, they would contribute to beta functions of the MSSM gauge 
coupling constants. Perturbative gauge coupling unification, 
one of the most important motivations of low-energy supersymmetry, 
would not be maintained any more. As long as the ``massless'' pairs 
have masses of the order of SUSY-breaking scale or higher,\footnote{
Since all the argument based on Calabi--Yau compactification 
preserves ${\cal N} = 1$ supersymmetry, a vector-like pair of 
multiplets that are massless at a supersymmetric limit may have masses 
when the supersymmetry is weakly broken.} they have 
not showed up in experiments, and moderate number of them are 
tolerable in phenomenology.\footnote{Constraints such as FCNC 
and nucleon decay depend on couplings that those extra particles have 
with the chiral multiplets in the MSSM. Although such constraints can be 
very severe, we do not discuss them in this article.} 

Superpotential of the Heterotic theory is given by \cite{Het-super}
\begin{equation}
W = c M_G^3 \int_X \Omega \wedge 
        \tr {}_{\rm adj.} \left(A dA - \frac{2}{3}i AAA \right) ,
\label{eq:Het-super}
\end{equation}
which is valid for all the Kaluza--Klein modes $\Phi_{J}$ in
(\ref{eq:KK-chiral}) \cite{AGW}.
Here, $c$ is a numerical constant of order unity,\footnote{Combined 
with a K\"{a}hler potential, 
\begin{equation}
\frac{K}{M_G^2} = - \ln \left(S + S^\dagger\right)
          - \ln \left(\int_X (T + T^\dagger)^3 \right) 
          - \ln \left( \int_X \Omega \wedge \overline{\Omega} \right), 
\label{eq:Het-Kahler}
\end{equation}
this superpotential reproduces a part of the gaugino kinetic term 
(among other things) with the right dependence on $\alpha'$ and $g_s$. } 
$M_G \simeq 2.4 \times 10^{18}$ GeV and $\Omega$ is a dimensionless 
holomorphic 3-form of a Calabi--Yau 3-fold $X$.
We treat $\Omega$ purely as a background, as we will only discuss 
what happens within the visible sector $E_8$ in this article. 

Kaluza--Klein masses of infinite chiral multiplets originate from $d - 2 i \vev{A}$,
second derivative of (\ref{eq:Het-super}) with respect to $A$. 
Writing the superpotential (\ref{eq:Het-super}) fully in terms of 
D = 4 chiral multiplets, we have bilinear (Kaluza--Klein mass) and 
trilinear terms:
\begin{eqnarray}
 W & \ni &  M_{\bf 5} \; ({\bf 10}, \bar{\bf 5})_* \;
                         (\overline{\bf 10}, {\bf 5})_*   
     + M_{\bf 10} \;   ({\bf 5},{\bf 10})_* \;
                       (\bar{\bf 5},\overline{\bf 10})_* 
     + M_{\bf 1} \; ({\bf adj.}, {\bf 1})_* \; ({\bf adj.}, {\bf 1})_*,  
\label{eq:vctlike-mass} \\
   & + &  y^u \; ({\bf 5},{\bf 10}) \; ({\bf 5},{\bf 10}) \;
  (\overline{\bf 10},{\bf 5})
    + y^d \; ({\bf 10},\bar{\bf 5}) \; ({\bf 5},{\bf 10}) \;
  ({\bf 10},\bar{\bf 5}),   \label{eq:udeYukawa} \\
   &  + & y^\nu \; ({\bf 10},\bar{\bf 5}) \; ({\bf adj.},{\bf 1}) \;
  (\overline{\bf 10},{\bf 5})
    + y'^{\nu} \; (\bar{\bf 5}, \overline{\bf 10}) \;
      ({\bf adj.}, {\bf 1}) \; ({\bf 5}, {\bf 10}), 
 \label{eq:nuYukawa} \\
 & + & y''^{\nu} \; ({\bf adj.}, {\bf 1}) \; ({\bf adj.}, {\bf 1}) \; 
     ({\bf adj.}, {\bf 1}).
 \label{eq:SSS}
\end{eqnarray}
Here, multiplets $(R',R)_*$ represent infinitely many massive chiral 
multiplets $(R', R)_I$ in the $R$ representation of $\SU(5)_{\rm GUT}$; 
we will use $(R', R)_0$ when we refer specifically to massless modes.
Those without any $_*$ or $_0$ stand for both. 
Chiral multiplets ({\bf adj.}, {\bf 1}) are $\SU(5)_{\rm GUT}$ singlets,
and ({\bf adj.}, {\bf 1})$_0$ correspond to gauge-field moduli.\footnote{
  Separation between the mass terms and the last two lines is not well-defined; 
  this is because continuous deformation of a gauge field background 
  along its moduli space corresponds to changing vev's 
  of massless chiral multiplets ({\bf adj.}, {\bf 1})$_0$, 
  and extra mass terms arise from the last two lines.} 
Mass matrices such as $M_{\bf 5}$ and $M_{\bf 10}$ are of infinite rank. 
Rank of $M_{\bf 5}$ may be reduced at certain subset of gauge field 
moduli space, where extra pairs of chiral multiplets in the 
$\SU(5)_{\rm GUT}$-${\bar{\bf 5}}+{\bf 5}$ representations 
(or possibly in the ${\bf 10}+\overline{\bf 10}$) 
are in the low-energy spectrum \cite{Penn5-2}. 

{\bf Yukawa Interactions}

The last three lines, (\ref{eq:udeYukawa}--\ref{eq:SSS}), 
are trilinear interactions involving massless and/or massive 
chiral multiplets, with coupling constants given by overlap 
integration of relevant mode functions. 
Trilinear interactions involving only massless modes are 
directly relevant to low-energy physics.
The first term of (\ref{eq:udeYukawa}) contains up-type Yukawa 
couplings $W \ni y^u \;{\bf 10}\;{\bf 10}\;H({\bf 5})$, and 
the second term down-type/charged-lepton Yukawa couplings 
$W \ni y^d \;\bar{\bf 5}\;{\bf 10}\;\bar{H}(\bar{\bf 5})$.
Neutrino Yukawa couplings 
$W \ni y^\nu \;\bar{\bf 5}\;\overline{N}\;H({\bf 5})$
can only be found in the first term of (\ref{eq:nuYukawa}), and 
hence the chiral multiplets for right-handed neutrinos are 
identified with (a subset of) gauge-field moduli,\footnote{Strictly
speaking, right-handed neutrinos do not have to be identified 
with zero modes.  See also the 3+2 model to be discussed 
in section~\ref{ssec:4132}.} ({\bf adj.},~{\bf 1})$_0$ \cite{WittenSU(3)}.  
% %
% \begin{equation}
%  W = y^u {\bf 10}. {\bf 10}. H({\bf 5}) +
%      y^{d/e} \bar{\bf 5} \cdot {\bf 10} \cdot \bar{H}(\bar{\bf 5}) + 
%      y^\nu \bar{\bf 5} \cdot {\bf 1} \cdot H({\bf 5}),
% \end{equation}
% %

As we have introduced no distinction between 
$H_d \subset \bar{H}(\bar{\bf 5})$ and $L \subset \bar{\bf 5}$'s, 
however, the second term in (\ref{eq:udeYukawa}) generically contains 
the trilinear R-parity violating operators (\ref{eq:dim4}) as well. 
Thus, a generic $(X,V)$ is not acceptable phenomenologically 
because of the dimension-4 proton decay problem.

{\bf R parity}

The most popular solution to this problem is to impose an R parity.
In terms of compactification of a super Yang--Mills theory on 9+1 dimensions,
this is to assume a $\Z_2$ symmetry in ($X$, $V$).
% so that the massless modes in $({\bf 5},{\bf 10})$
% [in $({\bf 10},\bar{\bf 5})$, $({\bf adj.},{\bf 1})$]
% split into those even and odd under the $\Z_2$ transformation.
%%%%%%%%%%%%%%%%%%%%%%%%%%%%%%%%%%%%%%%%%%%%%%%%%%%%%%%%%%%%%%%%
\begin{table}
\begin{center}
\begin{tabular}{c|c|c|c}
chiral mult. & repr. & \# of zero modes & zero modes in low energy \\
\hline
\rule[0pt]{0pt}{12pt}
${\bf 10}_0$, ${\bf 10}_*$ & $({\bf 5}, \wedge^2 {\bf 5})^-$ &
  3 & $Q_i,\bar{U}_i,\bar{E}_i$ $(i=1,2,3)$ \\
${{\bf 10}^c}_*$ & $(\bar{\bf 5}, \wedge^2 \bar{\bf 5})^-$
  & 0 & - \\
\hline 
${{\bf 10}'}_*$ & $({\bf 5}, \wedge^2 {\bf 5})^+$ & 0 & - \\
${{\bf 10}^{'c}}_*$ & $(\bar{\bf 5}, \wedge^2 \bar{\bf 5})^+$ 
  & 0 & - \\
\hline 
\rule[0pt]{0pt}{12pt}
$\bar{H}(\bar{\bf 5})_0$, ${\bar{H}(\bar{\bf 5})}_*$ & 
 $(\wedge^2 {\bf 5}, \bar{\bf 5})^+$ & $m=0 (\bar{\bf 3}),1({\bf 2})$ 
   & $H_d$ \\
$H({\bf 5})_0$, ${H({\bf 5})}_*$ &
 $(\wedge^2 \bar{\bf 5}, {\bf 5})^+$ & $m=0 ({\bf 3}),1({\bf 2})$ & $H_u$ \\
\hline
\rule[0pt]{0pt}{12pt}
$\bar{\bf 5}_0$, ${\bar{\bf 5}}_*$ & 
 $(\wedge^2 {\bf 5}, \bar{\bf 5})^-$ & 3 & $\bar{D}_i, L_i$ $(i=1,2,3)$ \\
${\bar{\bf 5}^c}_*$ &
 $(\wedge^2 \bar{\bf 5}, {\bf 5})^-$ & 0 & - \\
\hline 
\rule[0pt]{0pt}{12pt}
$\overline{N}_0$, ${\overline{N}}_*$ & ({\bf adj.},{\bf 1})$^-$ & 
 some & RH neutrinos \\
 & ({\bf adj.}, {\bf 1})$^+$ & 0 & - 
\end{tabular}
\caption{
  List of chiral multiplets in scenarios with an R parity.
  In the third column, $m=0$ for the $\SU(3)_C$-triplet parts
  and $m=1$ for the $\SU(2)_L$-doublet parts in the third row of this table. 
  We listed just the minimum number of chiral multiplets 
  required in effective theory on 3+1 dimensions, 
  ignoring a vector-like (almost) massless pair 
  that can exist without phenomenological problems.
  \label{tab:reprRparity}
}
\end{center}
\end{table}
%%%%%%%%%%%%%%%%%%%%%%%%%%%%%%%%%%%%%%%%%%%%%%%%%%%%%%%%%%%%%%%%
% Furthermore, all the three massless chiral multiplets in the 
% $\SU(5)_{\rm GUT}$-{\bf 10} representation become odd under 
% R parity if the topology of $(X, V)$ is arranged so that 
% all the zero modes turn out to be $\Z_2$-odd.
% Assignment of R parity on the chiral multiplets of the MSSM
% (three ${\bf 10}$'s, three $\bar{\bf 5}$'s, some $\overline{N}$'s,
%  one $H_u$ and one $H_d$)
% are translated into the number of $\Z_2$-even or -odd zero modes
% of various irreducible components.
% See the third column of Table~\ref{tab:reprRparity}. 
% Although vev's of $\SU(5)_{\rm GUT}$-singlet chiral multiplets
% $({\bf adj.}, {\bf 1})_0$ do not break $\SU(5)_{\rm GUT}$ symmetry,
% those of the $({\bf adj.}, {\bf 1})_0^-$ break the $\Z_2$ symmetry.
We assign odd R-parity for three chiral multiplets
${\bf 10}_i = (Q_i ,\bar{U}_i,\bar{E}_i)$ ($i=1,2,3$) in bottom-up model building,
but it corresponds to assuming that there are three massless states
in the $({\bf 5}, {\bf 10})^-$ irreducible component in this context.
One further need to assume that there are none
in the other irreducible components $({\bf 5}, {\bf 10})^+$,
$(\bar{\bf 5},\overline{\bf 10})^+$ and $(\bar{\bf 5},\overline{\bf 10})^-$.
Similar assumptions have to be made for irreducible components
that are in the ${\bf 5}$ and $\bar{\bf 5}$ representations of $\SU(5)_{\rm GUT}$.
See Table~\ref{tab:reprRparity} for more details.

Both moduli $({\bf adj.}, {\bf 1})^+_0$ and $({\bf adj.}, {\bf 1})^-_0$ 
may have non-vanishing vev's without breaking $\SU(5)_{\rm GUT}$ symmetry, 
but non-vanishing vev's of the latter break the $\Z_2$ symmetry. 
Thus, none of $({\bf adj.}, {\bf 1})_0^-$'s
should develop a vev in order to maintain an unbroken R-parity.
% This is what it takes to impose a matter parity (or R parity).

With such an unbroken $\Z_2$ symmetry of ($X$, $V$),
down-type/charged-lepton Yukawa couplings
\begin{equation}
 W \ni y^{d} \; ({\bf 10},\bar{\bf 5})^{-}_0 \; ({\bf 5},{\bf 10})^{-}_0 
   \; ({\bf 10}, \bar{\bf 5})^{+}_0
\end{equation}
do not vanish, yet the $\Z_2$-odd operators 
\begin{equation}
 W \ni \lambda ({\bf 10},\bar{\bf 5})^{-}_0 \; ({\bf 5},{\bf 10})^{-}_0 
   \; ({\bf 10}, \bar{\bf 5})^{-}_0
\end{equation}
are absent because the overlap integrations for the couplings vanish.
% because of the $\Z_2$-odd nature.

%%%%%%%%%%%%%%%%%%%%%%%%%%%%%%%%%%%%%%%%%%%%%%%%%%%%%%%%%%%%%
\subsection{4+1 Model and 3+2 Model}
\label{ssec:4132}
%%%%%%%%%%%%%%%%%%%%%%%%%%%%%%%%%%%%%%%%%%%%%%%%%%%%%%%%%%%%%

Gauge field configuration $V$ in $\SU(5)' \subset E_8$
on a Calabi--Yau 3-fold $X$ should not be generic, since 
generic configuration gives rise to the R-parity violating 
trilinear operators (\ref{eq:dim4}), leading to too rapid 
proton decay.
An R-parity preserving configuration ($X$, $V$) with a $\Z_2$ symmetry
is an example of non-generic cases.

Discrete symmetries other than R parity have also been discussed
in the literature as solutions to the proton decay problem.
Phenomenological consequences of these theory can be different 
from those of R-parity preserving ones; 
the LSP may not be stable, for example.
Despite the difference in phenomenology,
all the solutions based on discrete symmetries---whether 
preserving R parity or not---are quite similar in philosophy. 
Enhanced discrete (or continuous) symmetries are left unbroken 
only at special points (or subsets) of moduli space ($X$, $V$).
Therefore, solutions based on discrete symmetries are based on a belief
that some dynamics that we do not know today
will eventually select out vacua with enhanced symmetries,
and lift all the other part of moduli space.
Certainly this belief is not without reason.
CP symmetry does not have to be preserved in QCD,
but once we knew that QCD instanton effects generate a potential of axion, we understood that CP-preserving $\theta_{\rm eff.}=0$
is the minimum of axion potential. So, who can say that history does not repeat itself?

In this article, however, we neither resort to this belief,
nor assume an unknown dynamics for vacuum selection.
We will present an alternative solution to the dimension-4 proton decay problem,
which holds at generic points in the gauge-field moduli space.
This subsection describes a class of compactification that solves the problem.
Key ideas were already written in \cite{TW1, TW2},
but a few important clarifications are newly added in this article.
We then move on in the rest of section~\ref{sec:frame} 
to discuss what kind of operators are to be expected 
in low-energy effective theories on 3+1 dimensions.

{\bf 4+1 Model}

Let us first suppose that a gauge field background is restricted
to an $\SU(4)\times \U(1)_\chi$ subgroup of $\subset \SU(5)' \subset E_8$.
Then, the gauge symmetry is $(\SU(5)_{\rm GUT} \times \U(1)_\chi) / \Z_5$
in an effective theory.
$\U(1)_\chi / \Z_5$ contains matter parity
as a subgroup, $\Z_{10} / \Z_5 \simeq \Z_2$.
Although the matter parity is not broken at this moment,
we will see later that the restriction on the gauge field background
is relaxed and the matter parity is broken,
yet the trilinear matter parity violating operators (\ref{eq:dim4}) are absent.

Each irreducible component of the $E_8$-{\bf adj.} representation in 
(\ref{eq:55decomp}) is further split up as the group of gauge field
background (called structure group) is reduced from $\SU(5)'$ to 
$\SU(4) \times \U(1)_\chi$:
\begin{eqnarray}
 ({\bf 5},{\bf 10}) & \rightarrow &
    ({\bf 4},{\bf 10})^{-1} + ({\bf 1},{\bf 10})^{+4}, \\
 (\wedge^2 {\bf 5},\bar{\bf 5}) & \rightarrow & 
  (\wedge^2 {\bf 4},\bar{\bf 5})^{-2} + ({\bf 4},\bar{\bf 5})^{+3}, \\
 ({\bf adj.}, {\bf 1}) & \rightarrow  & ({\bf adj.},{\bf 1})^0 + 
  ({\bf 4},{\bf 1})^{-5} + (\bar{\bf 4},{\bf 1})^{+5} + ({\bf 1}, {\bf 1})^0.
\end{eqnarray}
Massless modes are identified with various chiral multiplets
of the MSSM as in Table~\ref{tab:repr41} in a compactification
with this class of gauge field background.\footnote{Here is our 
naming rule of various chiral multiplets. Since anti-chiral 
multiplets containing right-handed quarks and leptons are 
denoted by $\bar{U}^\dagger, \bar{D}^\dagger$ and $\bar{E}^\dagger$, 
we save $\overline{N}^\dagger$ for right-handed neutrinos.
Chiral multiplets $\Psi$ and $\Psi^c$ ($\Psi = {\bf 10}, 
{\bf 10}', \bar{\bf 5},\overline{N}$, for example) 
arise from a Hermitian conjugate pair of irreducible components 
$(R',R)$ and $(\overline{R'},\overline{R})$ in $E_8$. 
The same rule is also applied to the 3+2 model.
In the 4+1 model, therefore, $\bar{H}(\bar{\bf 5})=H({\bf 5})^c$, 
but this is not the case in the 3+2 model.
Since gauge-field moduli ({\bf adj.},{\bf 1})$^0$ is a vector-like representation 
of $\SU(4) \times \SU(5)_{\rm GUT} \times \U(1)_\chi$, 
there is no distinction between $\Phi$ and $\Phi^c$.}
%%%%%%%%%%%%%%%%%%%%%%%%%%%%%%%%%%%%%%%%%%%%%%%%%%%%%%%%%%%%
\begin{table}[t]
\begin{center}
\begin{tabular}{c|c|c|c}
  fields & representations & number of zero-modes & zero-modes at low energy \\
  \hline
  \rule[0pt]{0pt}{12pt}
  ${\bf 10}_0$, ${\bf 10}_*$ & $({\bf 4}, \wedge^2 {\bf 5})^{-1}$ & 3
& $Q_i, \bar{U}_i, \bar{E}_i$ ($i=1,2,3$) \\
  ${\bf 10}^c_*$ & $(\bar{\bf 4}, \wedge^2 \bar{\bf 5})^{+1}$ & 0 & $-$ \\
  \hline
  ${{\bf 10}'}_*$ & $({\bf 1}, \wedge^2 {\bf 5})^{+4}$ & 0 & $-$ \\
  ${{\bf 10}'}^c_*$ & $({\bf 1}, \wedge^2 \bar{\bf 5})^{-4}$ & 0 & $-$ \\
  \hline
  \rule[0pt]{0pt}{12pt}
  $\bar{H}_0(\bar{\bf 5})$, ${\bar{H}}_*(\bar{\bf 5})$
& $(\wedge^2 {\bf 4}, \bar{\bf 5})^{-2}$& 
   1(${\bf 2}$) / 0($\bar{\bf  3}$)   & $H_d$ \\
  $H_0({\bf 5})$, ${H}_*({\bf 5})$
& $(\wedge^2 \bar{\bf 4}, {\bf 5})^{+2}$& 1(${\bf 2}$) / 0({\bf 3}) & $H_u$ \\
  \hline
  \rule[0pt]{0pt}{12pt}
  $\bar{\bf 5}_0$, ${\bar{\bf 5}}_*$ & $({\bf 4}, \bar{\bf 5})^{+3}$
& 3 & $\bar{D}_i, L_i$ ($i=1,2,3$) \\
  ${\bar{\bf 5}}^c_*$ & $(\bar{\bf 4}, {\bf 5})^{-3}$ & 0 & $-$ \\
  \hline
  \rule[0pt]{0pt}{12pt}
  $\overline{N}_0$, ${\overline{N}}_*$ & $({\bf 4}, {\bf 1})^{-5}$
& some & heavy RH neutrinos \\
  $\overline{N}^c_0$, ${\overline{N}}^c_*$
& $(\bar{\bf 4}, {\bf 1})^{+5}$ & $1 \leq$ & absorbed by $\U(1)_\chi$ \\
  \hline
  $\Phi_0$, ${\Phi}_* $ &  $({\bf adj.}, {\bf 1})^0$ & $-$ & $-$
\end{tabular}
\caption{\label{tab:repr41}
  List of chiral multiplets in the 4+1 model. 
  Those with a subscript $_0$ are zero modes of Kaluza--Klein decomposition,
  and those with $_*$ infinitely many massive modes.
  Chiral multiplets without either one of the subscripts $_0$ or $_*$
  that appear in the text represent both of them.
  The second column shows how the chiral multiplets transform under
  $\SU(5)_{\rm GUT} \supset \SU(3)_C \times \SU(2)_L \times \U(1)_Y$
  as well as underlying broken symmetries $\SU(4) \times \U(1)_\chi$ in this model.
  Only the minimum number of chiral multiplets are shown in the third column.
  The slashes in the third column represent the doublet-triplet splitting.
  For more, see the caption of Table~\ref{tab:reprRparity} and footnote \ref{fn:noextra}.
}
\end{center}
\end{table}
%%%%%%%%%%%%%%%%%%%%%%%%%%%%%%%%%%%%%%%%%%%%%%%%%%%%%%%%%%%%%%%%%%%%%%%%%%%
% We use subscripts $_*$ to distinguish massive modes from massless modes
% with subscripts $_0$, as in Table~\ref{tab:reprRparity}.
$H_u \subset H({\bf 5})$ and $H_d \subset \bar{H}(\bar{\bf 5})$ are 
completely vector like, as in the scenario with an R parity. 
Massive modes---those that acquire masses through 
$W \ni \tr (A dA - 2i A \vev{A} A )$---have subscripts $_*$, and massless 
modes have $_0$, like in Table~\ref{tab:reprRparity}.
Topology of geometry and gauge field configuration on it, ($X$, $V$), 
should be chosen so that the right number of massless modes 
are obtained. 
%
% There are infinite chiral multiplets ${\bf 10}$ and infinite 
% ${\bf 10}^c$, but there are three more ${\bf 10}$'s corresponding 
% to the three generations of ${\bf 10}=(Q,\bar{U},\bar{E})$, 
% which are notated without underlines.
% Similarly, there are three more $\bar{\bf 5}$ chiral multiplets than 
% $\bar{\bf 5}^c$'s. 
% There is no net chirality between $H({\bf 5}) = H$ and 
% $\bar{H}(\bar{\bf 5})=H^c$, but the mass matrix $M_H$ should be such 
% that a pair of Higgs doublets remain in the low-energy spectrum.
%
% Thus, one needs to assume that they remain in the low-energy spectrum 
% either because gauge-field moduli are stabilized so that 
% the mass parameter of $H_u H_d$ almost vanishes, or because 
% the topology of $(X,V)$ is such that the minimum number of massless
% vector-like pair of chiral multiplets is $m=1$. [Refer to some papers]

Superpotential is obtained by rewriting (\ref{eq:Het-super}) in 
terms of chiral multiplets in the irreducible components:\footnote{
An equation (\ref{eq:Yukawa41-u}--\ref{eq:Yukawa41-nuN}) sets up  
notation for trilinear couplings in the 4+1 model. When we just 
simply use $y^u$ and $y^d$ in the text, however, they mean some of 
$y^u_{(1)\mbox{-}(4)}$ and $y^d_{(1)\mbox{-}(4)}$, respectively.
Similarly, $y^\nu$ stands for some of $y^\nu_{(1)\mbox{-}(4)}$,
$y^{\prime\nu}_{(1)\mbox{-}(4)}$ and $y^{\prime\prime\nu}_{(1)\mbox{-}(2)}$.} 
\begin{eqnarray}
 W & = & y^u_{(1)} {\bf 10}\;{\bf 10}\;H({\bf 5})
 + y^u_{(2)} {\bf 10}\;{{\bf 10}'}_*\;\bar{\bf 5}^c_*
  \label{eq:Yukawa41-u}\\
   & + & y^u_{(3)} \; {{\bf 10}}^c_*\;{{\bf 10}}^c_*\;\bar{H} + 
  y^u_{(4)} \; {\bf 10}^c_* \; {{\bf 10}'}^c_* \;
  \bar{\bf 5} \label{eq:Yukawa41-u-conj} \\
 & + & y^d_{(1)} \; \bar{\bf 5} \; {\bf 10} \; \bar{H}(\bar{\bf 5}) +
  y^d_{(2)} \; \bar{H}(\bar{\bf 5}) \; {{\bf 10}'}_* \; 
  \bar{H}(\bar{\bf 5}) \label{eq:Yukawa41-de} \\
 & + & y^d_{(3)} \; {\bar{\bf 5}}^c_* \; {{\bf 10}}^c_* \; H({\bf 5}) + 
  y^d_{(4)} \; H({\bf 5}) \; {{\bf 10}'}^c_* \; 
   H({\bf 5}) \label{eq:Yukawa41-de-conj} \\
 & + & y^\nu_{(1)} \; \overline{N} \; \bar{\bf 5} \;H({\bf 5}) + 
  y^\nu_{(2)} \; {\overline{N}}^c_* \; {\bar{\bf 5}}^c_* \; \bar{H}
 + y^\nu_{(3)} \; \bar{H}(\bar{\bf 5}) \; \Phi \; H({\bf 5}) 
 + y^\nu_{(4)} \; \bar{\bf 5} \; \Phi \; \bar{\bf 5}^c_*
   \label{eq:Yukawa41-nu} \\ 
 & + & y^{'\nu}_{(1)} \; \overline{N} \; {{\bf 10}'}_*  \; {\bf 10}^c_*
   +   y^{'\nu}_{(2)} \; {\overline{N}}^c_* \; {{\bf 10}'}^c_*  \; {\bf 10}
   +   y^{'\nu}_{(3)} \; {\bf 10} \; \Phi \; {\bf 10}^c_*
   +   y^{'\nu}_{(4)} \; {\bf 10}'_* \; \Phi \; {\bf 10}^{'c}_* 
   \label{eq:Yukawa41-nu10} \\ 
 & + & y^{''\nu}_{(1)} \bar{N} \; \Phi \; \bar{N}^c 
   +   y^{''\nu}_{(2)} \Phi \; \Phi \; \Phi
   \label{eq:Yukawa41-nuN} \\
 & + & M_{\bf 10} {{\bf 10}}_* {{\bf 10}}^c_* 
   +   M_{{\bf 10}'} {{\bf 10}'}_* {{\bf 10}'}^c_*
   +   M_H {H}_* {\bar{H}}_*
   +   M_{\bf 5} {\bar{\bf 5}}_* {\bar{\bf 5}}^c_* 
   +   M_N {\overline{N}}_* {\overline{N}}^c_*
   +   M_\Phi \Phi_* \Phi_*.
   \label{eq:mass41}
\end{eqnarray}
%
% If all the interactions involving heavy multiplets
% are just simply dropped from the superpotential above,
% we recover the Yukawa couplings of quarks and leptons.
Dropping all the interactions involving heavy multiplets,
the Yukawa couplings of quarks and leptons remain in the superpotential above.
On the other hand, trilinear R-parity violating operators (\ref{eq:dim4})
are not found anywhere in 
(\ref{eq:Yukawa41-u}--\ref{eq:Yukawa41-nuN}), because 
\begin{equation}
 ({\bf 4},\bar{\bf 5})^{+3} \otimes ({\bf 4},\wedge^2 {\bf 5})^{-1} 
 \otimes  ({\bf 4},\bar{\bf 5})^{+3}
\end{equation}
does not contain a singlet of $\SU(4) \times \SU(5)_{\rm GUT} \times 
\U(1)_\chi$.

The Fayet--Iliopoulos parameter of $\U(1)_\chi$ symmetry is given by \cite{Munich}
\begin{equation}
\xi_\chi = \frac{10 M_G^2}{32 \pi^2} \left[ 
   \frac{2\pi l_s^2}{{\rm vol}(X)} \int_X c_1(L) \wedge J \wedge J 
    - \frac{g_{\rm YM}^2 e^{2\tilde{\phi}_4}}{2} 
       \int_X c_1(L) \left(c_2(V) - \frac{1}{2}c_2(TX) \right)
                                     \right], 
\label{eq:FY-chi}
\end{equation}
which depends on K\"{a}hler moduli $J$ (volume moduli) in the 
tree-level contribution (the first term) and dilaton expectation value 
at the 1-loop level (the second term: a piece well-known since 1980's). 
$l_s = 2\pi \sqrt{\alpha'}$ is the string length. 
% , and $g_{\rm YM}$ 
% is the $\SU(5)_{\rm GUT}$ gauge coupling constant in the effective
% theory on 3+1 dimensions. 
See \cite{Munich, TW1} for details of the convention. 
As long as moduli fields $J$ and $\widetilde{\phi}_4$ are stabilized 
by potential other than the D-term of $\U(1)_\chi$ symmetry, 
there is no reason to believe that the tree and 1-loop contributions 
cancel one another. Thus, $\xi_\chi$ is not expected to vanish. 
If it is negative, then $+5|\overline{N}^c|^2$ in the D-term potential 
may absorb $\xi_\chi$ to restore supersymmetry.
This scenario is called 4+1 model.\footnote{
  The 4+1 model corresponds
  to turning on a rank-5 vector bundle $V$ that is given by an extension 
  of a rank-4 vector bundle $U_4$ by a line bundle $L$:
  \begin{equation}
    0 \rightarrow L \rightarrow V \rightarrow U_4 \rightarrow 0.
  \end{equation}
  $L$ is a sub-bundle of $V$, and $L \otimes V$ and 
  $\wedge^2 \overline{U_4}$ are sub-bundles of 
  $\wedge^2 V$ and $\wedge^2 \overline{V}$, respectively.
  \label{fn:subspace41}
}
The matter parity $\Z_{10} / \Z_5 \simeq \Z_2$
is broken by an expectation value of $\overline{N}^c$.

An order-of-magnitude estimate of $\U(1)_\chi$ breaking 
vev $\vev{\overline{N}^c}$ was obtained in \cite{TW1}.
We assume that there is no significant cancellation
between the tree and 1-loop terms in (\ref{eq:FY-chi}),
and that a Calabi--Yau manifold $X$ is ``isotropic'', that is,
its volume moduli are characterized by only one typical radius $R$.
Then we can roughly estimate the Fayet--Iliopoulos parameter $\xi_\chi$
in terms of Kaluza--Klein scale $M_{\rm KK} \sim 1/R$. 
$\xi_\chi$ in turn determines a vev of $\overline{N}^c$.
Canonically normalized $\overline{N}^c$ typically develops a vev of order
\begin{equation}
 \left| \vev{\overline{N}^c}\right|^2 \approx 
%   \frac{1}{8\pi}\frac{(M_G l_s)^2}{R^2},
    \frac{1}{4 \alpha_{\rm GUT}} \frac{1}{R^2},
\end{equation}
which is roughly around the Kaluza--Klein scale \cite{TW1}. 

Such a large vev of $\overline{N}^c$ generates Majorana mass 
terms of right-handed neutrinos $\overline{N}$.  
Once massive chiral multiplets $\Phi_*$'s are integrated out, 
an effective interaction 
\begin{equation}
 W \ni \frac{(y^{''\nu}_{(1)})^2}{M_\Phi} \; 
   \overline{N}^c \; \overline{N} \; \overline{N}^c \; \overline{N}
\label{eq:4N-2}
\end{equation}
is generated \cite{FaraggiB}. 
It is also known \cite{ws-instanton} that world-sheet instanton 
effects generate 
\begin{equation}
 W \ni e^{-T} \; \overline{{\bf 27}} \; \overline{\bf 27} \; 
                 {\bf 27} \; {\bf 27} 
% %  \equiv \frac{1}{M_*}
% %        \; \overline{{\bf 27}} \; \overline{\bf 27} \; 
% %                 {\bf 27} \; {\bf 27} 
%  \overline{N}^c \; \overline{N}^c \; \overline{N} \; \overline{N}
%  \equiv \frac{1}{M_*}
%    \overline{N}^c \; \overline{N}^c \; \overline{N} \;  \overline{N}
\label{eq:4N}
\end{equation}
in some compactifications of the Heterotic string theory 
with an unbroken $E_6$ gauge group, and (\ref{eq:4N}) contains 
an interaction of the form 
\begin{equation}
 W \ni \frac{1}{M_*} \; \overline{N}^c \; \overline{N}^c \; 
     \overline{N} \; \overline{N}.
\label{eq:4N-3}
\end{equation}
Once the interaction of this form is generated, either from 
(\ref{eq:4N-2}) or (\ref{eq:4N}), then non-vanishing vev of 
$\overline{N}^c$ provides Majorana mass terms of right-handed 
neutrinos \cite{DIN2}, with masses of order 
\begin{equation}
 M_{R} = \frac{\vev{\overline{N}^c}^2}{M_*}; \qquad 
   \frac{1}{M_*} \approx {\rm max} \left(
       \frac{ \left(y^{''\nu}_{(1)} \right)^2}{M_\Phi}, \; e^{-T}
                                   \right).
%  \simgt        \frac{ \left(y^{''\nu}_{(1)}
%	       \vev{\overline{N}^c}\right)^2}{M_\Phi}.
\label{eq:MR}
\end{equation}
Majorana right-handed neutrinos generate small masses 
of left-handed neutrinos through the see-saw mechanism.
At the same time, a flat direction 
$5|\overline{N}^c|^2-5|\overline{N}|^2 + \xi_\chi = 0$ is lifted 
because of the Majorana mass terms.
Without a fine-tuning of expectation values of gauge-field moduli, 
we can obtain $\vev{\overline{N}^c} \neq 0$ while
$\vev{\overline{N}}=0$; $\vev{\overline{N}^c} \neq 0$ is crucial for 
neutrino masses as we have seen above, and $\vev{\overline{N}}=0$ 
is crucial for the absence of trilinear R-parity violation \cite{TW1}.

In the presence of non-vanishing $\vev{\overline{N}^c}$,
trilinear terms of (\ref{eq:Yukawa41-nu}--\ref{eq:Yukawa41-nuN})
involving $\overline{N}^c$ gives rise to extra mass terms.
We have already seen how the mass matrix in 
the $\SU(5)_{\rm GUT}$-singlet sector is deformed; an interaction 
(\ref{eq:4N-2}) with $\overline{N}^c$ replaced by their vev's is
regarded as a mass term obtained after diagonalizing the mass matrix 
of $\Phi$ and $\overline{N}$ \cite{FaraggiB}.
Let us now look at how $\vev{\overline{N}^c} \neq 0$ modifies 
mass matrices of $\SU(5)_{\rm GUT}$-charged multiplets. 

In principle, mass eigenstates in a given representation 
of $\SU(5)_{\rm GUT}$ are mixture of states with different 
$\U(1)_\chi$ charges, because the $\U(1)_\chi$ symmetry 
is already broken spontaneously. 
In the $\SU(5)_{\rm GUT}$-$({\bf 10}+\overline{\bf 10})$ 
sector, the mass matrix becomes 
\begin{equation}
\left(
  \begin{array}{cc} 
    {\bf 10}^c_* & {{\bf 10}'}^c_*
  \end{array}
\right) 
\left(
  \begin{array}{cc|c}
     & M_{\bf 10} & \\ \hline
    y^{'\nu}_{(2)} \vev{\overline{N}^c} & y^{'\nu}_{(2)} \vev{\overline{N}^c} & M_{{\bf 10}'}
  \end{array}
\right)
\left(
  \begin{array}{c}
     {\bf 10}_0 \\ {\bf 10}_* \\ {{\bf 10}'}_*
  \end{array}
\right).
\label{eq:41-matrix10}
\end{equation}
By this $2 \times 3$ matrix, we actually mean an $(\infty + \infty) \times 
(3+ \infty + \infty)$ matrix for cases of practical interest.
% Massless eigenstates in the {\bf 10} representation are not exceptions.
%  ... exceptions of what? TW.
$\U(1)_\chi$ eigenstates and mass eigenstates are related by 
a basis transformation\footnote{Here, we are not very careful in 
defining the phase of mass eigenstates.}
\begin{equation}
\left(
  \begin{array}{c}
     {\bf 10}_0 \\ {\bf 10}_* \\ {{\bf 10}'}_*
  \end{array}
\right) = 
\left(
  \begin{array}{ccc}
    \frac{M_{{\bf 10}'} }
         {\sqrt{|M_{{\bf 10}'}|^2 + |y^{'\nu}_{(2)}
         \vev{\overline{N}^c}|^2}}
       & * & * \\
    0  & * & * \\
    - \frac{y^{'\nu}_{(2)}\vev{\overline{N}^c}}
           {\sqrt{|M_{{\bf 10}'}|^2 + |y^{'\nu}_{(2)}
	   \vev{\overline{N}^c}|^2}} 
       & * & * 
  \end{array}
\right)
\left(
  \begin{array}{c}
    \hat{\bf 10}_0 \\ \hat{\bf 10}_* \\ {\hat{\bf 10}}_*
  \end{array}
\right),
\label{eq:41-mix10}
\end{equation}
where $\hat{\bf 10}_*$ are massive mass eigenstates,
which consist mainly of linear combination of ${\bf 10}_*$ and ${\bf 10}'_*$,
and $\hat{\bf 10}_0$ are massless degrees of freedom in the presence of non-vanishing 
vev of $\overline{N}^c$. Both massive and massless eigenstates 
$\hat{\bf 10}_*$ and $\hat{\bf 10}_0$ are mixtures of states 
${\bf 10}$ and ${\bf 10}'$ with different $\U(1)_\chi$ charges.
% Massless eigenstates $\hat{\bf 10}_0$ indeed 
% have interactions that ${\bf 10}'_*$ have, as well as those 
% of ${\bf 10}_0$.

% More important to physics is a nature of mixing of chiral multiplets 
% in the $\SU(5)_{\rm GUT}$-$\bar{\bf 5}$ representation,
% since the mixing between $L_{i}$ and $H_{d}$
% can give rise to the R-parity violating trilinear terms (\ref{eq:dim4}).
Mass matrix in the $\SU(5)_{\rm GUT}$-$(\bar{\bf 5}+{\bf 5})$ sector 
is given by
\begin{equation}
\left(
  \begin{array}{ccc}
    H_0 & H_* & \bar{\bf 5}^c_*
  \end{array}
\right)
\left(
  \begin{array}{cc|cc}
    0 & & & \\
      & M_H & & \\ \hline
    y^\nu_{(2)} \vev{\overline{N}^c} & y^\nu_{(2)} \vev{\overline{N}^c}
    & 0 & M_{\bf 5}
  \end{array}
\right)
\left(
  \begin{array}{c}
    \bar{H}_0 \\ \bar{H}_* \\
    \bar{\bf 5}_0 \\ \bar{\bf 5}_*
  \end{array}
\right), 
\label{eq:41-matrix5}
\end{equation}
and one finds that the $\U(1)_\chi$ eigenstates contain massless 
(and massive) eigenstates as in 
\begin{eqnarray}
\left(
  \begin{array}{c}
    H_0 \\ H_* \\ \bar{\bf 5}^c_*
  \end{array}
\right) & = & 
\left(
  \begin{array}{ccc}
    1 & 0 & 0 \\ 0 & * & * \\ 0 & * & * 
  \end{array}
\right)
\left(
  \begin{array}{c}
    \hat{H}_0 \\ \hat{\bf 5}_* \\ \hat{\bf 5}_*
  \end{array}
\right), \\ % \qquad 
\left(
  \begin{array}{c}
    \bar{H}_0 \\ \bar{H}_* \\
    \bar{\bf 5}_0 \\ \bar{\bf 5}_*    
  \end{array}
\right) & = & 
\left(
  \begin{array}{cccc}
    \frac{M_{{\bf 5}}}
         {\sqrt{|M_{{\bf 5}}|^2 + |y^\nu_{(2)}\vev{\overline{N}^c}|^2}}
      & 0 & * & * \\
    0 & 0 & * & * \\
    0 & 1 & 0 & 0 \\
  - \frac{y^\nu_{(2)} \vev{\overline{N}^c}}
         {\sqrt{|M_{{\bf 10}'}|^2 + |y^\nu_{(2)}\vev{\overline{N}^c}|^2}}
      & 0 & * & *
  \end{array}
\right)
\left(
  \begin{array}{c}
    \hat{\bar{H}}_0 \\ \hat{\bar{{\bf 5}}}_0 \\
    \hat{\bar{\bf 5}}_* \\ \hat{\bar{\bf 5}}_*
  \end{array}
\right),
\label{eq:41-mix5bar}
\end{eqnarray}
where $\hat{\bar{\bf 5}}_*$ and $\hat{\bf 5}_*$ are massive mass eigenstates
and $\hat{H}_0$, $\hat{\bar{{\bf 5}}}_0$ and $\hat{\bar{H}}_0$ massless degrees of freedom.
All the massive mass-eigenstates are mixture of states with 
different $\U(1)_\chi$ charges, and so is a massless eigenstate 
$\hat{\bar{H}}(\bar{\bf 5})_0$. But, other massless eigenstates, 
$\hat{H}({\bf 5})_0$ and $\hat{\bar{{\bf 5}}}_0$, remain pure 
$\U(1)_\chi$ eigenstates, $H({\bf 5})_0$ and $\bar{\bf 5}_0$, 
respectively.\footnote{
The disparity between the natures of these massless modes 
stems from existence of well-defined subbundles 
$L$, $L \otimes U_4$ and $\wedge^2 \overline{U_4}$ 
that we mentioned in footnote \ref{fn:subspace41}. 
Massless eigenstates that remain pure $\U(1)_\chi$-eigenstates, namely, 
$\hat{\bar{{\bf 5}}}_0$ and $\hat{H}_0$, are characterized as 
$H^1(X; L \otimes V) \subset H^1(X; \wedge^2 V)$ and 
$H^1(X; \wedge^2 \overline{U_4}) \subset H^1(X; \wedge^2 \overline{V})$.
Other massless eigenstates such as $\hat{\bf 10}_0$ or $\hat{\bar{H}}_0$ 
do not have such characterization associated with subbundles \cite{TW1}.}

It is important to note that there is a strict rule  
on the mixing of massless eigenstates. 
Massless eigenstates have their own $\U(1)_\chi$ charges 
in a $\vev{\overline{N}^c} \rightarrow 0$ limit.  
When $\vev{\overline{N}^c}$ does not vanish,
they can pick up interactions of states with different $\U(1)_\chi$ charges,
only when holomorphic insertion of $\vev{\overline{N}^c}$
can supply the right $\U(1)_\chi$ charge \cite{TW1, TW2}.
An $\bar{H}$-like massless eigenstate $\hat{\bar{H}}_0$
have a non-vanishing $\bar{\bf 5}$ component---(4, 1) entry 
of the mixing matrix (\ref{eq:41-mix5bar})---because 
$\vev{\overline{N}^c} \; \bar{H}$
has the same $\U(1)_\chi$ charge as $\bar{\bf 5}$. 
On the other hand, $\bar{\bf 5}$-like massless eigenstates 
$\hat{\bar{{\bf 5}}}_0$ do not have $\bar{H}$ components---vanishing 
(2, 2) entry of (\ref{eq:41-mix5bar})---because 
$\vev{\overline{N}^c} \; \bar{\bf 5}$
does not have the same $\U(1)_\chi$ charge as $\bar{H}$.
Mixing of massless eigenstates $\hat{\bf 10}_0$ 
is also understood this way. 

Mixing matrices are important, because interactions 
of mass eigenstates are obtained by substituting 
(\ref{eq:41-mix10}) and (\ref{eq:41-mix5bar}) into 
(\ref{eq:Yukawa41-u}--\ref{eq:mass41}).
Since massive mass-eigenstates are generic mixtures of states with 
different $\U(1)_\chi$ charges, the $\U(1)_\chi$ symmetry is virtually 
powerless in controlling their interactions in the superpotential.
Trilinear terms in the superpotential that involve only massless 
eigenstates, however, are still controlled by the $\U(1)_\chi$ symmetry, 
because the mixing of massless eigenstates is under the rule 
above: $\U(1)_\chi$ charge can be supplied only through holomorphic 
insertion of $\vev{\overline{N}^c}$. Three point interactions of 
massless states do not exist, if sums of $\U(1)_\chi$ charges 
of these states in the $\vev{\overline{N}^c} \rightarrow 0$ limit 
are positive, because such interactions are not $\U(1)_\chi$ invariant 
even after allowing holomorphic insertion 
of positively charged $\vev{\overline{N}^c}$. 
Here, we have a selection rule in the superpotential.

The R-parity violating trilinear interactions 
$\bar{\bf 5}_0 \; {\bf 10}_0 \; \bar{\bf 5}_0$ have positive $(+5)$
$\U(1)_\chi$ charges, and holomorphic insertion of positively charged 
$\vev{\overline{N}^c}$ cannot make them neutral under $\U(1)_\chi$. 
Thus, such interactions do not exist, and the 4+1 model is a solution 
to the dimension-4 proton decay problem.
The $\U(1)_\chi$-charge counting allows an R-parity violating
interaction of the form $W \ni \vev{\overline{N}^c} \; \bar{H}_0 \; 
{\bf 10}_0 \; \bar{H}_0$. Indeed, by substituting (\ref{eq:41-mix10}) 
and (\ref{eq:41-mix5bar}) into (\ref{eq:Yukawa41-de}), we find 
\begin{equation}
W \ni y^d \frac{y^\nu \vev{\overline{N}^c}}
                  {\sqrt{M^2 + |y^\nu \vev{\overline{N}^c}|^2}}
 \hat{\bar{H}}_0 \; \hat{\bf 10}_0 \; \hat{\bar{H}}_0.
\label{eq:anotherdim4}
\end{equation}
But this operator vanishes because of anti-symmetric contraction
of $\SU(5)_{\rm GUT}$ indices and of the fact that
there is only one down-type Higgs doublet in the MSSM.
Therefore, trilinear R-parity violating operators are absent 
in the 4+1 model.

Finally, let us comment on what it takes in the 4+1 model 
to have two Higgs doublets in the low-energy spectrum, and 
a $\mu$-term in the superpotential. 
In the doublet part\footnote{
The doublet and triplet parts of $H(5)$ and $\bar{H}(\bar{5})$
have different KK towers, especially different numbers of massless modes,
as discussed in footnote \ref{fn:doublet-triplet}.
Mass matrix (\ref{eq:41-matrix5}) is intended to be for the doublet part.
There are no massless modes for the triplet part,
so the 1st raw and the 1st and 3rd columns are dropped from the mass matrix.
%
%  One should take $\mathrm{SU(5)_{GUT}}$ representations
%  in (\ref{eq:41-matrix5}--\ref{eq:41-mix5bar}) as the doublet parts of them.
%
%  Since the triplet parts of $H(5)$ and $\bar{H}(\bar{5})$ have 
%  no massless modes, we do not have to take care of their mixings 
%  with $\bar{D}_{i}$.
% ---> quite misleading TW.
%
  \label{fn:doublet-triplet_2}
} of the $\SU(5)_{\rm GUT}$-${\bf 5}+\bar{\bf 5}$ 
representations,
we need three $\hat{\bar{{\bf 5}}}_0$-like chiral multiplets $L_i$,
one $\hat{\bar{H}}_0$-like $H_d$ and one $\hat{H}_0$-like $H_u$
in the low-energy spectrum.
This means that there should be at least 1 and $m'+1$ zero modes 
in the doublet part of $H({\bf 5})$ and $\bar{H}(\bar{\bf 5})$ sectors, 
and $m'$ and 3 zero modes in the $\bar{\bf 5}^c$ and $\bar{\bf 5}$ 
sectors ($m' \geq 0$), respectively.
$m' > 0$ is allowed when $m'$ pairs of chiral multiplets 
$H_0$--$\bar{\bf 5}^c_0$ acquire large masses\footnote{
  Minimal choice $m'=0$ is assumed implicitly in the mass matrix 
  (\ref{eq:41-matrix5}) and Table~\ref{tab:repr41}.
  \label{fn:noextra}
}
through the second term in (\ref{eq:Yukawa41-nu}).
%no extra chiral multiplets remain in low-energy spectrum even for $m' > 0$.
Whether $m'=0$ or not, we need a pair of vector-like massless chiral
multiplets in the doublet part of $H({\bf 5})$--$\bar{H}(\bar{\bf 5})$.

The existence of this extra massless vector-like pair
is not guaranteed by topology.
% from the net chirality, which is a topological quantity.
This pair, essentially the two Higgs doublets of the MSSM,
can be in the low-energy spectrum for two possible reasons.
The first possibility is that there is at least one extra pair of 
massless multiplets in the doublet part of 
$H({\bf 5})$--$\bar{H}(\bar{\bf 5})$ sector for \textit{generic}
gauge-field configuration in a given topological class of ($X, V$).
See section~\ref{ssec:Het} for more about this case.
The second possibility is that the background gauge-field 
configuration of our world is somewhat \textit{special}\footnote{
  Possibly for an anthropic reason \cite{Donoghue}. 
  Special choice of gauge-field configuration could follow 
  as a consequence of moduli stabilization, in principle,
  but no such claim based on an explicit string construction 
  has been made so far.
}
and an extra pair of multiplets becomes almost massless for the 
choice of background. 

It is not hard for the second case for such two Higgs doublets 
to have a $\mu$-term.  Since they have trilinear couplings 
$W \ni \sum_I \; \Phi_{I; 0} \; \hat{H}_0 \; \hat{\bar{H}}_0$ with 
gauge field moduli $\Phi_{I; 0}$
(here the lower suffix $I$ denotes the $I$-th modulus),
$\mu$-term is generated once the 
vev's of these moduli fields shift by of the order of SUSY-breaking scale.
This is essentially the next-to-minimal SUSY Standard
Model (NMSSM).\footnote{Hence such moduli are free from moduli problem.}
In the first case, such a trilinear coupling is absent
(by definition).
Instead, 1-loop diagrams generate such terms as
%
%%%%%%%%%%%%%%%%%%%%%%%%%%%%%%%%%%%%%%%%%%%%%%%%%%%%%%%%%%%
%
\begin{figure}
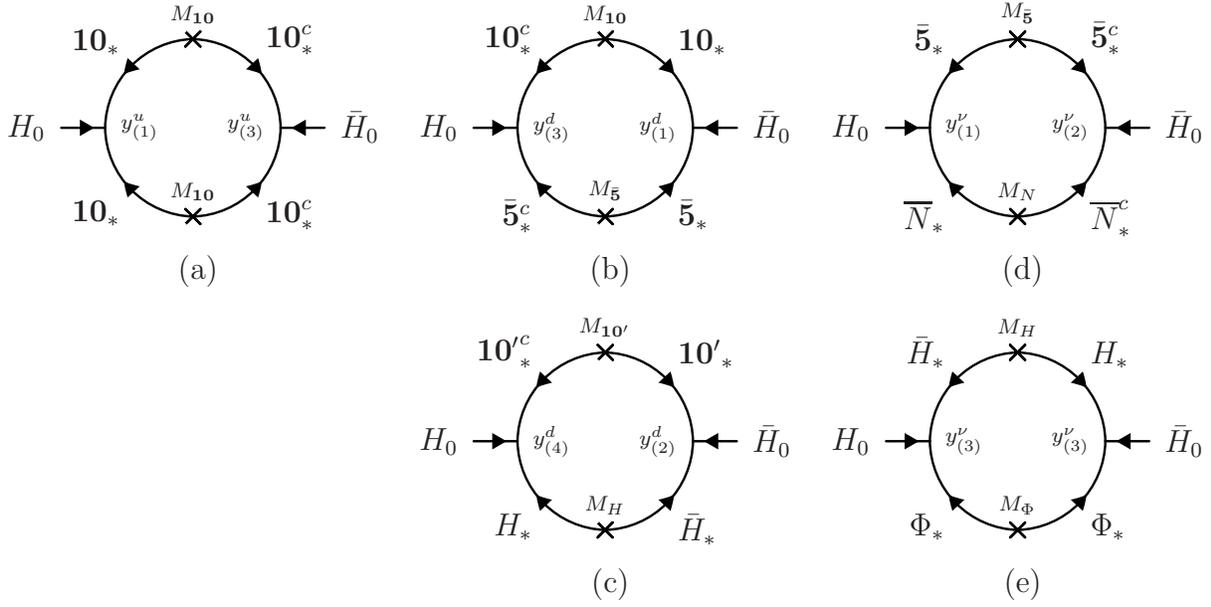

\begin{center}
\begin{tabular}{ccccc}
\input{figure2-1a} & \hspace{1cm} & \input{figure2-1b} & \hspace{1cm} & \input{figure2-1d} \\
(a) && (b) && (d) \\
&&&&\\
 && \input{figure2-1c} && \input{figure2-1e} \\
 && (c) && (e) \\
\end{tabular}
\end{center}
\caption{\label{fig:muterm}Super Feynman diagrams that generate 
$\mu$-term in the 4+1 model. There are three different kinds of graphs; 
a pair of multiplets in the {\bf 10} and {\bf 10} representations 
of $\SU(5)_{\rm GUT}$ are running in the loop (a), those in the
 $\overline{\bf 10}$ and {\bf 5} representations of
 the $\SU(5)_{\rm GUT}$ in the loop (b) (c), and finally, 
singlets and $\bar{\bf 5}$'s in the loop (d) (e).}
\end{figure}
%
%%%%%%%%%%%%%%%%%%%%%%%%%%%%%%%%%%%%%%%%%%%%%%%%%%%%%%%%%%%
%
\begin{equation}
 K \ni \frac{|y^u|^2}{16\pi^2}
       \frac{M_{\bf 10}^{*2}}{|M_{\bf 10}|^2}  \hat{H}_0 \hat{\bar{H}}_0.
\label{eq:mu!}
\end{equation}
This one comes from the first diagram of Fig.~\ref{fig:muterm}, 
and there are also similar contributions from loops with multiplets 
in other representations; see Fig.~\ref{fig:muterm}. 
Remembering that holomorphic mass parameters in the superpotential 
have non-vanishing $\theta^2$ components, at least by of order 
$M(1+\theta^2 \, m_{3/2})$ when $\vev{W}^*/M_G^2 \neq 0$, 
$\mu$-term and $B_\mu$-term are generated at 1-loop.
This is essentially the mechanism in \cite{Hall-mu}, except that 
the vector-like massive multiplets in the loop are identified 
with Kaluza--Klein towers here. 
Note, however, that the $\mu$- and $B_\mu$-terms generated in 
this way are known to have a fine-tuning problem \cite{DGP}.
Thus, the problem may be an indication that the Higgs sector is 
a little more complicated than we imagine here, and that 
a bit of model building is necessary as in gauge- and 
anomaly-mediation scenarios.
%
% There are some theoretical subtleties in this way of generating 
% a $\mu$-term in the context of ours, and we will come back to this 
% issue in section \ref{ssec:bilinear}.

{\bf 3+2 Model}

There are some variations of the 4+1 model \cite{TW1},
and one of them is called 3+2 model.
Instead of restricting the structure group of gauge field background
to $\SU(4) \times \U(1)_\chi \subset \SU(5)'$ at the beginning, one can choose 
$\SU(3) \times \SU(2) \times \U(1)_{\tilde{q}_7} \subset \SU(5)'$
as the structure group.
This restriction on the structure group is relaxed later
by $\U(1)_{\tilde{q}_7}$-breaking vev, just like in the 4+1 model.
Irreducible components in (\ref{eq:55decomp}) split into 
\begin{eqnarray}
 ({\bf 5}, \wedge^2 {\bf 5}) & \rightarrow &
    ({\bf 1}, {\bf 2}, \wedge^2 {\bf 5})^{-3}
  + ({\bf 3}, {\bf 1}, \wedge^2 {\bf 5})^{+2}, \\
 (\wedge^2 {\bf 5}, \bar{\bf 5}) & \rightarrow & 
    ({\bf 1}, \wedge^2 {\bf 2}, \bar{\bf 5})^{-6} 
  + ({\bf 3}, {\bf 2}, \bar{\bf 5})^{-1} 
  + (\wedge^2 {\bf 3}, {\bf 1}, \bar{\bf 5})^{+4}, \\
 ({\bf adj.}, {\bf 1}) & \rightarrow & 
    ({\bf adj.}, {\bf 1}, {\bf 1})^0 + ({\bf 1}, {\bf adj.}, {\bf 1})^0
  + (\bar{\bf 3}, {\bf 2}, {\bf 1})^{-5}
  + ({\bf 3}, \bar{\bf 2}, {\bf 1})^{+5}
\end{eqnarray}
under $\SU(3) \times \SU(2) \times \SU(5)_{\rm GUT} 
\times \U(1)_{\tilde{q}_7}$.
%
%%%%%%%%%%%%%%%%%%%%%%%%%%%%%%%%%%%%%%%%%%%%%%%%%%%%%%%%%%%%
\begin{table}[t]
\begin{center}
\begin{tabular}{c|c|c|c}
  fields & representations & number of zero-modes & zero-modes at low energy \\
\hline
\rule[0pt]{0pt}{12pt}
  ${\bf 10}_0$, ${\bf 10}_*$
& $({\bf 1},{\bf 2},\wedge^2 {\bf 5})^{-3}$
& 3
& $Q_i, \bar{U}_i, \bar{E}_i$ ($i = 1,2,3$) \\
  ${\bf 10}^c_*$
& $({\bf 1},{\bf 2},\wedge^2 {\bf 5})^{+3}$
& 0
& - \\
\hline
  ${{\bf 10}'}_*$
& $({\bf 3},{\bf 1},\wedge^2 {\bf 5})^{+2}$
& 0
& - \\
  ${{\bf 10}'}^c_*$
& $(\bar{\bf 3},{\bf 1}, \wedge^2 {\bf 5})^{-2}$
& 0
& - \\
\hline
\rule[0pt]{0pt}{12pt}
  $\bar{\bf 5}_0$, $\bar{\bf 5}_*$
& $({\bf 3},{\bf 2},\bar{\bf 5})^{-1}$
& $3\;(+b)$
& $\bar{D}_i, L_i$ ($i=1,2,3$) \\
  ${\bar{\bf 5}}^c_0$, ${\bar{\bf 5}}^c_*$
& $(\bar{\bf 3},{\bf 2},{\bf 5})^{+1}$
& $0\;(+a)$
& - \\
\hline
  $H_0({\bf 5})$, $H_*({\bf 5})$
& $({\bf 1},{\bf 1},{\bf 5})^{+6}$
& 1({\bf 2}) / 0({\bf 3}) $\; (+b)$
& $H_u$ \\
  $H^c_*(\bar{\bf 5})$
& $({\bf 1},{\bf 1},\bar{\bf 5})^{-6}$
& 0
& - \\
\hline
\rule[0pt]{0pt}{12pt}
  $\bar{H}_0(\bar{{\bf 5}})$, $\bar{H}_*(\bar{{\bf 5}})$
& $(\wedge^2 {\bf 3},{\bf 1},\bar{\bf 5})^{+4}$
& 1({\bf 2}) / 0($\bar{\bf 3}$) $\; (+a)$ 
& $H_d$ \\
  $\bar{H}^c_*({\bf 5})$
& $(\wedge^2 \bar{\bf 3},{\bf 1}, {\bf 5})^{-4}$
& 0
& - \\
\hline
\rule[0pt]{0pt}{12pt}
  $\overline{N}_0$, $\overline{N}_*$
& $(\bar{\bf 3},{\bf 2},{\bf 1})^{-5}$
& 1
& heavy RH neutrino \\
  $\overline{N}^c_*$
& $({\bf 3},{\bf 2},{\bf 1})^{+5}$
& 0
& - \\
\hline
  $\Phi_0$, ${\Phi}_*$
& $({\bf adj.}, {\bf 1}, {\bf 1})^0$
&
& \\
& $({\bf 1}, {\bf adj.}, {\bf 1})^0$
&
& 
\end{tabular}
\caption{
  \label{tab:repr32}
  List of fields for 3+2 model. 
  Fields with subscripts $_0$ or $_*$ can be understood as before.
  The second column shows how the chiral multiplets transform under
  $\SU(5)_{\rm GUT}$ as well as underlying broken symmetries
  $\SU(3) \times \SU(2) \times \U(1)_{\tilde{q}_7}$ in the 3+2 model.
  The slashes in the third column represent the doublet-triplet splitting.
  $a=b=0$ in the minimal choice. We will find in section~\ref{ssec:dim5} 
  that some effective operators can be enhanced depending on whether 
  such a non-minimal pair ($a \neq 0$ or $b \neq 0$) 
  of (eventually massive) chiral multiplets are in the spectrum or not.
}
\end{center}
\end{table}
%%%%%%%%%%%%%%%%%%%%%%%%%%%%%%%%%%%%%%%%%%%%%%%%%%%%%%%%%%%%%%%%%%%%%
Each irreducible component has its own towers of chiral and vector
multiplets. Topology of ($X$, $V$) should be arranged so that
the number of zero modes of each component is appropriate for 
a low-energy effective theory; Table~\ref{tab:repr32} shows 
the required number of zero modes as well as notation 
of chiral multiplets originating from each sector. 
% See Table~\ref{tab:repr32} for our notation.
Note that the tower containing the down-type Higgs doublet
$\bar{H}(\bar{\bf 5})$ is not the Hermitian conjugate
of that containing the up-type Higgs doublet $H({\bf 5})$
in the 3+2 model, unlike in the 4+1 model.

Superpotential of the 3+2 model is given by rewriting (\ref{eq:Het-super}):
%in terms of chiral multiplets in the irreducible components:
%
\begin{eqnarray}
W & = & y^u_{(1)} {\bf 10} \; {\bf 10} \; H 
  + y^u_{(2)} {\bf 10} \; {{\bf 10}'}_* \; \bar{\bf 5}^c_* 
  + y^u_{(3)} {{\bf 10}'}_* \; {{\bf 10}'}_* \;
  \bar{H}^c_* \label{eq:Yukawa32-u} \\
  & + &  y^u_{(4)} {\bf 10}^c_* \; {\bf 10}^c_* \; H^c_* 
       + y^u_{(5)} {\bf 10}^c_* \; {{\bf 10}'}^c_* \; \bar{\bf 5} 
      + y^u_{(6)} {{\bf 10}'}^c_* \; {{\bf 10}'}^c_*
      \; \bar{H} \label{eq:Yukawa32-u-conj} \\
  & + & y^d_{(1)} \bar{H}\;{\bf 10}\;\bar{\bf 5}
     + y^d_{(2)} \bar{H} \; {{\bf 10}'}_* \; H^c_* 
     + y^d_{(3)} \bar{\bf 5}\; {{\bf 10}'}_* \;\bar{\bf 5} 
     \label{eq:Yukawa32-de} \\
  & + & 
 y^d_{(4)} \bar{H}^c_* \; {\bf 10}^c_* \;\bar{\bf 5}^c_*
+ y^d_{(5)} \bar{H}^c_*\; {{\bf 10}'}^c_* \; H 
+ y^d_{(6)} \bar{\bf 5}^c_*\;{{\bf 10}'}^c_*\;
         \bar{\bf 5}^c_* \label{eq:Yukawa32-de-conj} \\
 & + &
  y^{\nu}_{(1)} \bar{\bf 5} \; \overline{N} \; H
   + y^{\nu}_{(2)} \bar{H} \; \overline{N} \; \bar{\bf 5}^c 
   +  y^{\nu}_{(3)} \bar{\bf 5}^c \; \overline{N}^c \; H^c_*
   + y^{\nu}_{(4)} \bar{H}^c_* \; \overline{N}^c \; \bar{\bf 5} 
    \label{eq:Yukawa32-nu} \\
 & + &
   y^{\nu}_{(5)} H\; \Phi \; H^c_*
 + y^{\nu}_{(6)} \bar{H}\; \Phi \; \bar{H}^c_*
 + y^{\nu}_{(7)} \bar{\bf 5}\; \Phi \; \bar{\bf 5}^c_*
    \label{eq:Yukawa32-nu5} \\    
 & + &
   y^{'\nu}_{(1)} \overline{N}\; {\bf 10}^c_* \; {{\bf 10}'}_* 
   + y^{'\nu}_{(2)} \overline{N}^c\; {\bf 10} \; {{\bf 10}'}^c_*
  + y^{'\nu}_{(3)} {\bf 10} \; \Phi \; {\bf 10}^c_*
  + y^{'\nu}_{(4)} {{\bf 10}'}_* \; \Phi \; {{\bf 10}'}^c_*
    \label{eq:Yukawa32-nu10} \\
 & + &
   y^{''\nu}_{(1)} \bar{N} \; \Phi \; \bar{N}^c 
  + y^{''\nu}_{(2)} \Phi \; \Phi \; \Phi
    \label{eq:Yukawa32-nuN} \\
 &  + &
  M_{\bf 10} {\bf 10}^c_* \;{\bf 10}_* 
+ M_{{\bf 10}'} {{\bf 10}'}^c_* \;{{\bf 10}'}_* 
+ M_{\bar{H}} \bar{H}^c_* \;\bar{H}_* 
+ M_{\bar{\bf 5}} \bar{\bf 5}^c_* \;\bar{\bf 5}_*
+ M_{H} H^c_*\;H_*\\
\label{eq:mass32_1}
 &  + &  
M_N \overline{N}^c_* \;\overline{N}_*
+ M_{\Phi} \Phi_* \;\Phi_*.
\label{eq:mass32_2}
\end{eqnarray}
%
% where the ellipses stand for terms with chiral multiplets without 
% subscripts of asterisk replaced by those with asterisks, 
% the same as before in the 4+1 model.
The trilinear R-parity violating operators (\ref{eq:dim4}) are 
absent because 
\begin{equation}
 ({\bf 3}, {\bf 2},\bar{\bf 5})^{-1} \otimes 
 ({\bf 1}, {\bf 2},\wedge^2 {\bf 5})^{-3} \otimes 
 ({\bf 3}, {\bf 2}, \bar{\bf 5})^{-1}
\end{equation}
does not contain a singlet of 
$\SU(3) \times \SU(2) \times \SU(5)_{\rm GUT} \times \U(1)_{\tilde{q}_7}$.

\begin{figure}
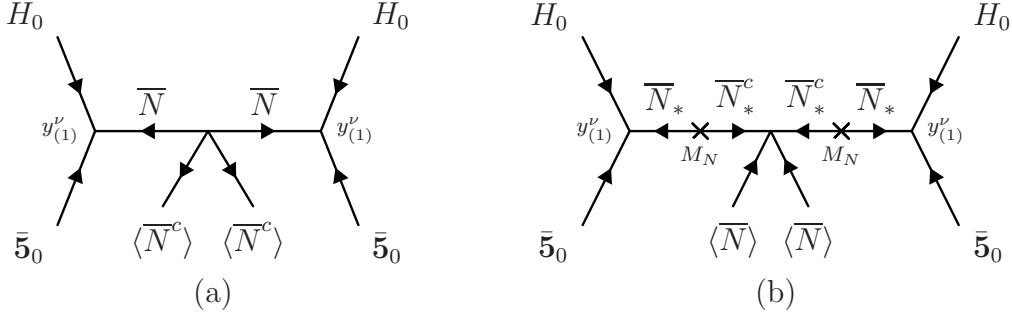

\begin{center}
\begin{tabular}{ccc}
\input{figure2-2a} & \hspace{20mm} & \input{figure2-2b} \\
&& \\
(a) && (b)
\end{tabular}
\end{center}
\caption{Diagrams which generate 
$W \ni H({\bf 5})\; H({\bf 5})\;\bar{\bf 5}\;\bar{\bf 5}$
for the 4+1 model (a) and for the 3+2 model (b).}
\label{fig:super-HH}
\end{figure}

If the Fayet--Iliopoulos parameter of $\U(1)_{\tilde{q}_7}$ is positive, 
only zero modes $\overline{N}_0$, which carry negative $\U(1)_{\tilde{q}_7}$
charge, develop non-vanishing expectation values. 
Majorana mass terms of $\overline{N}^c$ are generated in the 3+2 model, 
just like those of $\overline{N}$ are in the 4+1 model; see the
discussion following (\ref{eq:4N-2}, \ref{eq:4N}).
Small masses of left-handed neutrinos are generated through the 
double see-saw mechanism, since vector-like mass terms of 
$\overline{N}_* \overline{N}^c_*$ are available from the last term of 
(\ref{eq:mass32_2}) \cite{TW1}.
Super-diagram description of the double see-saw mechanism is 
the one on the right-hand side of Fig.~\ref{fig:super-HH}.
%%%%%%%%%%%%%%%%%%%%%%%%%%%%%%%%%%%%%%%%%%%%%%%%%%%%%%
%
%
%%%%%%%%%%%%%%%%%%%%%%%%%%%%%%%%%%%%%%%%%%%%%%%%%%%%%%%%%%%%
The effective mass scale $M_{R, {\rm eff.}}$ that characterizes 
the low-energy neutrino mass through 
$m_\nu \sim (y^\nu v)^2/M_{R,{\rm eff.}}$ is given by 
\begin{equation}
 M_{R, {\rm eff.}} \sim \frac{M_N^2}{M_R}
\label{eq:MReff}
\end{equation}
with $M_R$ given in (\ref{eq:MR}).

In the presence of non-vanishing vev's of
$\overline{N} \in (\bar{\bf 3}, {\bf 2}, {\bf 1})^{-5}$,
$(\infty + \infty) \times \left(\infty + 3 + \infty \right)$ mass matrix
of the $\SU(5)_{\rm GUT}$-${\bf 10}+\overline{\bf 10}$ sector is deformed into
\begin{equation}
\left(
  \begin{array}{cc} 
    {{\bf 10}'}^c_* & {\bf 10}^c_*
  \end{array}
\right) 
\left(
  \begin{array}{c|cc}
 M_{{\bf 10}'} & & \\
 \hline
 y^{'\nu}_{(1)} \vev{\overline{N}} & 0 & M_{\bf 10}
  \end{array}
\right)
\left(
  \begin{array}{c}
    {\bf 10}'_* \\ {\bf 10}_0 \\ {\bf 10}_*
  \end{array}
\right),
\label{eq:32-matrix10}
\end{equation}
which means that the $\U(1)_{\tilde{q}_7}$ eigenstates
are related to mass eigenstates as
\begin{equation}
\left(
  \begin{array}{c}
 {{\bf 10}'}_* \\ {\bf 10}_0 \\ {\bf 10}_* 
  \end{array}
\right) = 
\left(
  \begin{array}{ccc}
    0 & * & * \\
    1 & 0 & 0 \\
    0 & * & * 
  \end{array}
\right)
\left(
  \begin{array}{c}
    \hat{\bf 10}_0 \\ \hat{\bf 10}_* \\ \hat{\bf 10}_*
  \end{array}
\right).
\label{eq:32-mix10}
\end{equation}
Mass matrix in the $\bar{\bf 5}+{\bf 5}$ sector becomes
% \footnote{$a=0$ 
% is assumed here. For the cases with $a\neq 0$, the mass matrix would
% look like $6 \times 6$. [really ?? TW]} 
%
\begin{equation}
\left(
  \begin{array}{ccccc}
    \bar{H}^c_* & \bar{\bf 5}^c_0 & \bar{\bf 5}^c_* & H_0 & H_*
  \end{array}
\right)
\left(
  \begin{array}{cc|cc|c}
     & M_{\bar{H}} & & & \\ \hline
     y^\nu_{(2)} \vev{\overline{N}} & y^\nu_{(2)} \vev{\overline{N}} &
       & & \\
     y^\nu_{(2)} \vev{\overline{N}} & y^\nu_{(2)} \vev{\overline{N}} &
       & M_{\bar{\bf 5}} & \\ \hline 
     & & y^\nu_{(1)} \vev{\overline{N}} & y^\nu_{(1)} \vev{\overline{N}} & \\
     & & y^\nu_{(1)} \vev{\overline{N}} & y^\nu_{(1)} \vev{\overline{N}} & M_H
  \end{array}
\right)
\left(
  \begin{array}{c}
    \bar{H}_0 \\ \bar{H}_* \\
    \bar{\bf 5}_0 \\ \bar{\bf 5}_* \\
    H^c_* 
  \end{array}
\right), 
\label{eq:32-matrix5}
\end{equation}
and the $\U(1)_{\tilde{q}_7}$ eigenstates have massless eigenstates as 
components specified by 
\begin{equation}
\left(
  \begin{array}{c}
    \bar{H}_0 \\ \bar{H}_* \\
    \bar{\bf 5}_0 \\ \bar{\bf 5}_* \\ H^c_*   
  \end{array}
\right)\sim 
\left(
  \begin{array}{ccccc}
   \frac{M_{\bar{\bf 5}}}
      {\sqrt{M_{\bar{\bf 5}}^2 + \left|y^\nu_{(2)}\vev{\overline{N}}\right|^2}}
       & 0 & * & * & * \\ 
    0 & 0 & * & * & * \\
    0 & \frac{M_H}
             {\sqrt{|M_H|^2 + \left|y^\nu_{(1)}\vev{\overline{N}}\right|^2}}
       & * & * & * \\
  - \frac{y^\nu_{(2)}\vev{\overline{N}}}
      {\sqrt{M_{\bar{\bf 5}}^2 + \left|y^\nu_{(2)}\vev{\overline{N}}\right|^2}}
       & 0 & * & * & * \\
    {\cal O}\left(\frac{\vev{\overline{N}}^2}{M_{\bar{\bf 5}} M_H} \right)
    & -\frac{y^\nu_{(1)}\vev{\overline{N}}}
      {\sqrt{|M_{H}|^2 + \left|y^\nu_{(1)} \vev{\overline{N}}\right|^2}} 
      & * & * & * 
  \end{array}
\right)
\left(
  \begin{array}{c}
     \hat{\bar{H}}_0 \\ \hat{\bar{{\bf 5}}}_0 \\
     \hat{\bar{\bf 5}}_* \\ \hat{\bar{\bf 5}}_* \\ \hat{\bar{\bf 5}}_*
  \end{array}
\right).
\label{eq:32-mix5}
\end{equation}
%
% [this equation must be checked, since the mass matrix is changed]
% I think that's OK; TW 23/01
%
Using (\ref{eq:32-mix10}) and (\ref{eq:32-mix5}) in 
(\ref{eq:Yukawa32-de}), one finds that the trilinear R-parity 
violating operators (\ref{eq:dim4}) involving massless eigenstates 
are not generated. 
The essence is that U(1)$_{\tilde{q}_7}$-eigenstates 
$\bar{H}(\bar{\bf 5})$ do not have the massless eigenstates 
$\hat{\bar{{\bf 5}}}_0$ as a component, and 
$\U(1)_{\tilde{q}_7}$-eigenstates ${\bf 10}'$ do not contain massless 
eigenstates $\hat{\bf 10}_0$.\footnote{
  The 3+2 model corresponds to a rank-5 vector bundle $V$
  given by an extension of a rank-3 bundle $U_3$ by a rank-2 bundle $U_2$:
  \begin{equation}
    0 \rightarrow U_2 \rightarrow V \rightarrow U_3 \rightarrow 0.
  \end{equation}
  $\U_2$ is a sub-bundle of $V$, and $U_2 \otimes V$ that of $\wedge^2 V$.
  The massless eigenstates $\hat{\bf 10}_0$ and $\hat{\bar{{\bf 5}}}_0$
  are characterized as $H^1(X; U_2) \subset H^1(X; V)$
  and $H^1(X; U_2 \otimes V) \subset H^1(X; \wedge^2 V)$, respectively, 
  and have restricted interactions than general massless states in 
  $H^1(X; V)$ and $H^1(X; \wedge^2 V)$, respectively.
  \label{fn:subspace32}
}
Eigenvectors for massless eigenstates in 
(\ref{eq:32-mix10}, \ref{eq:32-mix5}) have non-vanishing entries 
only when holomorphic insertion of $\U(1)_{\tilde{q}_7}$-breaking 
vev of $\overline{N}$ can fill the gap of $\U(1)_{\tilde{q}_7}$-charges, 
just like in the 4+1 model. Therefore, trilinear interactions involving 
only massless eigenstates are subject to the selection rule based on 
holomorphic insertion of $\vev{\overline{N}}$ and 
$\U(1)_{\tilde{q}_7}$-charge counting.
%
% which is a consequence of the fact that the mixing of massless eigenstates
% into various $\U(1)_{\tilde{q}_7}$-eigenstates is possible
% only with holomorphic insertion of $\vev{\overline{N}}$.
% This is just like the case in the 4+1 model.
%

% The net chirality between $\SU(5)_{\rm GUT}$-$\bar{\bf 5}$ and 
% ${\bf 5}$ representations in the mass matrix (\ref{eq:32-matrix5}) 
% is three, both in the doublet and triplet parts; this automatically 
% follows from the anomaly argument, once topology of $(X, V)$ is set 
% so that there are three chiral multiplets 
% in the $\SU(5)_{\rm GUT}$-{\bf 10} representation at low energies. 
% In the triplet part, we just need three chiral multiplets at low
% energies in the $\SU(3)_C$-$\bar{\bf 3}$ representation and none 
% in the ${\bf 3}$ representation. 

In the doublet part\footnote{
See footnotes \ref{fn:doublet-triplet} and \ref{fn:doublet-triplet_2}
for discussions of doublet-triplet splitting.
} of $\mathrm{SU(5)_{GUT}}$-$\mathbf{5+\bar{5}}$,
one $H_u \subset \hat{H}_0$, one $H_d \subset \hat{\bar{H}}_0$,
and three $L_i \subset \hat{\bar{{\bf 5}}}_0$
should remain in the low-energy spectrum,
although $H_u$ and $\{ H_d, L_i \}$ 
% form vector-like pair of representations of the MSSM gauge group.
are in a pair of vector-like representations of the MSSM gauge group.
Thus, the mass matrix (\ref{eq:32-matrix5}) somehow has to have 
a reduced rank for the doublet part. 
Note that it is actually an $((\infty + a+\infty+ (1+b)+\infty) \times
((1+a)+\infty+(3+b)+\infty+\infty))$ matrix for the doublet
part,\footnote{For the triplet part, $(1+b)$ and $(1+a)$ are replaced by
$b$ and $a$, respectively. Rank reduction is not necessary.} 
although it looks like $5 \times 5$;
here $a$ and $b$ denote the number of massless modes
in $\bar{5}^c$-$\bar{H}$ and $H$-$\bar{5}$ vector-like pairs, respectively,
that become massive after spontaneous $U(1)_{\tilde{q}_7}$ breaking.
It is supposed to have rank 
$\infty + a + \infty + b+ \infty$, not 
$\infty + a + \infty + (1+b) + \infty$.

%The mass term of $H_0$--$\bar{\bf 5}_0$ is directly generated
%from the trilinear terms of the superpotential
%in the presence of $\vev{\overline{N}}$,
%thus forming $(1+b) \times (3+b)$ block in (\ref{eq:32-matrix5}).
%The mass term of $H_0$--$\bar{H}_0$ is not generated in this way,
%but obtained by integrating out massive modes $\bar{\bf 5}_*$--$\bar{\bf 5}^c_*$,
%which is $W \ni (y^\nu_{(1)} \; \overline{N}_0 \; H_0 )
%\; (y^\nu_{(2)} \; \overline{N}_0 \; \bar{H}_0) /M_{\bf 5}$.

First of all, in order for three $L_i$'s to remain in the low-energy spectrum,
the $(1+b) \times (3+b)$ submatrix---(4th, 3rd) block---of 
(\ref{eq:32-matrix5}) should have rank $b$, not $(1+b)$. 
Such rank reduction of mass matrix can happen for a generic point 
on the gauge-field moduli space,
as we have reviewed in section~\ref{ssec:Het},
and remarked as the first possibility
for the light two Higgs doublets in the 4+1 model.
% earlier in this section~\ref{ssec:4132}.
This means that there are one linear combination of $H_0$'s
and three of $\bar{\bf 5}_0$'s
that do not have couplings of the form
$W \ni y^\nu \; H_0 \; \overline{N}_0 \; \bar{\bf 5}_0$.

Secondly, another massless doublet $H_d \subset \hat{\bar{H}}_0$
also has to remain in the low-energy spectrum.
In general, $\hat{H}_0$-type zero modes and $\hat{\bar{H}}_0$-type
zero modes effectively have mass terms of the form 
$W \ni [(y^\nu_{(1)} \; \vev{\overline{N}_0})
(y^\nu_{(2)} \; \vev{\overline{N}_0} ) / M_{\bar{\bf 5}}]
\hat{H}_0 \; \hat{\bar{H}}_0$, and become massive.
It may be, however, that this mass matrix has a reduced rank 
for generic moduli field (including $\overline{N}$'s) value, 
just like we assumed above. 
Alternatively, the effective mass parameter happens to be small 
at the value of $\vev{\overline{N}}$'s where moduli are stabilized.
While the first possibility is aesthetically better,
the origin of $\mu$ parameter is not explained because the zero modes
$\hat{H}_0$ and $\hat{\bar{H}}_0$ are not coupled to moduli fields 
that might play a role of the singlet field of the NMSSM.
The second scenario seems to involve a fine-tuning
to get the $\mu$ parameter small;\footnote{There may be multiple 
$\overline{N}$'s that acquire vev's, and there are infinitely many 
Kaluza--Klein states $\bar{\bf 5}^c_*$--$\bar{\bf 5}_*$. 
Thus, the effective $\mu$-parameter being small does not necessarily
mean that $y^\nu \vev{\overline{N}}$ itself is small.}
some anthropic selection may be responsible for it \cite{Donoghue}.

{\bf Remark}

There are some other variations of the 4+1 and 3+2 models \cite{TW1},  
but they are essentially the same in the mechanism of eliminating the 
trilinear R-parity violating operators. 
% In the 4+1 and 3+2 models we have discussed so far, 
% the R parity violating condensation 
% $\vev{\overline{N}^c}$ or $\vev{\overline{N}}$ 
% does induce mixing between $\bar{\bf 5}$-like 
% and $\bar{H}(\bar{\bf 5})$-like multiplets in the mass terms. 
% But the mass matrix is such that three massless eigenstates remain 
% purely $\bar{\bf 5}$-like, and hence the operators (\ref{eq:dim4})
% are not generated. Other variations also adopt essentially the same 
% mechanism.
Thus, we use these two models as illustrative examples representing 
a theoretical framework that can be an alternative to R parity. 

%%%%%%%%%%%%%%%%%%%%%%%%%%%%%%%%%%%%%%%%%%%%%%%%%%%%%%%%%%%
\subsection{Dimension-5 Proton Decay Operator}
\label{ssec:dim5p}
%%%%%%%%%%%%%%%%%%%%%%%%%%%%%%%%%%%%%%%%%%%%%%%%%%%%%%%%%%%

The idea in section \ref{ssec:4132} was %to use the fact 
that the mixing of massless modes is governed by holomorphic insertion 
of vacuum expectation value of chiral multiplets.
The trilinear R-parity violating operators are absent and the dimension-4 
proton decay problem is solved, without an unbroken discrete symmetry 
such as R-parity. 
Supersymmetric extensions of the standard model, however, has another
problem. If the effective superpotential contains 
\begin{equation}
 \Delta W = \frac{1}{M_1} \; Q \; Q \; Q \; L 
          + \frac{1}{M_2} \; \bar{E} \; \bar{U} \; \bar{U} \; \bar{D}
   \equiv {\cal O}_1 + {\cal O}_2 \subset 
  \frac{1}{M_{\rm eff.}} \; 
 {\bf 10} \; {\bf 10} \; {\bf 10} \; \bar{\bf 5},
 \label{eq:op-dim5}
\end{equation}
these terms violate both baryon number and lepton number,
and hence proton decays. 
Thus, we discuss in this subsection whether or not 
the theoretical framework in section~\ref{ssec:4132} predicts 
the operators above in the low-energy effective superpotential.

{\bf 4+1 model}

In the minimal supersymmetric SU(5) GUT model, the dimension-5 
operators above are generated by integrating out colored Higgs 
multiplets. The essence is that the pair of colored Higgs 
multiplets in the $\SU(3)_C$-${\bf 3} + \bar{\bf 3}$ representations 
has a vector-like mass term. The 4+1 model in section~\ref{ssec:4132}
also shares the same property. 
The up-type and down-type Higgs doublets of the MSSM
originate from irreducible components $H$ and $\bar{H}$
in the $\SU(5)_{\rm GUT}$-${\bf 5}+\bar{\bf 5}$ representations,
and they form a Hermitian conjugate pair even in the $E_8$ gauge group. 
%%%%%%%%%%%%%%%%%%%%%%%%%%%%%%%%%%%%%%%%%%%%%%%%%%%%%%%%%%%%%%%%%%
\begin{figure}
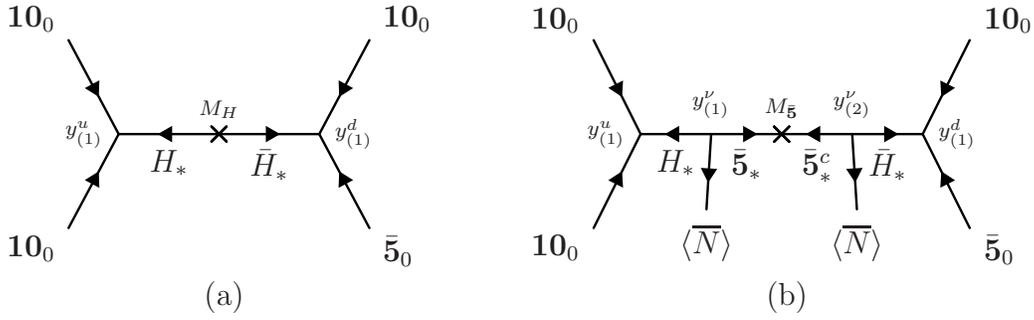

\begin{center}
\begin{tabular}{ccc}
\input{figure2-3a} & \hspace{2cm} & \input{figure2-3b} \\
&&\\
(a) && (b)
\end{tabular}
\end{center}
\caption{Dimension-5 proton decay operators in 
$W \ni {\bf 10}\;{\bf 10}\;{\bf 10}\;\bar{\bf 5}$ 
are generated by a diagram (a) in the 4+1 model. 
Although they appear to be generated in the 3+2 model 
as well through the diagram in (b), it turns out 
that they are actually not. See the text for explanation.
}
\label{fig:super-1010105}
\end{figure}
%%%%%%%%%%%%%%%%%%%%%%%%%%%%%%%%%%%%%%%%%%%%%%%%%%%%%%%%%%%%%%%%%%%%%%
Therefore, the superdiagram in Figure~\ref{fig:super-1010105}~(a)
generates the term $\Delta W = \mathbf{10 \; 10 \; 10 \; \bar{5}}$
in low-energy effective theory,
similarly to the minimal supersymmetric $\SU(5)$ GUT model.
In particular, the $QQQL + \bar{E}\bar{U}\bar{U}\bar{D}$ part 
relevant to proton decay is generated by colored Higgsino 
exchange.\footnote{Scalar colored Higgs exchange is also relevant 
to $QQQL + \bar{E}\bar{U}\bar{U}\bar{D}$ proton decay.}

There are a couple of differences, however, between 
the minimal supersymmetric SU(5) GUT model and the 4+1 model.
In the 4+1 model, propagating between the two vertices 
of Figure~\ref{fig:super-1010105}~(a) are not just one pair 
of colored Higgs multiplets. All of massive Kaluza--Klein 
excitations contribute to the amplitude. 
These infinite number of contributions add up coherently. 
Thus, we have an effective operator 
\begin{equation}
 W_{\rm eff.} \ni \sum_I \frac{y^u_I y^d_I}{M_{H; \; I}} 
  {\bf 10}_0 \; {\bf 10}_0 \; {\bf 10}_0 \; \bar{\bf 5}_0 
 \equiv \frac{1}{M_{\rm eff.}}
  {\bf 10}_0 \; {\bf 10}_0 \; {\bf 10}_0 \; \bar{\bf 5}_0,
\label{eq:dim5p-in-KKsum}
\end{equation}
with massless eigenstates in the external lines.
As this effective operator is generated in the 4+1 model without  
a mixing induced by $\vev{\overline{N}^c}$, we do not make an 
explicit distinction between the $\U(1)_\chi$-eigenstates and 
mass-eigenstates. 
Yukawa couplings $y_I$ are determined by overlap integration 
of mode functions of three relevant states;
\begin{equation}
 y_I = \frac{g_{\rm YM}}{{\rm vol}(X)} \int_X d^6 y \tr 
   \left( \bar{\chi}_0 \gamma^m \varphi_{0; m} \chi_I \right),
\label{eq:overlap}
\end{equation}
where $\chi_0$ and $\varphi_{0; \; m}$ are
wavefunctions of massless modes to be used in the external lines, 
and $\chi_I$ that of $I$-th massive Higgs multiplet in the internal 
line with mass $M_{H; I}$. The overall factor\footnote{
Mode functions should be normalized so that the kinetic terms 
of chiral multiplets are normalized canonically. 
The expression (\ref{eq:overlap}) contains $e^{\vev{K}/M_G^2}$, 
where $\vev{K}/M_G^2$ is a vev of (\ref{eq:Het-Kahler}).} 
$g_{\rm YM}$ originates 
from the normalization convention of the mode functions in 
(\ref{eq:gYM-normalize}). 

Sum of infinite contributions in (\ref{eq:dim5p-in-KKsum})
is treated better in an equivalent description
\begin{eqnarray}
 \frac{1}{M_{\rm eff}} & = &
  \frac{g_{\rm YM}}{{\rm vol}(X)} \int_X d^6 y 
         (\bar{\chi}_0 \gamma^m \varphi_{0; \; m})(y)
  \frac{g_{\rm YM}}{{\rm vol}(X)} \int_X d^6 y' 
         (\varphi_{0; \; n} \gamma^n \chi_0)(y') 
  \sum_I \frac{\chi_I (y) \bar{\chi}_I (y')}{M_{H; \; I}} \\
  & \sim & 
    \frac{g_{\rm YM}^2}{{\rm vol}(X)} 
    \int_X d^6 y 
         (\bar{\chi}_0 \gamma^m \varphi_{0; \; m})(y) \; 
\left[    \int \frac{d^6 p'}{(2\pi)^6} 
              \frac{e^{i p\cdot (y - y')}}{\sslash{p}}
\right] \; 
    \int_X d^6 y' 
         (\varphi_{0; \; n} \gamma^n \chi_0)(y'),  
\label{eq:dim5inXdim}
\end{eqnarray}
%
%where $'$ in the momentum integration means that %is meant to note that
%contributions of zero-modes should be dropped from the propagator. 
Discrete summation labeled by $I$ in the first line is approximated by continuous 
momentum integration in the second line.
This is a reasonable thing to do, as long as we are interested in contributions 
from highly excited Kaluza--Klein states. 
The factor in the square bracket in (\ref{eq:dim5inXdim}) is 
nothing but Green function over the internal space $X$, and its 
short-distance singularity 
$\gamma^k \partial_k \left(1/|y-y'|^4 \right)$ corresponds to 
summing up contributions from infinitely many Kaluza--Klein states.
% infinite contributions from highly excited Kaluza--Klein modes.
% Summation over KK modes with large KK masses
% has become integration over large $|p|$,
% which corresponds to the short-distance singularity 
% of the fermion Green function 
% $\gamma^m \partial_m \left(1/|y-y'|^4 \right)$  
% in the internal six dimensions. 
For zero-mode wavefunctions $\chi_0$ and $\varphi_{0; \; m}$ 
without particularly singular or rapidly varying behavior, 
integration over $|y-y'|$ in (\ref{eq:dim5inXdim}) is dominated 
by the long-distance region, $|y-y'| \approx R$, not 
by the short-distance region. Thus, $1/M_{\rm eff.}$ is finite.
We also learn here that contributions from low-lying triplet Higgsino 
exchange dominate the amplitude $1/M_{\rm eff.}$
of dimension-5 proton decay operators. 

% In terms of summation over $I$ or integration over $\vec{p}$, 
% the phase factor $e^{i \vec{p} \cdot (\vec{y} - \vec{y}')}$ oscillates 
% so rapidly for large $|\vec{p}|$ that infinite contributions in 
% (\ref{eq:dim5p-in-KKsum}) do not really ``add'' up. Furthermore, 
% for large $|\vec{p}|$, heavy Higgsino wavefunction $\chi_I(y)$ 
% oscillates so rapidly in (\ref{eq:overlap}) that $|y_I|$ themselves 
% become small. We see, therefore, that heavy Higgs multiplets 
% whose masses are above the Kaluza--Klein scale (typical separation 
% and/or width of zero-mode wavefunctions) do not contribute very much 
% for the dimension-5 proton decay operator in the effective theory. 

Experimental limits on the dimension-5 operators \eqref{eq:op-dim5} are set 
by several decay modes of proton.
$\tau(p \rightarrow K^+ + \bar{\nu}) \simgt 2.3 \times 10^{33}
 \; {\rm yrs.}$ \cite{SKnucleondecay}, one of the most important 
ones in the minimal supersymmetric SU(5)$_{\rm GUT}$ model, 
roughly corresponds to
\begin{equation}
  M_{\rm eff.} \simgt 10^{24} \,\GEV.
\end{equation}
Limits from other decay modes are somewhat different, 
but not by several orders of magnitude.
All kinds of GUT models are marginally in conflict with this constraint
\cite{MurayamaPierce,HarnikLarsonMurayamaThormeier}
as long as the two Higgs triplets are vector like in any
underlying symmetry groups. This property is shared also 
by the 4+1 model \footnote{The R-parity preserving scenario 
explained at the end of section~\ref{ssec:Het}. also does.}
If $1/M_{\rm eff.}$ is approximated by contributions from a few
 lightest triplet massive Higgsinos in the Kaluza--Klein
tower, $1/M_{\rm eff.} \approx y^u_I y^d_I /M_{\rm KK}$,  
the experimental limit is translated to
\begin{equation}
  \sqrt{y^u_I y^d_I} \simlt 10^{- 4} \times
    \left( \frac{M_{\rm KK}}{ 10^{16} \; \GEV } \right)^{1/2}.
\label{eq:dim5-limit}
\end{equation}
%
% If we refer yukawa couplings $y^c \simeq 10^{-2}$ and
% $y^s \simeq 10^{-3}$ of MSSM, which are usually used to calculate
% MSGUT proton lifetime, this  constraint seems relatively severe.
This constraint seems a little severe if the Kaluza--Klein scale 
$M_{\rm KK}$ is around the GUT scale $M_{\rm GUT} \sim 10^{16} \; \GEV$,
and the trilinear couplings are of the order of the Yukawa couplings 
of the second generation quarks, $y_c \sim 10^{-2}$ and $y_s \sim 10^{-3}$.
However, $y^{u,d}_I$'s in (\ref{eq:dim5p-in-KKsum}, \ref{eq:dim5-limit})
in the 4+1 model 
are not related to the Yukawa couplings of quarks and leptons 
by $\SU(5)_{\rm GUT}$ symmetry.
% , as we assume in calculation of proton decay with minimal $\SU(5)$
% SUSY GUT. 
$y_I$'s are calculated 
in (\ref{eq:overlap}) by using mode functions of Kaluza--Klein 
triplet Higgsinos $\chi_I(y)$, whereas the observable Yukawa 
couplings are calculated by using mode functions of 
two massless doublet Higgsinos. Since $y_I$'s are 
``Fourier transform'' of zero-mode wavefunctions 
$(\bar{\chi}_0 \gamma^m \varphi_{0; m})$ on a compact 
internal manifold $X$, they become very small 
for higher Kaluza--Klein states. 
Therefore, it will not be conservative to exclude the 4+1 model 
with $M_{\rm KK} \approx M_{\rm GUT}$.

% However, with regard to the fact that $y_I$'s that we used here are
% not yukawa couplings between zero-modes like ones in MSSM but
% ones between zero-modes and KK modes, we cannot assure us
% that  $y_I$'s are in the same order as $y$'s in MSSM, and so we
% still hesitate to exclude a scenario with $M_{\rm KK}$ around
% the scale of gauge coupling unification.

%Second, the ``Yukawa couplings'' $y^u_I$ and $y^d_I$ in (\ref{eq:overlap}) 
%are not directly related to the Yukawa couplings of ordinary 
%quarks and leptons. 
%All we know from quark and lepton masses are their couplings with 
%massless two Higgs doublets. Since the wavefunctions of massive
%colored Higgsinos, $\chi_I$ and $\bar{\chi}_I$, are different from those of 
%zero-mode Higgsinos, $y^{u/d}_I$'s involving massive states cannot 
%be determined from low-energy observables alone.
%In fact, for highly excited Kaluza--Klein modes of $H$ and $\bar{H}$, 
%``Yukawa couplings'' (\ref{eq:overlap}) are
%% approximately Fourier transforms of zero-mode wavefunctions, and they are
%expected to be very small even on a curved manifold.
%Without a knowledge on the ``Yukawa couplings'' involving excited
%states, no precise predictions on $M_{\rm eff.}$ can be made.

{\bf Remark}

% Two theoretical side remark is in order here,
There is a theoretical side remark here, 
before moving over to 
% discussion of the dimension-5 proton decay problem in 
the 3+2 model.
In the process going from (\ref{eq:Het-super})
and/or (\ref{eq:Yukawa41-u}--\ref{eq:mass41})
to (\ref{eq:dim5p-in-KKsum}),
we are rewriting 1PI effective action: 
from the one written in terms of (Kaluza--Klein decomposition of) 
Yang--Mills multiplets on 9+1 dimensions
to the one in terms only of Kaluza--Klein zero modes on 3+1 dimensions.
The dimension-5 operator (\ref{eq:dim5p-in-KKsum}) is \textit{necessary}
as a 1PI vertex when 1PI effective action
is written in terms only of zero modes,
although it was \textit{absent} in (\ref{eq:Het-super})
where all Kaluza--Klein modes appear in 1PI effective action.
It is therefore almost trivial that
the sum of infinite contributions in (\ref{eq:dim5p-in-KKsum}) is finite;
the amplitude described by (\ref{eq:dim5p-in-KKsum})
and by Figure~\ref{fig:super-1010105}~(a) is nothing more than
a {\it tree-level} scattering amplitude of super Yang--Mills theory
with the external states having wavepackets of Kaluza--Klein zero modes.
% and the tree-level amplitude is finite.

{\bf 3+2 model}

Let us now study whether the dimension-5 proton decay operators 
\eqref{eq:op-dim5} are generated in the 3+2 model. 
In this model, $\U(1)_{\tilde{q}_7}$-eigenstates $H$ and $\bar{H}$ 
are not Hermitian conjugate, and hence the dimension-5 proton 
decay operators are absent in the effective theory 
in the $\vev{\overline{N}} = 0$ limit. Thus, whether those operators 
exist in low-energy effective theory depends on how the vev's of 
$\overline{N}$ can be inserted in the effective superpotential.

$\U(1)_{\tilde{q}_7}$-breaking vev's $\vev{\overline{N}}$ induce 
mixing between states with different $\U(1)_{\tilde{q}_7}$ charges. 
The mixing can take place among massive states propagating in the 
internal line as in the diagram in Figure~\ref{fig:super-1010105}~(b).
On the other hand, the massless eigenstates $\hat{\bf 10}_0$ 
in the external lines remain pure $\U(1)_{\tilde{q}_7}$-eigenstates, 
as we saw in (\ref{eq:32-mix10}). 
Although $H^c_*$-type $\U(1)_{\tilde{q}_7}$-eigenstates contain 
the massless eigenstates $\hat{\bar{{\bf 5}}}_0$ as components,  
there is no interaction involving  ${\bf 10}$ and $H^c$ 
simultaneously in (\ref{eq:Yukawa32-de}).
Thus, mixing in the internal line is responsible, if the dimension-5 
proton decay operators are to be generated. 

Let $M_{IJ}$ denote the mass matrix in (\ref{eq:32-matrix5}), 
and $U$ and $V$ unitary matrices that diagonalize $M_{IJ}$; 
\begin{equation}
 U_{HI} \; M_{IJ} \; V_{JK} = \delta_{HK} \hat{M}_K.
\end{equation}
Then the dimension-5 proton decay operators 
in the effective theory are given by 
\begin{equation}
 W_{\rm eff.} \ni \sum_{I \in H({\bf 5}), \; J \in \bar{H}(\bar{\bf 5})}
  {\sum_{K}}'
  \frac{(y^d_J V_{JK}) \; (y^u_I \, U_{KI})}{\hat{M}_K} \; 
   \hat{{\bf 10}}_0 \; \hat{{\bf 10}}_0 \; \hat{{\bf 10}}_0 \; 
   \hat{\bar{{\bf 5}}}_0
  \equiv \frac{1}{M_{\rm eff.}} 
  \hat{{\bf 10}}_0 \; \hat{{\bf 10}}_0 \; \hat{{\bf 10}}_0 \; 
  \hat{\bar{{\bf 5}}}_0; 
\label{eq:dim5p-in-mixed-KKsum}
\end{equation}
the prime in the summation means that the label $K$ runs only over 
massive states. The effective coupling $1/M_{\rm eff.}$ can be expressed 
in a much simpler way. 
\begin{equation}
 \frac{1}{M_{\rm eff.}} = y^d_{(1)} \;
   \left(\sum_{K} V_{\bar{H}(\bar{\bf 5}) K} \hat{M}_K^{-1} 
                  U_{K H({\bf 5})}\right) \;
   y^u_{(1)} 
  = y^d_{(1)} \; \left(M^{-1}\right)_{\bar{H}(\bar{\bf 5})H({\bf 5})} 
    \; y^u_{(1)}.
\label{eq:dim5-KKsum-smplfd}
\end{equation}
Thus, although the unitary mixing matrices $U$ and $V$ are not given 
by rational functions of parameters in the superpotential, 
the coupling $1/M_{\rm eff.}$ is in the effective superpotential.  

Let us examine the matrix $M^{-1}$, first with the simplest case 
$a = b = 0$ in Table~\ref{tab:repr32}.
This corresponds to a case where there is no pair of chiral multiplets 
in the $\SU(5)_{\rm GUT}$-$\bar{\bf 5}+{\bf 5}$ representations
that acquire masses after $\vev{\overline{N}}$'s become non-zero.
First, we are concerned only about the ${\bf 3}+\bar{\bf 3}$ part of 
the mass matrix in the ${\bf 5}+\bar{\bf 5}$ sector, because the 
dimension-5 proton decay operators are generated only through 
triplet exchange diagrams. Second, three massless chiral multiplets 
$\bar{D}_i \subset \hat{\bar{{\bf 5}}}_0$ are taken out of the triplet 
part of the mass matrix (\ref{eq:32-matrix5}), because only massive 
states are integrated out.
Now, the mass matrix (\ref{eq:32-matrix5}) has become 
\begin{equation}
 M_{IJ} = \left(\begin{array}{ccc}
	   M_{\bar{H}} & 0 & 0 \\
           y^\nu_{(2)} \vev{\overline{N}} & M_{\bar{\bf 5}} & 0 \\
           0 & y^\nu_{(1)} \vev{\overline{N}} & M_H 
		\end{array}\right), 
\label{eq:3by3-Hc-mass}
\end{equation}
and we treat it as if it were a $3 \times 3$ matrix. 
The index $I$ runs over 
$\{\bar{H}^c({\bf 3}), \bar{\bf 3}^c, H({\bf 3})\} 
 \subset \{\bar{H}^c({\bf 5}), \bar{\bf 5}^c, H({\bf 5})\}$, and $J$ over 
$\{\bar{H}(\bar{\bf 3}), \bar{\bf 3}, H^c(\bar{\bf 3})\} 
 \subset \{\bar{H}(\bar{\bf 5}), \bar{\bf 5}, H^c(\bar{\bf 5})\}$.
There is no qualitative problem in approximating the summation over 
contributions from infinite Kaluza--Klein states by a sum over 
three distinct Kaluza--Klein towers, as we have discussed before 
for the 4+1 model. The entire contribution from a tower is finite, 
and only those from low-lying massive states have sizable contributions.
It is now easy to see that the cofactor of 
($H({\bf 3})$, $\bar{H}(\bar{\bf 3})$) element of
the matrix (\ref{eq:3by3-Hc-mass}) vanishes, and so is 
$(M^{-1})_{\bar{H}(\bar{\bf 3}) H({\bf 3})}$. Thus, the dimension-5 
proton decay operators are not generated in the 3+2 model with 
spectra characterized by $a= b = 0$. 

Individual contribution from a given mass-eigenstate 
in (\ref{eq:dim5p-in-mixed-KKsum}), $V_{JK} \hat{M}^{-1}_K U_{KI}$
without summation over $K$, does not vanish, as the Feynman diagram 
in Figure~\ref{fig:super-1010105}~(b) indicates. 
Some of those contributions are of order 
$M (y^\nu \vev{\overline{N}})^{*2}/|M|^4$, where 
$M \approx M_{\bar{H},\bar{\bf 5},H}$. But, all these contributions 
cancel, and the total amplitude $1/M_{\rm eff.}$ vanishes. 
The total $1/M_{\rm eff.}$ should be given by a rational function 
of parameters in the original superpotential, 
as we saw in (\ref{eq:dim5-KKsum-smplfd}). 
Vev's of anti-chiral multiplets $(y^\nu \vev{\overline{N}})^{*2}$ 
should not survive cancellation. 

Similar study can be carried out for cases with either/both 
$a$ or/and $b$ is/are non-zero.\footnote{In these cases, extra massless 
multiplets in the $\SU(5)_{\rm GUT}$-${\bf 5}+\bar{\bf 5}$
representations appear in the $\vev{\overline{N}} \rightarrow 0$
limit.} The $5 \times 5$ matrix (\ref{eq:32-matrix5}) without 
rank reduction is used instead of (\ref{eq:3by3-Hc-mass}) 
for the case with $a \neq 0$ and $b \neq 0$, and $4 \times 4$ matrices 
are used for the two other cases. The cofactor of the 
$(H({\bf 3}), \bar{H}(\bar{\bf 3}))$ element turn out to vanish 
for all these matrices, and hence the dimension-5 proton decay 
operators are not generated for any of these cases. 
Thus, the dimension-5 proton decay operators are absent 
in low-energy effective superpotential of the 3+2 model, 
independent of whether $\bar{\bf 5}^c$--$\bar{H}(\bar{\bf 5})$ like 
and/or $H({\bf 5})$--$\bar{\bf 5}$ like states with masses of order 
$y^\nu \vev{\overline{N}}$ exist at high energy or not. 

The absence of the operator in the 3+2 model can be understood 
in terms of $\U(1)_{\tilde{q}_7}$-charge counting. Requiring 
that the effective coupling $1/M_{\rm eff.}$ is a rational function 
of parameters and vev's of the original superpotential, and that 
the effective superpotential is invariant under 
the $\U(1)_{\tilde{q}_7}$ symmetry, the only possible form is
$1/M_{\rm eff.} \sim y^u y^d M /(y^\nu \vev{\overline{N}})^2$, 
where $M$ is some mass scale.  Unless there are massive states 
in the $\SU(5)_{\rm GUT}$-$\bar{\bf 5}+{\bf 5}$ representations 
with their masses given by 
$\approx {\cal O}(('y^\nu \vev{\overline{N}})^2/M)$, 
such terms are not generated in the low-energy effective 
superpotential.
This could have been the case if there were extra pairs 
of zero modes of $H({\bf 5})$-type and $\bar{H}(\bar{\bf 5})$-type, 
in addition to those listed in Table~\ref{tab:repr32}. We have 
ignored such a possibility so far, because the dimension-5 proton 
decay operators with $1/M_{\rm eff.} \sim y^u y^d M_{\bar{\bf 5}}/
(y^\nu \vev{\overline{N}})^2$ seem too large compared with 
the experimental limits, 
if $M_{\bar{\bf 5}} \sim M_{\rm KK} \sim M_{\rm GUT}$ and 
$y^\nu \vev{\overline{N}} \ll M_{\rm KK}$, despite large 
uncertainties associated with trilinear couplings 
$y^u$ and $y^d$ that involve Kaluza--Klein states.

\subsection{Bilinear R-parity Violation}
\label{ssec:bilinear}
%%%%%%%%%%%%%%%%%%%%%%%%%%%%%%%%%%%%%%%%%%%%%%%%%%%%%%%%%%%

Let us now study consequences of R-parity violation 
in the framework presented in section \ref{ssec:4132}. 
We begin with a study of mixing between  
$L_i \subset \hat{\bar{{\bf 5}}}_0$ and $H_d \subset \hat{\bar{H}}(\bar{\bf 5})_0$; 
they are in the same representation of the MSSM gauge group 
$\SU(2)_L \times \U(1)_Y \subset \SU(5)_{\rm GUT}$, and have 
distinct symmetry charges only in the presence of an unbroken 
R-parity or $\U(1)$ symmetry.
Any interactions involving R-parity violation (and hence a vev of 
$\overline{N}^c$ (in the 4+1 model) or $\overline{N}$ (in the 3+2 model)) 
have a chance to induce mixing between $L_i$'s and $H_d$.

We have already seen in section \ref{ssec:4132} that the vev's of 
$\overline{N}^c$ [resp. $\overline{N}$] deform mass matrices 
in the superpotential in the 4+1 [resp. 3+2] model. Consequences 
of the deformed mass matrices, however, were quite limited 
in the superpotential at the renormalizable level. Mixing of 
massless eigenstates are under strict control of the spontaneously 
broken $\U(1)_\chi$ [resp. $\U(1)_{\tilde{q}_7}$] symmetry, and 
the trilinear R-parity violating operators (\ref{eq:dim4}) are not 
generated for massless modes. Although the $\U(1)$ symmetry does not rule out
the other trilinear R-parity violating operator of the 
form (\ref{eq:anotherdim4}), it is absent in the MSSM
because there is only one down-type Higgs doublet. Yukawa couplings 
of quarks and leptons will have different values from those 
in (\ref{eq:Yukawa41-u}--\ref{eq:Yukawa41-nu}) 
[resp. (\ref{eq:Yukawa32-u}--\ref{eq:Yukawa32-nu})], because the massless
eigenstates may not be exactly the same as the original 
$\U(1)$-eigen zero modes. Those are all the consequences. 

\begin{figure}
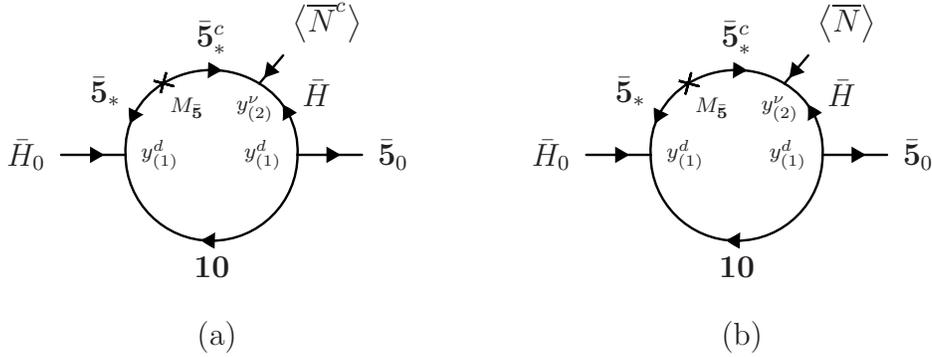

\begin{center}
\begin{tabular}{ccc}
\input{figure2-4a} & \hspace{2cm} & \input{figure2-4b} \\
&&\\
(a) && (b)
\end{tabular}
\end{center}
\caption{Diagrams which produce $K \ni \bar{\bf 5}^\dag\; \bar{H}$ for the 4+1 model (a)
and for the 3+2 model (b).}
\label{fig:kahler-5dagh}
\end{figure}

Vev's of $\overline{N}^c$ or $\overline{N}$, however, generate mixing 
in the {\it K\"{a}hler potential} as well. 
Kinetic mixing % \footnote{This is 
% in the $\SU(5)_{\rm GUT}$ notation just for simplicity. } 
%
\begin{equation}
 K_{\rm eff.} \ni \epsilon \; \bar{\bf 5}^\dagger\; \bar{H}(\bar{\bf 5})
                  + {\rm h.c.}
\label{eq:kahler-5Hbar}
\end{equation}
is generated in both the 4+1 and 3+2 model at 1-loop 
by super Feynman diagrams in Figure~\ref{fig:kahler-5dagh}. 
%%%%%%%%%%%%%%%%%%%%%%%%%%%%%%%%%%%%%%%%%%%%%%%%%%%%%%%%%%%%%%%%%%%%
%
%
%%%%%%%%%%%%%%%%%%%%%%%%%%%%%%%%%%%%%%%%%%%%%%%%%%%%%%%%%%%%%%%%%%%%%%%
Those diagrams give rise to operators, 
\begin{eqnarray}
\text{4+1 model} & & 
  K_{\rm eff.}  \sim \frac{|y^d|^2}{16\pi^2} 
          \frac{M_{\bar{\bf 5}}^* y^{'\nu} \vev{\overline{N}^c}}{|M|^2} \; 
         \bar{\bf 5}^\dagger \; \bar{H} + {\rm h.c.},
  \label{eq:5HbarKahler-41} \\
\text{3+2 model} & & 
 K_{\rm eff.} \sim \frac{|y^d|^2}{16\pi^2} 
        \frac{M_{\bar{\bf 5}}^* y^\nu \vev{\overline{N}} }{|M|^2}
           \bar{\bf 5}^\dagger \; \bar{H} + {\rm h.c.}
  \label{eq:5HbarKahler-32}
\end{eqnarray}
$|M|^2$'s in the denominators stand for the largest among 
$|M_{\bar{\bf 5}}|^2$, $|M_{\bf 10}|^2$ and $|M_{\bar{H}}|^2$, 
because that is where the dominant contribution comes from 
in loop momentum integration. 
There are two other kinds of contributions as well, but they are quite 
similar to those above; see the caption of Figure~\ref{fig:muterm}.
We treat 1-loop amplitudes here, as if only finite number of massive 
particles ran in the loop.

It is true in the context of string compactification 
that infinitely many Kaluza--Klein particles and 
stringy states are also running in the loop. 
The argument above estimates contributions only from 
low-lying Kaluza--Klein multiplets. Unless the 
remaining UV contributions exactly cancels the IR 
contributions above, however, the total 1-loop amplitude 
does not vanish. Since the IR and UV contributions 
are likely to depend on geometry differently, 
it is unlikely that they cancel. Thus, as long as 
there are non-vanishing IR contributions, it is likely 
that the total amplitude does not vanish. 
This is what we can guess from the argument using Feynman 
diagrams. 
 
Although the UV contributions to $\epsilon$ have not been 
discussed yet, it is sufficient to treat it symbolically 
in seeing that its effects disappear from renormalizable 
interactions of the massless modes of the MSSM. 
By a non-unitary basis transformation 
\begin{equation}
 \left( \begin{array}{c}
  \bar{H} \\ \bar{\bf 5}
	\end{array}\right)
 = \left( \begin{array}{cc}
    1 & \\ - \epsilon & 1 
	  \end{array}\right)
  \left( \begin{array}{c}
   \bar{H}' \\ \bar{\bf 5}'
	 \end{array}\right), 
% \bar{\bf 5}' & = & \bar{\bf 5} + \epsilon \bar{H}(\bar{\bf 5}), \\
% \bar{H}(\bar{\bf 5})' & = & \bar{H}(\bar{\bf 5}), 
\label{eq:absorb-e}
\end{equation}
the kinetic terms are diagonalized, 
yet the R-parity violating trilinear operators (\ref{eq:dim4}) 
are absent when the superpotential is rewritten in terms of 
newly defined chiral multiplets $\bar{H}'$ and $\bar{\bf 5}'$.
Although operators of the form (\ref{eq:anotherdim4}) now exist 
when written in the new chiral multiplets, there is no such term 
that consists only of massless multiplets in the MSSM because of 
anti-symmetric contraction of $\SU(5)_{\rm GUT}$ indices \cite{TW2}.
Jacobian of this field redefinition in path integral is trivial, and there
are no other effects from this redefinition. 

Similarly, 1-loop amplitudes generate kinetic mixing among chiral
multiplets in the $\SU(5)_{\rm GUT}$-${\bf 10}+\overline{\bf 10}$ 
representations. An appropriate field redefinition, however, can 
diagonalize the kinetic terms without generating R-parity violating trilinear
operators that involve only massless multiplets in the MSSM. 
In the 3+2 model, for example, an effective K\"{a}hler potential term
\begin{equation}
 K_{\rm eff.}^{(3+2)} \sim \epsilon \; {\bf 10}^\dagger \;
    {\bf 10}' + {\rm h.c.}
%
%\epsilon^* \; {\bf 10}^{'\dagger} \; {\bf 10}
%  + {\rm h.c.}
\end{equation}
is generated, with a coefficient $\epsilon$ proportional to 
$y^\nu \vev{\overline{N}}$. 
A new basis $({\bf 10}', {\bf 10})'$ is chosen by 
\begin{equation}
 \left(\begin{array}{c}
  {\bf 10}' \\ {\bf 10}
       \end{array}\right) = 
 \left(\begin{array}{cc}
  1 &  \\ -\epsilon & 1
       \end{array}\right)
 \left(\begin{array}{c}
  {\bf 10}' \\ {\bf 10}
       \end{array}\right)', 
\label{eq:absorb-e10}
\end{equation}
so that the kinetic terms become diagonal.
Note that the redefinitions of chiral multiplets (\ref{eq:absorb-e})
and (\ref{eq:absorb-e10}) use parameters $\epsilon$, which involve 
only holomorphic vev's of either $\overline{N}^c$ or $\overline{N}$.
This is why the basis transformation matrices 
% holomorphicity leads to the basis transformation matrices 
(\ref{eq:absorb-e}) and (\ref{eq:absorb-e10}) are lower triangular.
It thus follows that the mass matrices 
(\ref{eq:41-matrix5}, \ref{eq:32-matrix5}) and (\ref{eq:32-matrix10}) 
remain lower triangular, and whole argument we have had so far 
in sections~\ref{ssec:4132} and \ref{ssec:dim5p} does not have to be 
changed qualitatively.\footnote{We will make an order of magnitude estimate of 
$\epsilon$ at the end of this section, and 
argue that the additional deformation to mass matrices is 
quantitatively unimportant compared with the tree-level deformation 
discussed in section~\ref{ssec:4132}.}

\begin{figure}
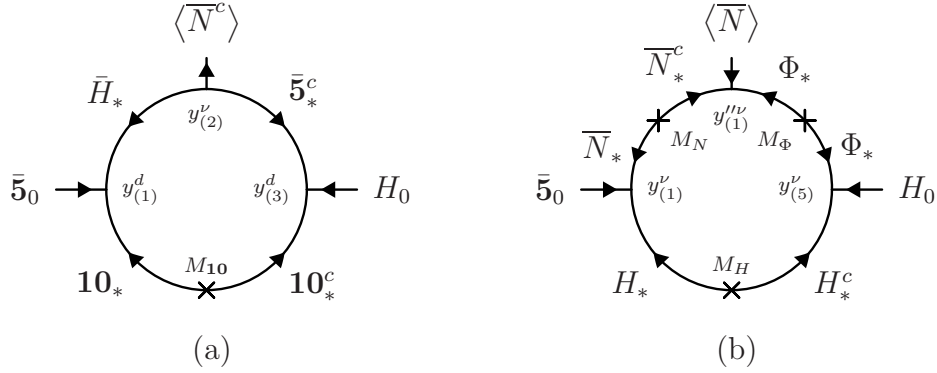

\begin{center}
\begin{tabular}{ccc}
\input{figure2-5a} & \hspace{2cm} & \input{figure2-5b} \\
(a) && (b)
\end{tabular}
\end{center}
\caption{(a) is a typical diagram for \eqref{eq:5HKahler-41} in 4+1 model.
 (b) is a typical diagram for \eqref{eq:5HKahler-32} in 3+2 model.}
\label{fig:5HKahler}
\end{figure}

The K\"{a}hler potential of the effective theory may also have a bilinear 
term 
\begin{equation}
 K_{\rm eff.} \ni c \; \bar{\bf 5}\; H({\bf 5}) + {\rm h.c.}
\label{eq:5H}
\end{equation}
%
% in the $\SU(5)_{\rm GUT}$ terminology.
1-loop super Feynman diagrams in Figure~\ref{fig:5HKahler} generate 
\begin{eqnarray}
\text{4+1 model} & & 
  K_{\rm eff.} \ni \frac{y^{d \; 2}}{16\pi^2}
  \frac{(y^\nu \vev{\overline{N}^c})^* M_{\bf 10}^* }{|M|^2} \; 
  \bar{\bf 5}\; H   + {\rm h.c.},
  \label{eq:5HKahler-41}  \\
\text{3+2 model} & & 
 K_{\rm eff.} \ni \frac{y^{\nu \; 2} }{16\pi^2} 
       \frac{(y^\nu \vev{\overline{N}}) M_N^* M_\Phi^* M_H^* }{|M|^4} \;
       \bar{\bf 5}\; H + {\rm h.c.}, 
  \label{eq:5HKahler-32}
\end{eqnarray}
where $|M|$'s in the denominator are the scale where the dominant 
contribution of the loop momentum integration comes from field 
theory on 3+1 dimensions. Similar contributions come from two other 
kinds of diagrams with particles in different $\SU(5)_{\rm GUT}$ 
representations running in the loop, 
just like several kinds of diagrams in Figure~\ref{fig:muterm}. 
%%%%%%%%%%%%%%%%%%%%%%%%%%%%%%%%%%%%%%%%%%%%%%%%%%%%%%%%%%%%%%%%%%%%%%%%%%%
%
%
%%%%%%%%%%%%%%%%%%%%%%%%%%%%%%%%%%%%%%%%%%%%%%%%%%%%%%%%%%%%%%%%%%%%%%%%%%%%%%
%
In both models, the coefficient $c$ in (\ref{eq:5H}) is roughly 
of the form 
\begin{equation}
 c \sim \frac{y^2}{16 \pi^2} \frac{y \vev{N}}{M},
\label{eq:estimate-c}
\end{equation}
where phases are ignored, and $\vev{N}$ means 
$\vev{\overline{N}^c}$ in the 4+1 model and $\vev{\overline{N}}$ 
in the 3+2 model. 
The estimate \eqref{eq:estimate-c} takes account only of 1-loop amplitudes 
with low-lying Kaluza--Klein multiplets in the loop, and 
there will also be UV contributions where higher Kaluza--Klein 
states and stringy states are in the loop. Although we have 
not estimated the UV contributions, it is unlikely that 
the UV and IR contributions cancel almost exactly, as we 
discussed for the kinetic mixing 
$K \ni \bar{\bf 5}^\dagger\; \bar{H}$.
Thus, it is likely that $c$ does not vanish in low-energy 
effective theory.

Holomorphic mass parameters $M$ are expected to have 
$\theta^2$ component in a vacuum with broken supersymmetry, at least 
of order $M(1 + \theta^2 m_{3/2})$. Therefore, the anti-holomorphic 
mass parameters in (\ref{eq:5HKahler-41}, \ref{eq:5HKahler-32}) 
have $\bar{\theta}^2$ components, and hence the 1-loop amplitudes
generate an R-parity violating bilinear term
\begin{equation}
 W_{\rm eff.} \ni c m_{3/2} \bar{\bf 5}\; H({\bf 5}).
\end{equation}

In the $\SU(3)_C$ triplet part, this additional 
mass term $\Delta W = \mu_i \bar{D}_i H({\bf 3})_*$ is not significant.
It deforms the mass matrix including  
$W \ni M_{H} \; H^c(\bar{\bf 3})_* H({\bf 3})_*$ 
in the $\SU(3)_C$-$\bar{\bf 3}+{\bf 3}$ sector, and induces mixing 
between $H^c(\bar{\bf 3})_*$ and $\bar{D}$. Massless eigenstates 
$\hat{\bar{D}}_i$ pick up interactions of 
$H^c(\bar{\bf 3})_* \subset H^c(\bar{\bf 5})_*$ with mixing coefficients 
$- \mu_i/M_{H}$. Thus, the massless eigenstates 
have trilinear interactions 
\begin{equation}
%  \text{4+1~model} \qquad 
 W_{\rm eff.}^{(4+1)} \ni - y^d_{(1)} \frac{\mu_i}{M_{H}} \; 
  \hat{\bar{D}}_i \; \bar{U} \; \bar{D} % \equiv {\cal O}''_0  
   - y^d_{(1)} \frac{\mu_i}{M_H} \; \hat{\bar{D}}_i \; Q \; L,
\label{eq:Op0''}
\end{equation}
in the 4+1 model, where $H^c(\bar{\bf 5}) = \bar{H}(\bar{\bf 5})$. 
When $M_{H} \sim M_{\rm KK}$ is around the GUT scale, 
effective R-parity violating coupling 
$\lambda'' \sim - y^d \mu_i/M_H$ is very small, and 
irrelevant to phenomenology except in nucleon decay processes 
discussed in section~\ref{ssec:proton}. 
This coupling is not even generated in the 3+2 model, 
because $H^c$ is different from $\bar{H}$.

In the doublet part, however, the additional mass terms, 
\begin{equation}
 W_{\rm eff.} \ni c_i m_{3/2} L_i H_u \equiv \mu_i L_i H_u  
\label{eq:mui}
\end{equation}
with $c_i$ roughly given by (\ref{eq:estimate-c}), are important 
in phenomenology. Since $H_u$ only has a small mass term 
$\Delta W = \mu_0 \; H_d H_u$, with the $\mu$-parameter $\mu_0$ 
of the order of the electroweak scale, 
the mixing angle of $L_i$--$H_d$ mixing,  $\sim \mu_i/\mu_0$,
is not so small as $\mu_i/M_H$.
Although this mixing angle is quite smaller than ${\cal O}(1)$ because of the suppression 
factor $c$ in (\ref{eq:estimate-c}), 
it is much larger than $\mu_i/M_H$ and can be phenomenologically significant.

In addition to $L_i$--$H_d$ mixing in massless states, 
$\hat{L}_i$ have mixing of the order $\mu_i/M_H$ with massive states in the doublet part 
of $H^c(\bar{\bf 5})_*$.
In the 4+1 model, this mixing generates trilinear R-parity violating interactions 
\begin{equation}
 W_{\rm eff.} \ni - y^d_{(1)} \frac{\mu_i}{M_H} \; 
   \hat{L}_i \; \bar{E} \; L 
   - y^d_{(1)} \frac{\mu_i}{M_H} \; \hat{L}_i \; Q \; \bar{D},
 \label{eq:Op0+Op0'}
\end{equation}
since $H^c(\bar{\bf 5}) \equiv \bar{H}(\bar{\bf 5})$ in the 4+1 model.
If $M_H \sim M_{\rm KK}$ is around the GUT scale, 
however, effective trilinear R-parity violating couplings 
$\lambda \sim \lambda' \sim - y^d_{(1)} \mu_i/M_H$  
are so small that it is negligible in phenomenology compared 
with $\mu_i/\mu_0$ contribution from \eqref{eq:mui}. 
Thus, even in the 4+1 model, 
trilinear R-parity violating couplings (\ref{eq:Op0''},
\ref{eq:Op0+Op0'}) are
too small to have phenomenological significance, 
and the bilinear mass term (\ref{eq:mui}) is virtually 
the only phenomenologically relevant R-parity violation 
at renormalizable level.\footnote{In section~\ref{ssec:proton},
however, we will see that the operator (\ref{eq:Op0''}) can be 
important for some choice of parameters. } 

% This additional mass term is not significant in the $\SU(3)_C$ 
% triplet part, as $c m_{3/2}$ is not more than the electroweak scale, 
% while all the colored Higgs multiplets are supposed to have much 
% larger mass terms. In the doublet part, however, this can be significant. 
% We have 
%
%\begin{equation}
% W \ni \mu_i L_i H_u \equiv c_i m_{3/2} L_i H_u,
%\label{eq:mui}
%\end{equation}
%
%with $c_i$ rougly given by (\ref{eq:estimate-c}) if there is no
%cancellation or extra contributions that are much larger.

We have seen so far that the trilinear R-parity violating operators 
\eqref{eq:dim4} are either absent [in the 3+2 model] or negligibly small[in the 4+1 model]
 in this framework. This framework predicts bilinear-dominated R-parity 
violation at renormalizable level, which was assumed in \cite{HS}. 
The dimension-4 proton decay problem is absent in bilinear-dominated 
R-parity violation, as baryon number symmetry is preserved 
(apart from a small breaking in (\ref{eq:Op0''})). 
So, it is easier to satisfy other phenomenological constraints as well in 
bilinear-term domination, comparing to the case of trilinear-term domination; as long as $\mu_i/\mu$'s (and bilinear
R-parity violating terms in the SUSY-breaking potential) remain 
sufficiently small, everything is fine. The only question is 
why $\mu_i/\mu$'s are small.

In our framework, coefficients $c_i$'s are all suppressed by 
$(y^\nu \vev{N})/M$ as in \cite{GR, DESY-Rparity}.
$\mu_i = c_i m_{3/2}$ becomes even smaller for smaller gravitino 
mass \cite{GR}.\footnote{Possible contributions from messenger sector 
fields in gauge mediation models have to be studied separately. 
This issue is not covered in this article, since such contributions 
will depend very much on details of models of gauge mediation.}
Even if gravitino mass is not very small, $c_i$'s involve an extra 
1-loop factor in (\ref{eq:estimate-c}) since those operators 
are generated at 1-loop.
$c_i$'s may be suppressed further by some ratio of mass parameters 
that we find in (\ref{eq:5HKahler-41}, \ref{eq:5HKahler-32}).
Therefore, there are many reasons for $\mu_i$ to be small in this framework. 

As we have made it clear, however, the rough estimate 
of the 1-loop amplitude (\ref{eq:estimate-c}) only accounts for 
a partial contribution. 
To see if we can rely on the estimate, let us first discuss 
if there are not any tree-level contributions at all. 
Secondly, we also need to see if the infinite Kaluza--Klein
particles and stringy states in the 1-loop amplitude could give 
significantly larger contributions than (\ref{eq:estimate-c}). 
% that comes only from low-lying Kaluza--Klein states. 

In order to find out whether there is a tree-level contribution, 
it is desirable, in principle, to calculate a sphere amplitude
on a Calabi--Yau background. In reality, though, such a calculation 
is rarely available except in orbifold limits of Calabi--Yau 
3-folds. However, tree-level 1PI effective action has been 
calculated for the Heterotic $E_8 \times E'_8$ string theory 
on a flat 9+1 dimensional spacetime in $\alpha'$-expansion.
Here, $\alpha' = 1/M_s^2$, and $M_s$ is the string scale. 
If $\mu_i \; L_i \; H_u$ mass terms are obtained from dimensional 
reduction of the tree-level 1PI effective action on 9+1 dimensions, 
it is likely that they are obtained also in sphere amplitudes 
for compactified models. If $\mu_i$'s obtained from dimensional 
reduction are proportional to a positive power of $M_s$, it 
is likely that they are, too, in compactified models, because 
we expect that the process of compactification introduces 
only $M_{\rm KK}$-dependence, not extra $M_s$ dependence. 
Thus, we content ourselves with guessing whether $\mu_i$'s are 
generated at tree-level, and if they are generated, how they 
depend on $M_s$, by using dimensional reduction of 1PI effective 
action on the flat 9+1 dimensional spacetime.

In order to find out whether dimensional reduction gives rise to 
% the $\bar{\bf 5}_i$--$H({\bf 5})$ mass 
$\mu_i$'s proportional to gravitino mass, 
we need to assume an origin of non-vanishing $\vev{W^*}$. 
It will depend on the assumption whether there are tree-level
contributions or not. 
In this article, we study only one possibility for the origin of 
$\vev{W^*}$, just to illustrate what one should consider 
to guess the tree-level contributions to $\mu_i$'s.

Let us suppose that $\vev{W^*}$ originates from a 3-form flux 
$\int_X \overline{\Omega} \wedge H \neq 0$ \cite{DIN1}. 
If 
% the $\bar{\bf 5}$--$H({\bf 5})$ 
a mass term of left-handed chiral fermions is to come 
% out of a term of effective action on 9+1 dimensions 
through dimensional reduction, then 
there should be 
% such a term should contain at least 
two gauge fermions $\Psi_{\bar{\alpha}} \equiv 
\sum_{a=1,2,3} \; \Psi^a \; e_{\bar{\alpha} a}$ 
in a term of effective action on 9+1 dimensions.
Since the mass parameter is supposed to be proportional to gravitino mass, 
a 3-form vev $\vev{H_{\alpha\beta\gamma}}$ should also be involved. 
In the 4+1 model, we further need a vev of an anti-chiral multiplet 
$\overline{N}^{c\, \dagger}$, 
which originates from a vector field $A_{\alpha}$ vev.
In order to write down such a term that is at least invariant under 
the holonomy $\SU(3)$, it must contain at least three derivatives; 
candidates of such terms that appear first in $\alpha'$-expansion 
are of the form 
\begin{equation}
 \frac{1}{\alpha' g_s^2} 
  \tr \left(D_\alpha \Psi_{\bar{\alpha}}
            D_\beta \Psi_{\bar{\beta}} 
            D_{\bar{\gamma}} A_{\gamma} \right)
  H_{\delta \epsilon \zeta} \; d^{10}y,
\label{eq:10D-HpspsA-41} 
\end{equation}
with $\SU(3)$ indices contracted by metric $h_{\alpha \bar{\alpha}}$, 
holomorphic 3-form $\Omega_{\alpha\beta\gamma}$ and its Hermitian 
conjugate $\overline{\Omega}_{\bar{\alpha}\bar{\beta}\bar{\gamma}}$.
This term is proportional to $1/g_s^2$ because it is suppose to be 
in the tree-level effective action. A dimensionful coefficient 
$1/\alpha'$ was supplied based on dimensional analysis. 

One can see, however, that such terms cannot be made 
$\SO(6) \simeq \SU(4)$ invariant. To see this, note first 
that the SU(3) indices in (\ref{eq:10D-HpspsA-41}) can be 
contracted by one 
$\overline{\Omega}_{\bar{\delta} \bar{\epsilon} \bar{\zeta}}$ 
and some metrics $h_{\alpha \bar{\alpha}}$. 
Since $\SU(4)$ contains $\SU(3) \times \U(1)$, (\ref{eq:10D-HpspsA-41})
should be neutral under the U(1), if it is to be made invariant 
under $\SO(6) \sim \SU(4)$. 
Remembering how spinor and vector objects transform\footnote{
A spinor field $\Psi_{\bar{\alpha}}$ is in the $\bar{\bf 3}^{1}$ 
representation of $\SU(3) \times \U(1)$, vector-type objects such as 
$D_\alpha$ and $A_\alpha$ transform as ${\bf 3}^{+2}$, and 
anti-vector-type objects such as $D_{\bar{\alpha}}$ as 
$\bar{\bf 3}^{-2}$.} under $\SU(3) \times \U(1)$, 
one finds that (\ref{eq:10D-HpspsA-41}) with 
an $\overline{\Omega}_{\bar{\delta}\bar{\epsilon}\bar{\zeta}}$ 
has $+6$ units of the $U(1)$ charge. 
Thus, (\ref{eq:10D-HpspsA-41}) cannot be made invariant under 
$\SU(3) \times \U(1) \subset \SU(4)$, no matter how the indices 
are contracted by metric. Needless to say, there is no term 
of the form (\ref{eq:10D-HpspsA-41}) that is invariant under 
$\SO(9,1)$. 

There is no tree-level contribution even from higher order 
terms in ${\cal O}(\alpha' D^2)$ expansion. 
This is because $D_{\alpha} D_{\bar{\alpha}}$ is neutral 
under U(1), and so is $(D_{\bar{\alpha}})^3$ accompanied by 
$\Omega_{\alpha\beta\gamma}$.

In the 3+2 model, on the other hand, it is vev's of chiral 
multiplets $\overline{N}$ that are inserted in (\ref{eq:5HKahler-32}).
Thus, a Lorentz invariant operator can exist in 9+1 dimensions at tree level, 
\begin{equation}
 \frac{1}{\alpha^{'3} g_s^2} \tr \left( 
  \overline{\Psi}_{\bar{\alpha}} \left(\Gamma^{\bar{\beta}} A_{\bar{\beta}}
			    \right)^{\bar{\alpha} \bar{\gamma}} 
  \Psi_{\bar{\gamma}}\right),
\label{eq:gaugino-kin-10D}
\end{equation}
which is nothing but a part of the kinetic term of gauginos 
in 9+1 dimensions. This term has become a part of the superpotential 
(\ref{eq:Het-super}). But it was a part of the assumptions of the 3+2
model that the massless doublet $\hat{H}_0$ remains in the low-energy 
spectrum because of rank reduction in the coupling 
$W \ni y^\nu \hat{H}_0 \; \overline{N}_0 \; \bar{\bf 5}_0 $. 
Therefore, the existence of (\ref{eq:gaugino-kin-10D}) 
in 9+1 dimensions is irrelevant to the question of whether 
the $L_i$--$H_u$ mass term is generated at tree level; even if the vev of 
gauge field moduli $\overline{N}_0$'s shift by of order $m_{3/2}$ 
in the presence of SUSY-breaking, nothing happens because 
the coupling is absent. 
Higher order terms in $\alpha'$-expansion may give rise to the 
tree-level $\hat{\bar{{\bf 5}}}_0$--$\hat{H}_0$ mass term in the 3+2 model 
as well;
\begin{equation}
 \frac{1}{\alpha' g_s^2} \tr \left( 
         D_\alpha \overline{\Psi}_{\bar{\alpha}}
         D_\beta A_{\bar{\beta}} 
         D_\gamma \Psi_{\bar{\gamma}}  
                             \right) 
         H_{\delta\epsilon\zeta} 
\label{eq:10D-HpspsA-32}
\end{equation}
can be made invariant under the holonomy $\SU(3)$ after the indices 
are contracted by $h_{\alpha \bar{\alpha}}$,
$\Omega_{\alpha\beta\gamma}$ and 
$\overline{\Omega}_{\bar{\alpha}\bar{\beta}\bar{\gamma}}$.
Following the same argument as in the 4+1 model, however, 
there is no way making (\ref{eq:10D-HpspsA-32}) invariant 
under $\SU(3) \times \U(1) \subset \SU(4) \simeq \SO(6)$.
There are no higher order terms in ${\cal O}(\alpha' D^2)$ expansion 
in the SO(9,1)-invariant effective action whose dimensional reduction 
gives rise to the tree-level contribution, just like in the 4+1 model.

% but this is not invariant under  Following the same argument as in the 4+1 model, 
%the leading contribution comes from an ${\cal O}(\alpha^{'2})$ term 
%or even higher order ones in $\alpha'$-expansion. Thus, the estimate 
%of the upper bound on the tree-level contribution is also the same 
%as (\ref{eq:mui-tree-expect}) in the 3+2 model, and the tree-level 
%contribution is unlikely to be as large as the 1-loop contribution.

The other issue is the contributions to the 1-loop amplitude 
with infinite Kaluza--Klein particles and stringy states in the loop.
The 1-loop amplitudes (\ref{eq:5HKahler-41}, \ref{eq:5HKahler-32}) 
are obtained by treating Kaluza--Klein towers as if they contained 
one (or a finite number of) Kaluza--Klein excitation(s) in each tower. 
Such amplitudes are UV-finite, because a superficial degree of divergence is 
$-2$ and $-4$ in the 4+1 and 3+2 model, respectively. 
But Kaluza--Klein towers contain infinite degrees of freedom 
on 3+1 dimensions, and their contributions are approximated by 
integrating loop momentum in 10D, not in 4D. Thus, the superficial 
degree of divergence becomes $+4$ and $+2$, respectively.
These UV divergence is tamed and made finite 
in string theory. The most naive guess is that the estimate 
(\ref{eq:estimate-c}) is multiplied by a factor of 
$(M_s / M_{\rm KK})^4$ and $(M_s/M_{\rm KK})^2$, respectively, 
expecting that the UV divergence is tamed at least by stringy states 
at an energy scale around $M_s$.

It is too naive, however, to make a guess based only on the superficial 
degree of divergence. Let us take 1-loop correction
to a gauge coupling constant as an example. 1-loop correction to 
$1/g_{\rm YM}^2$ is expected naively to be of order 
$(M_s/M_{\rm KK})^6$, since the superficial degree of divergence 
is $+6$ for amplitudes with only two gauge fields 
in the external lines in 9+1 dimensional spacetime. The 1-loop
correction, however, is known in orbifold calculations and is of order 
$(M_s/M_{\rm KK})^2$ \cite{DKL}. This explicit example clearly indicates that 
the guess based on the superficial degree of divergence is too naive.

1-loop 1PI effective action on flat 9+1 dimensional spacetime can explain 
why the 1-loop threshold correction is of order $(M_s/M_{\rm KK})^2$, 
not $(M_s/M_{\rm KK})^6$. 
If 1-loop 1PI effective action had a term $1/\alpha^{'3} \tr (|F|^2)$, 
then its dimensional reduction would give rise to a threshold correction  
of order $(R^6/\alpha^{'3})=(M_s/M_{\rm KK})^6$. But, it is known 
from worldsheet calculation \cite{GHMR,Yahikozawa} that 
there is no such correction.
% the coefficient of the 
% kinetic term of the super Yang--Mills multiplet on 9+1 dimensions, 
% which is proportional to $1/\alpha^{'3}=M_s^6$, 
% does not receive 1-loop corrections. 
Absence of an ${\cal O}(M_s^6)$ term in the 1-loop effective action 
on flat 9+1 dimensional spacetime corresponds to the absence of 
an ${\cal O}(M_s^6)$ term in the 1-loop threshold corrections to the gauge coupling 
constant of compactified models. 
On the other hand, the 1-loop 1PI effective action on flat 
9+1 dimensional spacetime contains a term \cite{EJM}
\begin{equation}
 d^{10} y \frac{1}{\alpha'} \tr (|F|^2) \tr(|F|^2), 
\label{eq:F4-1loop}
\end{equation}
and dimensional reduction of this term explains the threshold
correction of order $(M_s/M_{\rm KK})^2$ in compactified models. 

String-scale dependence of observables (such as the gauge coupling constant)
in compactified models may be determined by dimensional reduction of 
1PI effective action of string theory on flat 9+1 dimensions. 
This prescription is known to work for threshold correction of 
gauge couplings. 
Although it is desirable to check with orbifold calculations 
whether this prescription is correct for other observable quantities, 
yet it does not seem to be terribly wrong. 
Flat 9+1 dimensional spacetime and its compactification share 
spacetime structure (including an extended supersymmetry) and 
interactions on it at short distance. Thus, the string scale 
dependence may be the same for both.

Let us use this prescription for the 1-loop amplitude of 
the $\bar{\bf 5}$--$H({\bf 5})$ mass term; a brute force world sheet 
calculation would only be possible for orbifold compactification, 
and even in orbifold, it would take some time.
If we are to use known results on 1PI effective action on 
9+1 dimensions instead, we have already seen in the context 
of tree-level contributions that the mass term of chiral fermions 
in question cannot be obtained through dimensional reduction. 
The same is true for the 1-loop 1PI effective action as well, 
since the result at tree-level is based only on $\SO(9,1)$ invariance.
%
% the leading term in the Lorentz invariant effective action 
% on 9+1 dimensions start at ${\cal O}(\alpha^{'2})$ or even higher order 
% in $\alpha'$-expansion. 
% The same should be true for the 1-loop effective action as well, 
% since the absence of relevant terms up to ${\cal O}(\alpha')$ 
% is based only on $\SO(9,1)$ invariance. 
% Dimensional reduction of ${\cal O}(\alpha^{'2})$ terms 
% do not contribute to $\mu_i$ with a positive power in $M_s$. 
%
Therefore, the 1-loop amplitudes with higher Kaluza--Klein states 
and stringy states in the loop are not likely to contribute 
to the total amplitude.
On the other hand, contributions from low-lying Kaluza--Klein states 
are not captured by the world-sheet calculations on 
a flat 9+1 dimensional spacetime, since particle spectra 
on compactified space are quite different from those 
on a flat space at that scale. 
These IR contributions have already been estimated 
earlier in this subsection. 
Thus, the rough estimate (\ref{eq:estimate-c}) is valid, 
providing our prescription is justified.

The same prescription can be applied to the 1-loop contribution 
to the $\mu$-term in the 4+1 model that we described in 
section~\ref{ssec:4132}. Assuming that the non-vanishing gravitino 
mass is from the 3-form flux, we find that no such term can be 
written down in effective action on 9+1 dimensional spacetime with a 
coefficient of positive power of $1/\alpha' = M_s^2$.
Thus, the 1-loop contributions from around the Kaluza--Klein scale 
dominate the amplitude, and hence it is likely that 
the estimate following from (\ref{eq:mu!}) remains valid. 

This prescription is also applied to the R-parity violating 
coefficients $\epsilon$ for the kinetic mixing 
in (\ref{eq:kahler-5Hbar}).
This R-parity violating effect can be erased away from 
the kinetic terms by field redefinition such as (\ref{eq:absorb-e}) 
and (\ref{eq:absorb-e10}), but it reappears in mass matrices 
and trilinear interactions that involve massive states.
Certainly no R-parity violating trilinear interactions are generated 
among massless modes of the MSSM after the redefinition.  
It also turned out in section~\ref{ssec:dim5p} that mixing 
of massless eigenstates in the external lines are irrelevant 
to the effective couplings of the dimension-5 proton decay 
operators. This additional R-parity violating mixing, 
in principle, contributes to R-parity violating non-renormalizable 
operators in the effective theory, once the massive states are 
integrated out. 
Since we will discuss 
R-parity violating dimension-5 operators in section~\ref{ssec:dim5}, 
we need an order-of-magnitude estimate of the additional R-parity 
violating mixing from the 1-loop amplitudes. 
UV-finite 1-loop amplitudes in (\ref{eq:5HbarKahler-41}, 
\ref{eq:5HbarKahler-32}) take account only of contributions 
where low-lying Kaluza--Klein modes are running in the loop. 
Until now, we have postponed studying whether contributions from 
higher Kaluza--Klein states and stringy states dominate over those 
from low-lying states. We will now use the prescription to 
make an estimate of the UV contributions.

The 1-loop kinetic mixing (\ref{eq:kahler-5Hbar}) can be obtained 
through dimensional reduction, if the 1-loop 1PI effective action 
contains a term of the form 
\begin{equation}
 \frac{1}{\alpha^{'2-n}} \; 
   \tr ( F_{\mu\bar{\alpha}} (D_{\gamma} D_{\bar{\gamma}})^n 
         F^{\mu}_{\alpha} F_{\beta\bar{\beta}})  \; d^{10}y
 \qquad (n \geq 0) , 
\end{equation}
where vector indices $\alpha, \beta, \gamma, \bar{\alpha}, 
\bar{\beta}$ and $\bar{\gamma}$ are contracted by metric.
In fact, there are no terms cubic in field strengths 
in the 1-loop 1PI effective action~\cite{GHMR,Yahikozawa}. 
Although the 1PI effective action has a term quartic in field 
strengths, the leading ${\cal O}(1/\alpha')$ term  
is totally symmetric in the four field strengths \cite{EJM}, 
and is factorized into two $E_8$-singlets, as in (\ref{eq:F4-1loop}). 
Since none of $\bar{\bf 5}^\dagger \; N \; \bar{H}$ cannot be 
factorized into two singlets of $\U(1)_{\chi, \tilde{q}_7} \subset E_8$, 
the 1-loop kinetic mixing amplitudes are not obtained 
from dimensional reduction of quartic terms. 
% factorized into two $E_8$-singlets.
Thus, there is not even a term proportional to $\alpha^{'-1}$. 
All other terms with a coefficient in negative power of $\alpha'$
do not give rise to the kinetic mixing terms through dimensional 
reduction. Thus, we expect that the contribution from higher 
Kaluza--Klein states and stringy states is not proportional 
to a positive power of string scale, possibly because of some 
cancellation due to extended supersymmetry of those particles. 
The R-parity violating coefficients $\epsilon$, therefore, 
will not be much larger than the contributions from low-lying 
Kaluza--Klein states in (\ref{eq:5HbarKahler-41},
\ref{eq:5HbarKahler-32}), and we have an estimate 
\begin{equation}
  \epsilon \sim \frac{(y^{d})^2}{16\pi^2} \; 
                \frac{y^\nu \vev{N}}{M_{\rm KK}};
\label{eq:estimate-e}
\end{equation}
see comments after (\ref{eq:estimate-c}) for the meaning of $\vev{N}$.

R-parity violating vev's deform mass matrices at tree-level 
as in (\ref{eq:41-matrix10}, \ref{eq:41-matrix5}, 
\ref{eq:32-matrix10}, \ref{eq:32-matrix5}), and the mixing angles 
are of the order $y^\nu \vev{N}/M_{\rm KK}$. 
The additional mixing due to the 1-loop kinetic mixing is of order 
$\epsilon$. Since the estimate (\ref{eq:estimate-e}) contains 
an extra 1-loop factor, this additional mixing (\ref{eq:absorb-e},
\ref{eq:absorb-e10}) is quantitatively negligible. 
%
%Therefore, the order-of-magnitude estimates of 
%dimension-5 proton decay amplitudes in section~\ref{ssec:dim5p} 
%remain valid. 
%
Since the 1-loop mixing is qualitatively the same 
as the tree-level mixing, and negligible quantitatively, 
we will not make a distinction between the two different 
bases of (\ref{eq:absorb-e}, \ref{eq:absorb-e10}) in the following.

%%%%%%%%%%%%%%%%%%%%%%%%%%%%%%%%%%%%%%%%%%%%%%%%%%%%%%%%%%%
\subsection{R-parity Violating Dimension-5 Operators}
\label{ssec:dim5}
%%%%%%%%%%%%%%%%%%%%%%%%%%%%%%%%%%%%%%%%%%%%%%%%%%%%%%%%%%%

Low-energy effective theories are described only by light 
degrees of freedom, with all heavy states integrated out. 
Numerous non-renormalizable operators are generated as 
heavy states are integrated out, and naturally R-parity 
violation will also manifest itself in the non-renormalizable 
operators. 
There are four dimension-5 operators that are $\SU(5)_{\rm GUT}$ 
invariant and are odd under the R parity:
\begin{equation}
 W_{\rm eff.} \ni \frac{1}{M_{\rm eff.}}
   \hat{\bf 10}_0 \; \hat{\bf 10}_0 \; \hat{\bf 10}_0 \; \hat{\bar{H}}({\bf 5})_0 
 + \frac{1}{M_{\rm eff.}} 
   \hat{H}({\bf 5})_0 \; \hat{\bar{H}}(\bar{\bf 5})_0 \; 
   \hat{H}({\bf 5})_0 \; \hat{\bar{{\bf 5}}}_0
\label{eq:dim5-RPV-super}
\end{equation}
in the superpotential and 
\begin{equation}
 K_{\rm eff.} \ni \frac{1}{M_{\rm eff.}}
    \hat{{\bf 10}}_0 \; \hat{{\bf 10}}_0 \; \hat{\bar{{\bf 5}}}^\dagger_0
 + \frac{1}{M_{\rm eff.}}
  \hat{{\bf 10}}_0 \; \hat{\bar{H}}(\bar{\bf 5})_0 \;
  \hat{H}({\bf 5})^\dagger_0 + {\rm h.c.}
\label{eq:dim5-RPV-Kahler}
\end{equation}
in the K\"{a}hler potential\footnote{Although these operators are
trilinear in low-energy degrees of freedom, we only mean renormalizable 
trilinear ones in the superpotential by trilinear R-parity violation.} 
\cite{Allanach}. 
We will discuss in this subsection % ~\ref{ssec:dim5} 
whether these four operators are generated. 
For operators that are generated, the effective mass parameters 
$M_{\rm eff.}$ of these operators are expressed in terms of 
parameters of microscopic descriptions 
such as $M_{\rm KK}$, $y^\nu \vev{N}$ etc.

\subsubsection{Dimension-5 Operators in the Superpotential}

\begin{figure}
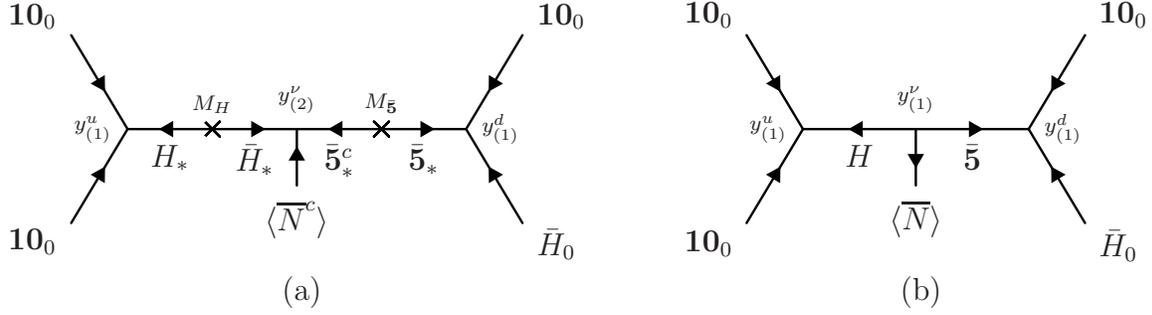

\begin{center}
\begin{tabular}{ccc}
\input{figure2-6a} &\hspace{2cm} & \input{figure2-6b} \\
&&\\
(a) && (b)
\end{tabular}
\end{center}
\caption{Typical Diagrams which generate $W \ni {\bf 10}\;{\bf 10}
\;{\bf 10}\; \bar{H}({\bf 5})$
for the 4+1 model (a) and for the 3+2 model (b).}
\label{fig:super-101010H}
\end{figure}

Let us begin with the first operator in (\ref{eq:dim5-RPV-super}).
This operator is generated by exchanging massive states in the 
$\SU(5)_{\rm GUT}$-$({\bf 5}+\bar{\bf 5})$ representations, as in 
the super Feynman diagrams in Figure~\ref{fig:super-101010H}. 
%
%%%%%%%%%%%%%%%%%%%%%%%%%%%%%%%%%%%%%%%%%%%%%%%%%%%%%%%
%%%%%%%%%%%%%%%%%%%%%%%%%%%%%%%%%%%%%%%%%%%%%%%%%%%%%%%%%%
%
Since this operator breaks R parity, R-parity violating vev's 
$\vev{\overline{N}^c}$ or $\vev{\overline{N}}$ have to be 
inserted either in the internal line or external lines in the 
diagram. Propagating in the internal line are massive states 
in the $\SU(5)_{\rm GUT}$-${\bf 5}+\bar{\bf 5}$ representations. 
Following the argument in (\ref{eq:dim5-KKsum-smplfd}), 
one finds that diagrams with the mixing in the internal line yield 
\begin{equation}
 \frac{1}{M_{\rm eff.}} \sim 
    y^d_{(1)} \; (M^{-1})_{\bar{\bf 5} H({\bf 5})} \; y^u_{(1)}
\label{eq:101010Hbar-EZ}
\end{equation}
in both the 4+1 and 3+2 model.

In the 4+1 model, one finds that the contribution from diagrams with 
the mixing in the internal line becomes  
\begin{equation}
 \frac{1}{M_{\rm eff.}} \sim y^d_{(1)}  y^u_{(1)}
  \frac{y^\nu_{(2)} \vev{\overline{N}^c}}{M_{\bar{\bf 5}} M_H}, 
\label{eq:temp}
\end{equation}
treating as if $M_{IJ}$ were a $2 \times 2$ or $3 \times 3$ matrix. 
We have seen in section~\ref{ssec:dim5p} that this finite rank 
treatment is sufficient in making a rough estimate of $M_{\rm eff.}$, 
although there are inifinite Kaluza--Klein particles.
Contributions due to $\hat{\bf 10}_0$--${{\bf 10}'}_*$ 
in the external lines are of order 
\begin{equation}
 \frac{1}{M_{\rm eff.}} \sim
%  y^u_{(1)} y^d_{(1)} 
%  \frac{y^\nu_{(2)} \vev{\overline{N}^c}}{M_{\bf 5} M_H}, \qquad 
  y^u_{(1)} y^d_{(2)} 
  \frac{y^{'\nu}_{(2)} \vev{\overline{N}^c}}{M_{{\bf 10}'} M_H}, \qquad
  y^u_{(2)} y^d_{(1)}
  \frac{y^{'\nu}_{(2)} \vev{\overline{N}^c}}{M_{{\bf 10}'} M_{\bar{\bf 5}}}. 
\end{equation}
$\hat{\bar{H}}(\bar{\bf 5})_0$--$\bar{\bf 5}_*$ mixing 
in the external line yields an amplitude of order (\ref{eq:temp}).
If all the massive states are around the Kaluza--Klein scale, 
$M_{\rm KK}$, then 
$1/M_{\rm eff.} \sim y^u y^d (y^\nu \vev{\overline{N}^c})/M_{\rm KK}^2$.
% in the 4+1 model. 
Since $1/M_{\rm eff.} \sim y^u y^d / M_{\rm KK}$ for the dimension-5 
proton decay operators (\ref{eq:op-dim5}), it is likely 
for $y^\nu \vev{\overline{N}^c} \ll M_{\rm KK}$ that 
the R-parity violating interaction 
$\Delta W = \hat{{\bf 10}}_0 \; \hat{{\bf 10}}_0 \; 
\hat{{\bf 10}}_0 \; \hat{\bar{H}}_0$ is weaker than the dimension-5 
proton decay operators in the 4+1 model. 

\begin{figure}
\begin{center}
% \begin{tabular}{c}
\input{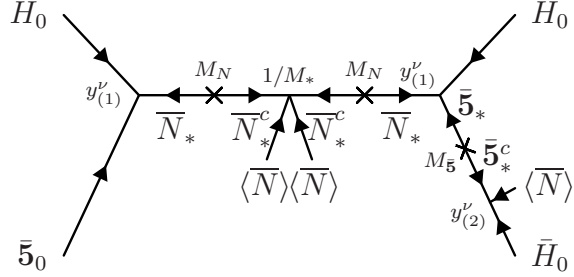} % &\hspace{14mm} & \input{figure_hhh5_4} \\
% &&\\
% (a) && (b)
% \end{tabular}
\end{center}
\caption{An example of super Feynman diagrams which generate
 $W \ni \hat{H}_0({\bf 5})\;\hat{\bar{H}}_0(\bar{\bf 5})\;
\hat{H}_0({\bf 5})\;\hat{\bar{{\bf 5}}}_0$ in the 3+2 model.
% The diagram (a) gives rise to the first one in
%  \eqref{eq:super-HHbarH5-32} and (b) to the second one in 
% \eqref{eq:super-HHbarH5-32} .
}
\label{fig:super-HHbarH5-32}
\end{figure}

In the 3+2 model, note first of all that mixing in the external lines 
is irrelevant. Massless eigenstates $\hat{\bf 10}_0$ remain pure 
$\U(1)_{\tilde{q}_7}$-eigenstates, and there is no mixing for these 
external states. The mixing of $\hat{\bar{H}}_0$ into 
$\U(1)_{\tilde{q}_7}$-eigenstates $\bar{\bf 5}_*$ or $H^c_*$ 
does not generate the opeartor 
$\Delta W = \hat{\bf 10}_0 \; \hat{\bf 10}_0 \; \hat{\bf 10}_0 
\; \hat{\bar{H}}_0$, since we have seen in section~\ref{ssec:dim5p} 
that there is no way to generate either 
$\Delta W = \hat{\bf 10}_0 \; \hat{\bf 10}_0 \; \hat{\bf 10}_0 
\; \bar{\bf 5}$ or 
$\Delta W = \hat{\bf 10}_0 \; \hat{\bf 10}_0 \; \hat{\bf 10}_0 
\; H^c$ in the 3+2 model.
We further notice that there is no contribution from mixing 
in the internal line by using (\ref{eq:101010Hbar-EZ}) 
and a $3 \times 3$ matrix (\ref{eq:3by3-Hc-mass}). The same is 
true for the case with $a > 0$ pairs of $\bar{\bf 5}^c$--$\bar{H}$ like 
states with masses of order $y^\nu \vev{\overline{N}}$.
On the other hand, if there are $b > 0$ pairs of zero-modes 
$H({\bf 5})_0$--$\bar{\bf 5}_0$ that become massive in the presence of 
$y^\nu_{(1)} \vev{\overline{N}} \neq 0$, the operator 
$W \ni \hat{{\bf 10}}_0 \; \hat{{\bf 10}}_0 \; 
\hat{{\bf 10}}_0 \; \hat{\bar{H}}(\bar{\bf 5})_0$ is generated, with 
\begin{equation}
 \frac{1}{M_{\rm eff.}} \sim y^u_{(1)} y^d_{(1)}
  \frac{1}{y^\nu_{(1)} \vev{\overline{N}}}.
\label{eq:in-denom}
\end{equation}
\begin{figure}
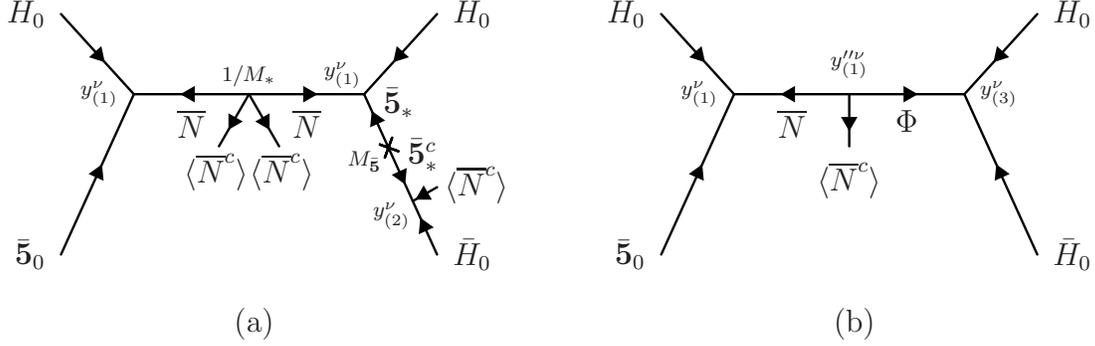

\begin{center}
\begin{tabular}{ccc}
 \input{figure_hhh5_2} &\hspace{2cm} & \input{figure_hhh5_1} \\
&&\\
(a) && (b)
\end{tabular}
\end{center}
\caption{Examples of super Feynman diagrams which generate
 $W \ni \hat{H}_0({\bf 5})\;\hat{\bar{H}}_0(\bar{\bf 5})\;\hat{H}_0({\bf 5})\;\hat{\bar{{\bf 5}}}_0$ in the 4+1 model.
% The diagram (a) gives rise to the first contribution in 
% \eqref{eq:super-HHbarH5-41} and (b) to the second one 
% in \eqref{eq:super-HHbarH5-41} .
}
\label{fig:super-HHbarH5-41}
\end{figure}
%%%%%%%%%%%%%%%%%%%%%%%%%%%%%%%%%%%%%%%%%%%%%%%%%%%%%%%%%%
%
The second R-parity violating operator in (\ref{eq:dim5-RPV-super})
is generated through diagrams in Figure~\ref{fig:super-HHbarH5-41} 
in the 4+1 model and Figure~\ref{fig:super-HHbarH5-32} in the 3+2 model.
There, massive $\SU(5)_{\rm GUT}$-singlets are integrated out.

In the 4+1 model, the effective mass scale from the diagram 
Figure~\ref{fig:super-HHbarH5-41}~(a) is approximately 
\begin{equation}
 \frac{1}{M_{\rm eff.}} \sim 
% y^\nu_{(1)} y^{\nu}_{(3)}
%   \frac{y^{''\nu}_{(1)} \vev{\overline{N}^c}}{M_N M_{\Phi}}, \qquad 
 (y^{\nu}_{(1)})^2 
  \frac{y^\nu_{(2)} \vev{\overline{N}^c}}{M_{\bar{\bf 5}}}
  \frac{1}{M_R} 
 \sim \frac{(y^\nu)^2}{M_{\rm KK}} 
      \frac{(y^\nu \vev{\overline{N}^c})}{M_{\rm KK}} 
      \frac{M_{\rm KK}}{M_R} ,  
  \label{eq:super-HHbarH5-41}
\end{equation}
where $M_R$ is defined in (\ref{eq:MR}).
Thus, the operator $\hat{H}_0 \; \hat{\bar{H}}_0 \; \hat{H}_0 \;
\hat{\bar{{\bf 5}}}_0$ is larger or smaller than the other 
R-parity violating dimension-5 operator 
$\hat{\bf 10}_0 \; \hat{\bf 10}_0 \; \hat{\bf 10}_0 \; 
\hat{\bar{H}}_0$ in the 4+1 model, depending on 
whether $M_R < M_{\rm KK}$ or not,\footnote{If the low-energy neutrino 
masses come from the see-saw mechanism involving right-handed neutrinos,
then $M_R$ cannot be larger than $10^{15} \; \GEV$. If $M_{\rm KK}$ is 
around $M_{\rm GUT} \sim 10^{16} \; \GEV$ or even larger, 
then $M_R < M_{\rm KK}$.} if difference among various trilinear 
couplings and mass spectra are ignored. 
If there are Kaluza--Klein zero modes in $\overline{N}$ 
(as assumed\footnote{They are necessary if masses of low-energy
neutrinos are to be generated from the seesaw mechanism involving 
right-handed neutrinos. Majorana masses from (\ref{eq:4N-3}) 
are not sufficient without a zero mode of $\overline{N}$
\cite{Nakajima}.} 
in Table~\ref{tab:repr41}), 
a diagram Figure~\ref{fig:super-HHbarH5-41}~(b) also contributes and 
\begin{equation}
 \frac{1}{M_{\rm eff.}} \sim y^\nu_{(1)} y^\nu_{(3)} 
    \frac{1}{y^{''\nu} \vev{\overline{N}^c}},
\end{equation}
which tends to be larger than (\ref{eq:super-HHbarH5-41}); 
see (\ref{eq:MR}). 
Therefore, in a crude approximation that all the Kaluza--Klein towers 
begin at a common scale $M_{\rm KK}$ and that 
$y^\nu \vev{\overline{N}^c}$ is somewhat smaller than $M_{\rm KK}$, 
R-parity violating $\Delta W = \hat{H}_0 \; \hat{\bar{H}}_0 \; \hat{H}_0 \;
\hat{\bar{{\bf 5}}}_0$ is enhanced by 
$((y^\nu \vev{\overline{N}})/M_{\rm KK})^{-1}$ relatively to 
the dimension-5 proton decay operators, and the other R-parity violating 
operator $\Delta W = \hat{\bf 10}_0 \; \hat{\bf 10}_0 \; \hat{\bf 10}_0 
\; \hat{\bar{H}}_0$ suppressed by 
$((y^\nu \vev{\overline{N}})/M_{\rm KK})^{-1}$ in the 4+1 model.
Here, we assume that there are zero modes of $\overline{N}$, and
difference among all the trilinear couplings are ignored. 

In the 3+2 model, an amplitude corresponding to the super Feynman diagram 
in Figure~\ref{fig:super-HHbarH5-32} is 
\begin{equation}
 \frac{1}{M_{\rm eff.}} \sim 
% % (y^{\nu})^2
% %  \frac{(y^\nu \vev{\overline{N}})^3}{M_5^3} \frac{1}{M_{N/\Phi}}, \qquad 
%  (y^\nu)^2 \frac{(y^\nu \vev{\overline{N}})^3}{M_{\rm KK}^4}, \qquad 
  (y^{\nu})^2 \frac{y^\nu \vev{\overline{N}}}{M_{\bar{\bf 5}}}
% %  \frac{\vev{\overline{N}}^2}{M_* M_N^2}, 
  \frac{1}{M_{R, {\rm eff.}}}, 
 \label{eq:super-HHbarH5-32}
\end{equation}
where $M_{R, {\rm eff.}}$ is defined in (\ref{eq:MReff}). 
Thus, we see that an R-parity violating dimension-5 operator 
is generated in the 3+2 model, independent of high-energy spectrum 
in the $\SU(5)_{\rm GUT}$-$\bar{\bf 5}+{\bf 5}$ sector [i.e., 
whether $b = 0$ or not]. 
For cases with $b > 0$, the second term in (\ref{eq:dim5-RPV-super}) 
has larger coefficient if $M_{R, {\rm eff.}} <  
(y^\nu\vev{\overline{N}})^2/M_{\rm KK}$, and 
otherwise, the first one has a larger coupling (\ref{eq:in-denom}).
%
%%%%%%%%%%%%%%%%%%%%%%%%%%%%%%%%%%%%%%%%%%%%%%%%%%%%%%%

\subsubsection{Dimension-5 Operators in the K\"{a}hler Potential}

\begin{figure}
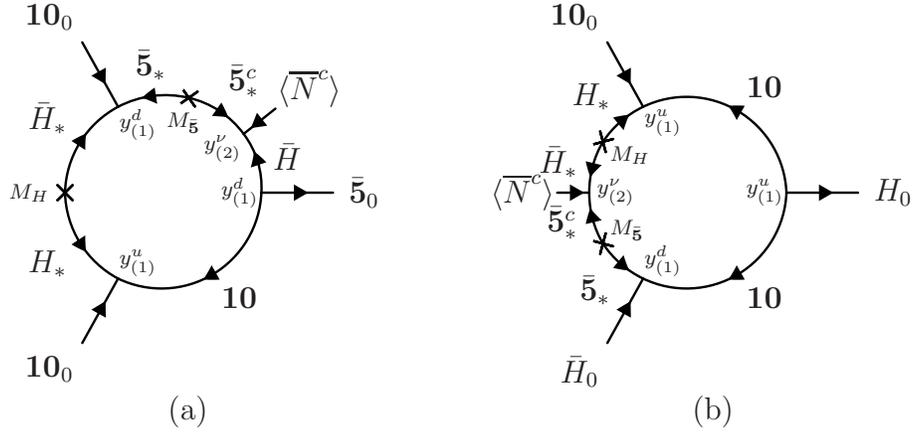

\begin{center}
\begin{tabular}{ccc}
\input{figure2-7a} & \hspace{2cm} & \input{figure2-7b} \\
&&\\
(a) && (b)
\end{tabular}
\end{center}
\caption{One-loop diagrams in the 4+1 model that generate R-parity
 violating dimension-5 operators in the K\"{a}hler potential.
% ${\bf 10}\;{\bf 10}\;\bar{\bf 5}^\dag$ (a)
%  and ${\bf 10}\;\bar{H}(\bar{\bf 5})\; H({\bf 5})^\dag$(b)  for the
% 4+1 model.
\label{fig:kahler-loop-41}}
\end{figure}

\begin{figure}
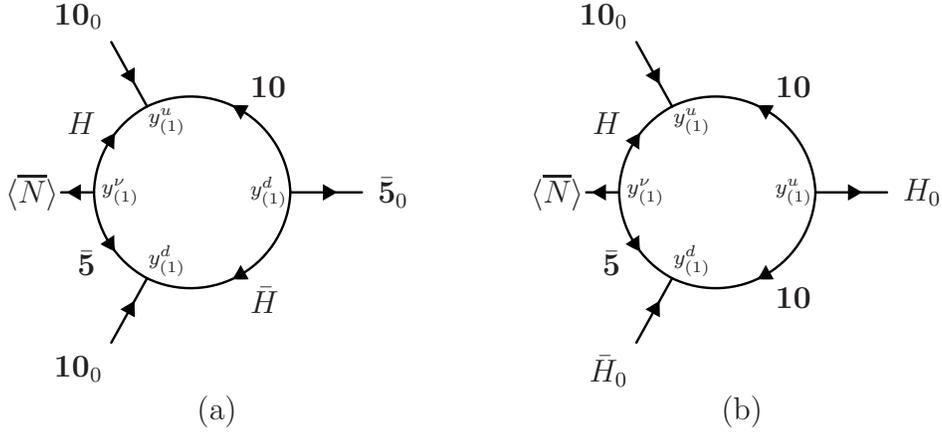

\begin{center}
\begin{tabular}{ccc}
\input{figure2-7c} & \hspace{2cm} & \input{figure2-7d} \\
&&\\
(a) && (b)
\end{tabular}
\end{center}
\caption{One-loop diagrams which generate ${\bf 10}\;{\bf 10}\;\bar{\bf 5}^\dag$ (a)
 and ${\bf 10}\;\bar{H}(\bar{\bf 5})\; H({\bf 5})^\dag$ (b) for the 3+2 model.}
\label{fig:kahler-loop-32}
\end{figure}

Two R-parity violating dimension-5 operators in the K\"{a}hler potential 
(\ref{eq:dim5-RPV-Kahler}) are also generated through 1-loop diagrams 
in Figure~\ref{fig:kahler-loop-41} (in the 4+1 model) and those 
in Figure~\ref{fig:kahler-loop-32} (in the 3+2 model). 
%
%%%%%%%%%%%%%%%%%%%%%%%%%%%%%%%%%%%%%%%%%%%%%%%%%%%%%%%%%%
%%%%%%%%%%%%%%%%%%%%%%%%%%%%%%%%%%%%%%%%%%%%%%%%
%
%
Only one Feynman diagram is shown for each operator in each model, 
although there are many others. We roughly estimate contribution
%  the coefficients of these operators 
from a given diagram, by treating Kaluza--Klein towers as if they 
were finite number of massive states in the 3+1 dimensions.
Estimates of the UV contributions are postponed until 
the end of this subsection.
This is just like what we did in sections~\ref{ssec:4132} and 
\ref{ssec:bilinear} for other 1-loop amplitudes. 
To take the operator 
$K \ni \hat{{\bf 10}}_0 \; \hat{{\bf 10}}_0\; 
\hat{\bar{{\bf 5}}}^\dagger_0$ as an example, contributions from 
% the 1-loop 
diagrams in Figure~\ref{fig:kahler-loop-41}~(a) and 
Figure~\ref{fig:kahler-loop-32}~(a) are roughly of order 
\begin{eqnarray}
 \text{4+1 model} & & \frac{1}{M_{\rm eff.}} \sim 
   \frac{|y^d|^2 y^u}{16\pi^2} 
   \frac{M_{\bf 5}^* M_H^* (y^\nu \vev{\overline{N}^c})}{|M|^4}, 
  \label{eq:estimate-10105Kahler-41}\\
 \text{3+2 model} & & \frac{1}{M_{\rm eff.}} \sim 
   \frac{|y^d|^2 y^u}{16\pi^2} 
  \frac{(y^\nu \vev{\overline{N}})^*}{|M|^2},
 \label{eq:estimate-10105Kahler-32}
\end{eqnarray}
where $|M|$'s in the denominators are energy scales where 
dominant contributions come from in integrations of loop momenta 
in field theory on 3+1 dimensional spacetime.
%
%%%%%%%%%%%%%%%%%%%%%%%%%%%%%%%%%%%%%%%%%%%%%%%%%%%%%%%%%%
%%%%%%%%%%%%%%%%%%%%%%%%%%%%%%%%%%%%%%%%%%%%%%%%
%
% Similar results will be obtained for the second 
% operator of (\ref{eq:dim5-RPV-Kahler}) as well. 
All the contributions, including those above, are roughly 
of order $1/M_{\rm KK}$ with a 1-loop factor $y^3/(16\pi^2)$ and 
an extra suppression factor $(y^\nu \vev{N})/M_{\rm KK}$.
Here, we assume that all the mass parameters appearing in the loop 
amplitudes are around a common Kaluza--Klein scale $M_{\rm KK}$.
We do not go into detailed discussion of whether those 
operators have enhanced contributions when states with 
masses of order ${\cal O}(y^\nu \vev{N})$ exist. 

The rough estimates above only account for 1-loop contributions 
from low-lying Kaluza--Klein particles. As we discussed 
in section~\ref{ssec:bilinear}, however, such contributions may 
not dominate in $1/M_{\rm eff.}$ in principle. 
The same operators in effective theories may also 
have tree-level contributions. It is also possible that 
the 1-loop amplitudes are dominated by contributions from 
higher Kaluza--Klein modes and stringy modes. 

\begin{figure}
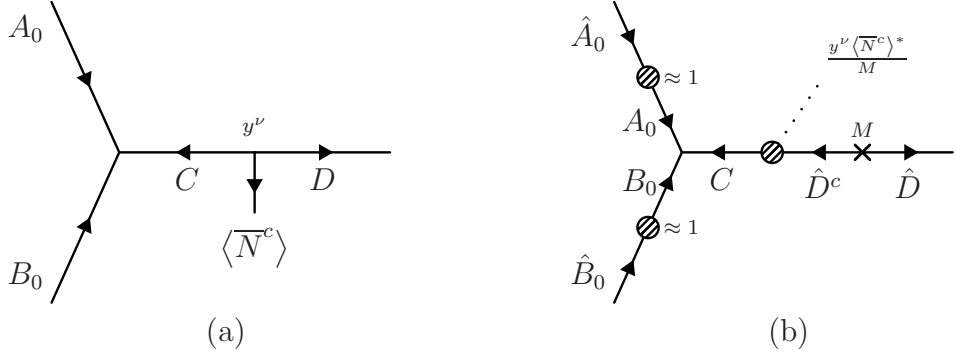

\begin{center}
\begin{tabular}{ccc}
\input{figure_kohler_1} &\hspace{2cm} & \input{figure_kohler_2} \\
(a) && (b)
\end{tabular}
\end{center}
\caption{
Starting with a trilinear superpotential 
$W \ni A_0 B_0 C + y^\nu C D_0 \vev{N}$, one might think that 
an effective operator $K \ni A_0B_0D^\dag_0$ could be generated in 
the K\"{a}hler potential as in the diagram (a).
However, such a trilinear term does not exist when the state going 
to the right is a massless eigenstate.
% However, such a trilinear term exists only when the state going 
% to the righ is massive. 
We are saying that the amplitudes in Figure~\ref{fig:super-HHbarH5-41} 
and \ref{fig:super-HHbarH5-32} with mixing in the external lines do not 
vanish for massless states, while the amplitude in
% Figure~\ref{fig:kohler-tree-explanation}
(a) does.
To understand this subtlety better, it helps to draw Feynman diagrams 
in terms of mass eigenstates as in (b). It is clear that a mass flip 
is necessary on the external line going to the right in (b), whereas 
it was not necessary in Figure~\ref{fig:super-HHbarH5-41} and 
Figure~\ref{fig:super-HHbarH5-32}.
This means that K\"{a}hler terms of the form $K \ni \hat{A}_0 \;
 \hat{B}_0 \; \hat{D}^\dag_0$, with all of external fields being
 massless, cannot be generated from a diagram like (a).
}
\label{fig:kohler-tree-explanation}
\end{figure}

First, it is easy to see that super Yang--Mills theory on 9+1 
dimensions does not give rise to the effective operators 
(\ref{eq:dim5-RPV-Kahler}) at tree-level; 
here, by super Yang--Mills theory, we only mean all the interactions 
that are derived from (Kaluza--Klein decomposition of) 
the leading order action on 9+1 dimensions, 
$(1/(\alpha^{'3} g_s^2))\tr (|F|^2) d^{10}y$. 
Supergraphs like those 
in Figure~\ref{fig:kohler-tree-explanation}~(a) exist only when one of 
chiral multiplets in the external lines is a massive mode. 
If all the external states are massless eigenmodes, such amplitudes 
vanish. For more detailed explanation, see the caption of 
Figure~\ref{fig:kohler-tree-explanation}. 
%
%%%%%%%%%%%%%%%%%%%%%%%%%%%%%%%%%%%%%%%%%%%%%%%%%%%%%%%
%%%%%%%%%%%%%%%%%%%%%%%%%%%%%%%%%%%%%%%%%%%%%%%%%%%%%%%%%%
%

Tree-level contributions to the effective interactions 
(\ref{eq:dim5-RPV-Kahler}) may also come from dimensional reduction 
of higher-order terms in $\alpha'$-expansion in the tree-level 
effective action on 9+1 dimensions. Let us take 
\begin{eqnarray}
 {\rm 4+1~model}: & & \bar{\bf 5}^\dagger \; {\bf 10} \;
             {\bf 10} \; \vev{\overline{N}^c}|_D =
     \partial_\mu \bar{\bf 5}^\dagger \; \partial^\mu {\bf 10} \; 
     {\bf 10} \; \vev{\overline{N}^c} + \cdots, \label{eq:eff-Kahler41}\\
 {\rm 3+2~model}: & & \bar{\bf 5}^\dagger \; \vev{\overline{N}}^* \; 
                                  {\bf 10} \; {\bf 10}|_D = 
     \partial_\mu \bar{\bf 5}^\dagger \; \vev{\overline{N}}^* \; 
     \partial^\mu {\bf 10} \; {\bf 10} + \cdots \label{eq:eff-Kahler32}
\end{eqnarray}
in the effective theory on 3+1 dimensions as an example for
concreteness; all the fields in the right-hand sides of 
(\ref{eq:eff-Kahler41}, \ref{eq:eff-Kahler32}) are complex scalars 
in given representations. Such interactions can be obtained through 
dimensional reduction, if string theory effective action has a term 
of the form 
\begin{equation}
\frac{1}{\alpha' g_s^2} \;  d^{10} y \; 
 \tr \left(F_{\mu \alpha} \; F^{\mu}_{\;\; \bar{\alpha}} \; 
           F_{\beta \bar{\beta}} \; F_{\gamma \bar{\gamma}} \right),
\end{equation}
with indices $\alpha$--$\gamma$ and $\bar{\alpha}$--$\bar{\gamma}$ 
contracted covariantly by metric $h_{\alpha \bar{\alpha}}$.
% if such a term exists in the effective action of string theory. 
It is important that all the four field strength tensors are 
in a single trace, not factorized into two. This is because 
there is no way separating four representations 
${\bf 5}^{-3}$, two ${\bf 10}^{-1}$'s and ${\bf 1}^{+5}$ in 
(\ref{eq:eff-Kahler41}) [resp. ${\bf 5}^{+1}$, 
two ${\bf 10}^{-3}$'s and ${\bf 1}^{+5}$ in (\ref{eq:eff-Kahler32})] 
into two singlets of $\SU(5)_{\rm GUT} \times \U(1)_\chi \subset E_8$ 
[resp. $\SU(5)_{\rm GUT} \times \U(1)_{\tilde{q}_7} \subset E_8$].
In fact, it is known that the ${\cal O}(1/\alpha')$ $F^4$ term 
of the sphere amplitude is factorized into two $E_8$ singlets in the 
Heterotic $E_8 \times E'_8$ string theory \cite{GrSl}.
Thus, the effective operators
(\ref{eq:eff-Kahler41}, \ref{eq:eff-Kahler32}) 
on 3+1 dimensions are not obtained through dimensional reduction of 
tree-level amplitudes, at least from ${\cal O}(1/\alpha')$ 
operators.
%
% If all the higher-order terms in $\alpha'$-expansion come in 
% powers of $\tr (|F|^2)$, as suggested in \cite{Bergshoeff}, then 
% we can safely say that there is no tree-level contribution obtained 
% from dimensional reduction at all order in $\alpha'$.
%
If the effective action on 9+1 dimensions contains a term 
$(1/g_s^2) \tr (F^5) \; d^{10}y$, which comes at the next order 
in $\alpha'$-expansion, then its dimensional reduction may give 
rise to a largest possible tree-level contribution
\begin{equation}
 \frac{1}{M_{\rm eff.}} \sim  
    \frac{g_{\rm YM}^2 \vev{N}}{M^2_{\rm KK}} \; 
    \left(\frac{M_{\rm KK}}{M_s}\right)^6. 
\end{equation}
Here, suppression due to small overlap of wavefunctions is 
ignored, and hence $y$'s and $g_{\rm YM}$ are much the same 
in this crude approximation. 
Rough estimates of 1-loop amplitudes (\ref{eq:estimate-10105Kahler-41}, 
\ref{eq:estimate-10105Kahler-32}) are 
$1/M_{\rm eff.} \sim g_{\rm YM}^2/(16\pi^2) \times 
(g_{\rm YM}^2 \vev{N} / M_{\rm KK}^2)$ at this level of crude
approximation. Thus, although tree-level contributions do not 
have a 1-loop factor $g_{\rm YM}^2/(16\pi^2)$, they are suppressed by 
$(M_{\rm KK}/M_s)^6$ or even more, and it is not very likely that 
they dominate over the estimates following from 
(\ref{eq:estimate-10105Kahler-41}, \ref{eq:estimate-10105Kahler-32}).

Let us now turn to 1-loop amplitudes. Rough estimates 
(\ref{eq:estimate-10105Kahler-41}, \ref{eq:estimate-10105Kahler-32}) 
take account only of contributions with low-lying Kaluza--Klein modes 
running in the loop. This is why the estimates based on field theory 
on 3+1 dimensions are UV finite; superficial degree 
of divergence is $-4$ and $-2$ in the 4+1 and 3+2 model, respectively.
When all the Kaluza--Klein states run in the loop, however, 
loop momentum is effectively integrated on a 10-dimensional space, 
and the superficial degree of divergence becomes $+2$ and $+4$, 
respectively. Naive guess would be, then, to multiply 
$(M_s/M_{\rm KK})^2$ and $(M_s/M_{\rm KK})^4$, respectively, to 
the rough estimates in (\ref{eq:estimate-10105Kahler-41}, 
\ref{eq:estimate-10105Kahler-32}), because stringy states may 
set in around the string scale $M_s$, taming the UV divergence.

We have already seen in section~\ref{ssec:bilinear}, however, that 
this guess is too naive. 
% 1-loop string amplitudes are UV finite, 
% and the question here is 
We introduced a prescription in section~\ref{ssec:bilinear} for 
how to guess whether the amplitudes $1/M_{\rm eff.}$ 
are enhanced by a positive power of $(M_s/M_{\rm KK})$:
read off $M_s = 1/\sqrt{\alpha'}$ dependence of terms in the 1-loop 
1PI effective action on 9+1 dimensions whose dimensional reduction 
give rise to low-energy effective interactions of our interest.
% , assuming 
% that no extra $M_s$-dependence is introduced through compactification.
%
% % A better way to see the contributions from infinite Kaluza--Klein states
% % and stringy states is to look at 1-loop effective action of string 
% % theory on 9+1 dimensions, as we did in section~\ref{ssec:bilinear}.
This prescription is based on a belief that Calabi--Yau compactification 
and a flat 9+1 dimensional spacetime have a common short-distance
spacetime and interactions on it, and that compactification introduces 
only $M_{\rm KK}$-dependence, but no extra $M_s$-dependence. 

1-loop effective action of the Heterotic $E_8 \times E'_8$ string 
theory on a flat 9+1 dimensional spacetime has been calculated from 
torus amplitudes. ${\cal O}(1/\alpha')$ terms quartic in gauge field 
strength are symmetric for four field strengths for the torus 
amplitude \cite{EJM}, and are factorized into two $E_8$-singlets, 
\begin{equation}
 \frac{1}{\alpha'} \tr \left(|F|^2 \right) \tr \left(|F|^2 \right) 
  \; d^{10} y. 
\end{equation} 
Dimensional reduction of this term does not give rise to the effective 
interactions (\ref{eq:eff-Kahler41}, \ref{eq:eff-Kahler32}) 
for the same reason explained for the tree-level amplitudes. 
All other operators that appear 
at higher order in the $\alpha'$-expansion do not have 
coefficients in negative power of $\alpha'$.
Therefore, the 1-loop amplitudes for the effective operators 
(\ref{eq:eff-Kahler41}, \ref{eq:eff-Kahler32}) will not have 
coefficients in a positive power of $M_s$, and in particular, the estimates 
(\ref{eq:estimate-10105Kahler-41}, \ref{eq:estimate-10105Kahler-32}) 
do not need to be changed.

%%%%%%%%%%%%%%%%%%%%%%%%%%%%%%%%%%%%%%%%%%%%%%%%%%
\subsection{Brief Summary}
\label{ssec:summary-frame}
%%%%%%%%%%%%%%%%%%%%%%%%%%%%%%%%%%%%%%%%%%%%%%%%%%

The minimal supersymmetric standard model has four independent 
chiral multiplets in the $({\bf 2},-1/2)$ representation of 
the $\SU(2)_L \times \U(1)_Y$ gauge group.
The dimension-4 proton decay problem implies that there must be 
some structure in the vector space ${\cal L}$ spanned by the 
four chiral multiplets. 
The most popular approach has been to introduce an 
unbroken discrete symmetry such as R parity. The vector space splits 
into two (or more) subspaces that transform differently under the discrete 
symmetry transformation. In the case of an unbroken $\Z_2$ symmetry, 
for example, 
\begin{equation}
 {\cal L} \simeq {\cal L}^+ \oplus {\cal L}^- 
          \equiv
        {\rm Span} \left\{ H_d \right\} \oplus 
        {\rm Span} \left\{ L_1, L_2, L_3 \right\}.
\end{equation}
Chiral multiplets in the two subspaces have different interactions, 
and hence down-type Yukawa couplings can exist, while trilinear 
R-parity violating couplings do not. 

An alternative approach presented in this article does not need to 
assume that a discrete symmetry remains unbroken. In order to 
solve the dimension-4 proton decay problem, the vector space 
${\cal L}$ does not need to split. It is sufficient to assume 
that ${\cal L}$ has a subspace
\begin{equation}
 {\cal L} \supset {\cal L}^{L}; 
\label{eq:subspace5}
\end{equation}
chiral multiplets that belong to the subspace ${\cal L}^{L}$ 
have restricted interactions. Three independent degrees of freedom 
in ${\cal L}^{L}$ are identified with the lepton doublets, 
and the remaining one degree of freedom in 
${\cal L} / {\cal L}^{L} \simeq {\cal L}^{H_d}$ with the
down-type Higgs doublet. It is not strictly necessary that 
${\cal L}$ has a subspace like ${\cal L}^+$ for $H_d$ that is 
characterized by its restricted interactions. 
Suppose that a theory has a U(1) gauge symmetry with a negative 
[resp. positive] Fayet--Iliopoulos parameter, and 
there are chiral multiplets $\overline{N}^c$ with a positive charge 
and $\overline{N}$ with a negative charge. Then, $\overline{N}^c$ 
[resp. $\overline{N}$] develops a non-vanishing expectation value, 
absorbing the Fayet--Iliopoulos parameter in the U(1) D-term potential. 
If there is an interaction (\ref{eq:4N-3}) in the superpotential, 
then the vev of $\overline{N}$ [resp. $\overline{N}^c$] is set 
to zero dynamically because of F-term potential coming from (\ref{eq:4N-3}).
The U(1) symmetry broken only by positively charged [resp. negatively
charged] chiral multiplets leaves a structure (\ref{eq:subspace5}) 
in ${\cal L}$, and similar ones in all the other vector 
spaces of massless modes of given representations of 
$\SU(3)_C \times \SU(2)_L \times \U(1)_Y$ as well. 
See (\ref{eq:41-mix5bar}, \ref{eq:32-mix10}, \ref{eq:32-mix5}),  
and discussions that follow.
Trilinear R-parity violating operators (\ref{eq:dim4}) are absent, 
no matter how large the Fayet--Iliopoulos parameter and 
$\vev{\overline{N}^c}$ [resp. $\vev{\overline{N}}$] are.
At the same time, the interaction (\ref{eq:4N-3}) gives rise to 
Majorana masses for right-handed neutrinos. Therefore, the absence 
of trilinear R-parity violating terms of massless modes is closely 
related to the Majorana masses. 

If the Heterotic string theory is compactified with a vector bundle 
given by an extension, then one always has the U(1) gauge
symmetry, Fayet--Iliopoulos parameter and interaction (\ref{eq:4N-3}).
Thus, dimension-4 proton decay problem is solved (see 
footnotes \ref{fn:subspace41} and \ref{fn:subspace32}), and 
Majorana masses are generated for right-haded neutrinos.
One does not need to choose a vacuum by hand only from special 
points in the moduli space, since an extra unbroken 
symmetry is not necessary any more.

Since the U(1) symmetry is broken spontaneously, it does not 
have a control over the K\"{a}hler potential of low-energy 
effective theory. Kinetic mixing terms between $L_i$ and $H_d$ 
can be generated, but they can be erased away by field redefinition, 
while keeping the structure (\ref{eq:subspace5}). See discussion 
that follows (\ref{eq:kahler-5Hbar}) for more detail.
Thus, the structure (\ref{eq:subspace5}) and the likes 
for other representations remain to be a valid solution 
to the dimension-4 proton decay problem.

The broken U(1) symmetry does not leave a structure like 
(\ref{eq:subspace5}) for massive fields. 
% This is the case, for example, for Kaluza--Klein multiplets.
Mass matrices are deformed by the vev's $\vev{N}$ 
(either $\vev{\overline{N}^c}$ or $\vev{\overline{N}}$). 
Once massive states are integrated out to obtain low-energy 
effective theory, U(1)-symmetry-breaking vev's can enter 
in the denominator of coefficients in the low-energy effective 
superpotential. Since positively [resp. negatively] charged 
vev's in the denominator supply negative [resp. positive] U(1)
charges, simple selection rule for the superpotential 
based on U(1)-charge counting ceased to be valid.
One cannot claim that an operator with positive [resp. negative] 
U(1) charges is not allowed in the effective superpotential, 
even when the U(1) symmetry is broken by vev's of positively 
[resp. negatively] charged fields.

This argument does not mean that there is no rule at all 
in the low-energy effective superpotential. Let us suppose 
that we have a superpotential of a fundamental theory with a 
typical mass scale $M$. 
Find vev's for all the fields, expand fluctuations around 
the vev's, diagonalize mass matrices, and integrate out 
massive states to obtain low-energy effective superpotential. 
There is nothing special in this process. 
But, it is important to note that terms trilinear in massless 
fluctuations remain unaffected in the last step of integrating 
out massive states. Thus, the selection rule using U(1)-charge 
counting can be used to constrain trilinear terms 
in the low-energy effective superpotential. 
Combined with an observation above that U(1)-breaking kinetic mixing 
in the K\"{a}hler potential can be absorbed be redefinition of chiral
multiplets while keeping the lower triangular nature of mass matrices, 
we see that the selection rule can be used in eliminating 
the dimension-4 proton decay operators (\ref{eq:dim4}).
It is not that this idea has not been presented 
anywhere else~\cite{GR,TW1,TW2,DESY-Rparity}. 
But we consider that this article strengthens theoretical basis 
for using the U(1)-charge selection rule for this purpose, by 
clarifying the limit of the rule as well as some subtleties and 
logical steps that we tend to overlook. 

When it comes to dimension-5 operators, first of all, 
there is a good chance that operators with negative [resp. positive] 
U(1) charges are in the low-energy effective superpotential, if 
holomorphic insertion of positively [resp. negatively] charged vev's 
can make them neutral under the U(1) symmetry.  
Such operators may be in the superpotential of a fundamental theory 
from the beginning, and even if this is not the case, they tend to 
be generated in the process of integrating out heavy particles. 

Dimension-5 operators with $n > 0$ [resp. $n < 0$] U(1) charges 
still have a chance of being generated, if 
some massive states in an appropriate pair of vector-like 
representations % (e.g., ${\bf 5}+\bar{\bf 5}$ of $\SU(5)_{\rm GUT}$) 
have mass parameters carrying $m \geq n$ [resp. $m \leq n$] U(1)
charges. Those operators may be generated when those states are 
integrated out.
If a dimension-5 operator does not satisfy this condition, 
it is not generated in the low-energy effective superpotential. 
This observation was used in one of the models described 
in this paper in eliminating the dimension-5 proton decay operators. 

Since R-parity is broken by vev's $\vev{N}$, it is interesting 
how R-parity violation appear in low-energy effective theory.
Particle contents at low-energy and their charges under the 
broken U(1) symmetry alone do not have enough power to answer 
to this question. This is because, for example,  
we never know whether dimension-5
operators with positive [resp. negative] U(1) charges are generated 
without knowing types of particles around the mass scale $M$ and 
interactions that those particles have. 
Since those operators tend to have couplings enhanced by some positive 
power of $(M/\vev{N})$, if they are generated, such operators tend 
to be the most important among dimension-5 operators. 
Thus, we need a (well-motivated) theoretical framework that 
specifies types of massive particles and their interactions. 

We turned to a class of Heterotic string compactification that 
we mentioned above. Two models (called 4+1 model and 3+2 model) 
that belong to this class were analyzed in this article.
In the 4+1 model, R-parity violating operator 
$\Delta W = H({\bf 5})_0 \; \bar{H}(\bar{\bf 5}) \; H({\bf 5})_0 \;
\bar{\bf 5}_0$ tends to have an enhanced coupling than the dimension-5
proton decay operators. The other R-parity violating dimension-5
operator $\Delta W = {\bf 10}_0 \; {\bf 10}_0 \; {\bf 10}_0 \;
\bar{H}(\bar{\bf 5})_0$ is also generated with a suppressed coefficient.
Typical effective couplings of those dimension-5 operators are of order 
\begin{equation}
 \frac{1}{M_{\rm eff.}} \sim y_{\rm 00H}^2 \frac{1}{M_{\rm KK}},
\label{eq:2summary_1}
\end{equation}
where $M_{\rm KK}$ is the Kaluza--Klein scale and 
$y_{\rm 00H}$'s are trilinear couplings among two light states and one 
massive state. Enhancement or suppression factor is roughly given 
by an appropriate power of $(y \vev{N} / M_{\rm KK})$.
In the 3+2 model, the dimension-5 proton decay operators are absent
in the effective theory. On the other hand, both R-parity violating 
dimension-5 operators can be generated.\footnote{An exception is the 
operator $\Delta W = {\bf 10}_0 \; {\bf 10}_0 \; {\bf 10}_0 \;
\bar{H}(\bar{\bf 5})_0$ for the case with spectrum characterized by
$b=0$.} Thus, the dimension-5 operator 
in the effective superpotential (apart from $H({\bf 5})\bar{\bf 5}H({\bf
5})\bar{\bf 5}$ that we may have already seen) with the largest coupling 
breaks R parity in both models. 

Since the theoretical framework in this section is realized in 
string theory, we can discuss 
i) R-parity violating operators generated by loop diagrams in the 
effective K\"{a}hler potential, as well as 
ii) those that originate from higher order terms suppressed by 
gravitational scale (or string scale).
Bilinear R-parity violation $W \ni \mu_i L_i H_u$ comes from a 1-loop 
correction to the K\"{a}hler potential, with $\mu_i$ proportional 
to the SUSY breaking. 
We found that tree-level contribution is absent when the gravitino mass 
originates from a 3-form flux. 
%
%Although there may be tree-level contributions, 
%we found by dimensional analysis that they are suppressed in a very 
%large power of string scale, and is not likely to be larger than the 
%1-loop contributions. 
%
Thus, $\mu_i$'s are suppressed by 1-loop factor and 
$(y\vev{N}/M_{\rm KK})$ relatively to gravitino mass, so that 
it can be very small compared with the electroweak scale.

Both dimension-5 R-parity violating operators in the effective
K\"{a}hler potential (\ref{eq:dim5-RPV-Kahler}) are also generated 
at 1-loop level.
At the crudest level of approximation, coefficients of these 
dimension-5 operators are of order 
\begin{equation}
 \frac{1}{M_{\rm eff.}} \approx \frac{y^3_{\rm 0HH}}{16\pi^2} \; 
  \frac{1}{M_{\rm KK}},
\label{eq:2summary_2}
\end{equation}
where now $y_{\rm 0HH}$'s are trilinear couplings among two heavy states 
and one light state.
Each operator is further suppressed or enhanced in an odd power 
of $(y \vev{N})/M_{\rm KK}$.
% , where $\vev{N}$ is a U(1) symmetry breaking vev. 
See section~\ref{ssec:dim5} for more details.

Sections~\ref{ssec:Het}--\ref{ssec:dim5} are written as if 
the theoretical framework is based on the Heterotic $E_8 \times E'_8$,
string theory. But absence of (or negligible) trilinear R-parity 
violation and presence of all other R parity violation rely 
only on a few ingredients: an extra U(1) symmetry, its charge assignment, 
non-vanishing Fayet--Iliopoulos parameter and interactions 
(\ref{eq:Yukawa41-u}--\ref{eq:mass41}) and 
(\ref{eq:Yukawa32-u}--\ref{eq:mass32_2}).
The algebra of $E_8$ alone is sufficient in justifying all these assumptions.
Therefore, M-theory dual and F-theory dual vacua 
that share the $E_8$ algebra should also share the same qualitative 
conclusion (c.f. \cite{TW1}). 
However, we have also used some properties 
specific to the Heterotic string theory, not just algebra of $E_8$. 
That is where we argued that tree-level contributions to 
(\ref{eq:kahler-5Hbar}, \ref{eq:5H}) are absent and 
those to (\ref{eq:dim5-RPV-Kahler}) can be negligible compared 
with those from 1-loop, and that 
the 1-loop contributions are not proportional to a positive power 
of string scale. Thus, one has to study such questions separately 
for non-Heterotic vacua.

%%%%%%%%%%%%%%%%%%%%%%%%%%%%%%%%%%%%%%%%%%%%%%%%%%%%%%%%%
\section{Phenomenology}
\label{sec:phen}
%%%%%%%%%%%%%%%%%%%%%%%%%%%%%%%%%%%%%%%%%%%%%%%%%%%%%%%%%

Most of phenomenological constraints on R-parity violating 
couplings that have been discussed in the literature 
are on the trilinear couplings in the superpotential 
(see \cite{Barbieri} for example). 
Since the theoretical framework in the previous section predicts 
that trilinear R-parity violation is either absent or highly suppressed, 
most of them, including the dimension-4 proton decay problem, 
are no longer a problem. 
R-parity violating interactions dominantly come from bilinear terms 
and dimension-5 operator in this framework. 
In the (virtual) absence of trilinear R-parity violation, 
constraints and predictions on nucleon decay are totally different, 
and some other constraints are expressed in much simpler ways 
because of fewer R-parity violating parameters. 
It is one of the purposes of this section to summarize
phenomenological constraints on R-parity violating couplings 
in the (virtual) absence of trilinear R-parity violation. 
In section~\ref{ssec:BBN}, we study lower bounds on bilinear 
and dimension-5 R-parity violating couplings, requiring that 
the LSP in the visible sector decay before the period of BBN. 
Recent developments in understanding of constraints on hypothetical 
particles from the BBN are reflected in the analysis. 
In section~\ref{ssec:proton}, we discuss nucleon decay processes 
induced by squark exchange diagrams combining dimension-5 and 
bilinear R-parity violating operators. 

The theoretical framework in the previous section not only 
predicts which operators are generated in low-energy 
effective theories. It also provided order-of-magnitude 
estimates of those R-parity violating couplings.   
This means that we can analyze whether or not it can survive 
various constraints. The order-of-magnitude estimates are 
compared with a constraint from low-energy neutrino masses 
in section~\ref{ssec:nu}, and with those from washout of baryon/lepton 
asymmetry in section~\ref{ssec:washout}.

We are by no means the first to study phenomenology of R-parity
violation in the absence of trilinear R-parity violating couplings 
in the superpotential. Indeed, there are a plenty of literature, 
especially those focusing on bilinear R-parity violation. 
Since R-parity violation is dominated by bilinear terms 
at the renormalizable level in low-energy effective theory 
of our framework, technical results in those literature are
quite useful. The appendix provides a quick summary of such 
results used in this section.

\subsection{Neutrino Mass}
\label{ssec:nu}
%%%%%%%%%%%%%%%%%%%%%%%%%%%%%%%%%%%%%%%%%%%%%%%%%%%%%%%%%%%%%%

Bilinear R-parity violation in (\ref{eq:mui}) introduces
neutrino--higgsino mixings, and hence the neutrino--neutralino 
system gives rise to an extra see-saw contribution to the 
neutrino masses \cite{HS}. As the neutrino masses 
are bounded from above by cosmological observations, 
there is an upper limit on bilinear R-parity violation. 
Using $m_\nu < 1 \; \EV$, the limit is placed on a misalignment 
parameter $|\epsilon'|$ \cite{Banks:1995by, Barbieri}:
\begin{equation}
 |\epsilon'| \simlt 3.0 \times 10^{-6} \times
   \frac{1}{\cos \beta} \times 
   \left(\frac{m_\nu}{1 \; \EV}\right)^{\frac{1}{2}}.
\label{eq:nu-constraint}
\end{equation} 
$|\epsilon'|$ measures the difference between  
$\mu_i/\mu_0$ and $v_i/v_d$;
its definition is found in \cite{Banks:1995by}
(and also (\ref{eq:epsi-def}) in the appendix).
The limit \eqref{eq:nu-constraint} also means that 
bilinear R-parity violation can provide 
the dominant part of neutrino masses 
in the case that \eqref{eq:nu-constraint} is marginally satisfied.

Since we have obtained a crude order-of-magnitude estimate of 
bilinear R-parity violation $\mu_i$ in section~\ref{sec:frame},
it is interesting to see if it satisfies the constraint (\ref{eq:nu-constraint}). 
$|\epsilon'|$ in (\ref{eq:nu-constraint}) is roughly of order ${\cal O}(|\mu_i/v|)$
in the absence of alignment (see the appendix and discussion below);
here, $v$ is the electroweak scale.
Using (\ref{eq:estimate-c}) in $\mu_i = c_i m_{3/2}$, 
\begin{equation}
 \frac{\mu_i}{v} \sim 10^{-8} \times
 \left(\frac{y}{10^{-2}}\right)^2 
 \left(\frac{y \vev{N}/M_{\rm KK}}{10^{-2}}\right)
 \left(\frac{m_{3/2}}{100 \; \GEV}\right).
\label{eq:mui2v}
\end{equation}
In gauge mediation scenario, it is very easy to satisfy 
the bound if $\mu_i$ is proportional to gravitino mass, but 
at the same time, it is very unlikely that the neutrino mass 
comes from bilinear R-parity violation.
In the case that $m_{3/2}\sim 100\,\GEV$,
the neutrino mass bound (\ref{eq:nu-constraint}) is also 
satisfied with a safe margin for a reasonable choice 
$y \sim 10^{-2}$ and 
$y^\nu \vev{N}/M_{\rm KK} \sim 10^{-2}$.
In this case, however, it is hard to say whether or not 
the neutrino--higgsino see-saw mechanism provides one of 
dominant contributions to the low-energy neutrino masses,
because of large theoretical uncertainties associated with 
(\ref{eq:mui2v}).

The misalignment parameter $|\epsilon'|$ (\ref{eq:epsi-def})
can be smaller than $\sqrt{\sum_i |\mu_i/\mu_0|^2}$ for some 
scenarios of SUSY breaking \cite{Hempfling, NP}. 
As we see in the appendix, each misalignment parameter
$\epsilon'_i$ $(i = 1,2,3)$ can be as small as 
$10^{-2} \times (\mu_i/\mu_0) $ in 
minimal SUGRA scenario, but it is unlikely that they are even 
smaller than that. Thus, the neutrino mass bound on $\mu_i/v$ 
may be relaxed by about two orders of magnitude.

In anomaly mediation scenario, $B_i \sim m_{3/2} \times \mu_i$,
so the misalignment parameter $|\epsilon'|$ can be larger than
naive expectation ${\cal O}(|\mu_i/\mu|)$.
If $B_0$ is somehow of order ${\cal O}(v^2)$ and $\mu_0 \sim {\cal
O}(v)$, then the misalignment parameters $\epsilon'_i$ 
are approximately $\epsilon'_i \sim - B_i/B_0 \approx 
- (m_{3/2}/v) \times (\mu_i/v) \sim 10^{+3} \times (\mu_i/v)$.
Thus, in this case, the limit on $\mu_i/v$ is stronger by three 
orders of magnitude.

%%%%%%%%%%%%%%%%%%%%%%%%%%%%%%%%%%%%%%%%%%%%%%%%%%%%%%%
\subsection{Washout of Baryon/Lepton Asymmetry}
\label{ssec:washout}
%%%%%%%%%%%%%%%%%%%%%%%%%%%%%%%%%%%%%%%%%%%%%%%%%%%%%%%

Any interactions violating either baryon or lepton number symmetry 
could wash out the baryon/lepton asymmetry of the universe that is 
once generated after inflation.
Thus, such couplings should not be too large. 
All the bilinear R-parity violating operators, 
\begin{equation}
 \Delta W = \mu_i H_u L_i
\end{equation}
and 
\begin{equation}
 \Delta V_{\rm soft} = - B_i H_u \tilde{l}_i 
+ m^2_{L\; 0i} H_d^\dagger \tilde{l}_i + {\rm h.c.}, 
\end{equation}
break lepton number symmetry. 
When all the dimension-5 R-parity violating operators are rewritten 
in terms of MSSM chiral multiplets, 
\begin{eqnarray}
& & \!\!\!\!\!\!\!\!\!\!\!\!\!\! \frac{1}{M_{\rm eff.}}
       \left[ {\bf 10}\;{\bf 10}\;{\bf 10}\;
              \bar{H}(\bar{\bf 5}) \right]_F 
     %  + {\rm h.c.}
      \rightarrow 
       \frac{1}{M_{3}} \left[ Q Q Q H_d \right]_F 
     + \frac{1}{M_{4}} \left[\bar{U} Q \bar{E} H_d\right]_F 
     %  + {\rm h.c.}
       \equiv {\cal O}_3 + {\cal O}_4,
       \label{eq:Op34} \\
 & & \!\!\!\!\!\!\!\!\!\!\!\!\!\! \frac{1}{M_{\rm eff.} }
      \left[ H({\bf 5}) \; \bar{\bf 5} \; H({\bf 5}) \; 
             \bar{H}(\bar{\bf 5}) \right]_F
        \rightarrow 
       \frac{1}{M_{6}} \left[H_u L H_u H_d \right]_F 
        \equiv {\cal O}_6,
       \label{eq:Op6} \\
 & & \!\!\!\!\!\!\!\!\!\!\!\!\!\! \frac{1}{M_{\rm eff.}} 
       \left[{\bf 10} \; {\bf 10} \; \bar{\bf 5}^\dagger \right]_D
     % + {\rm h.c.}
        \rightarrow  \left[
       \frac{1}{M_{7}} Q Q \bar{D}^\dagger % \right]_D
     + \frac{1}{M_{9}} % \left[
                \bar{U} Q L^\dagger % \right]_D
     + \frac{1}{M_{10}} %\left[
                \bar{U} \bar{E} \bar{D}^\dagger \right]_D
     % + {\rm h.c.},
        \equiv {\cal O}_7 + {\cal O}_9 + {\cal O}_{10},
       \label{eq:Op7910} \\
 & & \!\!\!\!\!\!\!\!\!\!\!\!\!\! \frac{1}{M_{\rm eff.}}
       \left[\bar{H}(\bar{\bf 5}) \; {\bf 10} \; 
             H({\bf 5})^\dagger \right]_D
        \rightarrow 
       \frac{1}{M_{8}} \left[ H_d \bar{E} H_u^\dagger \right]_D
    % + {\rm h.c.}
        \equiv {\cal O}_8,
       \label{eq:Op8}
\end{eqnarray}
it is easy to see that each operator breaks either baryon 
or lepton number symmetry.
Requiring that they do not wash out baryon/lepton asymmetry, 
upper bounds on these R-parity violating couplings are obtained~\cite{washout}.

Tiny trilinear R-parity violation 
(\ref{eq:Op0''}, \ref{eq:Op0+Op0'}) in the 4+1 model 
\begin{equation}
 - y^d \frac{\mu_i}{M_{\rm H}} \; 
  \left[L\; \bar{E} \; L\right]_F 
 - y^d \frac{\mu_i}{M_{\rm H}} \; 
  \left[L\; Q\; \bar{D}\right]_F 
   - y^d \frac{\mu_i}{M_{\rm H}} \; 
  \left[\bar{D}\; \bar{U} \; \bar{D}\right]_F 
\equiv {\cal O}_0 + {\cal O}'_0 + {\cal O}''_0
  \label{eq:Op0-def}
\end{equation}%
also breaks lepton and baryon number symmetry.
However, the coupling constants are so small for $M_{\rm H} \simgt M_{\rm GUT}$
($M_{\rm GUT} \sim 10^{16}$GeV) that they are irrelevant
to the washout of baryon/lepton asymmetry.

Bilinear R-parity violating operators are most dangerous at late time,
% \footnote{
%   Although trilinear R-parity violation (\ref{eq:Op0''}) 
%   that can exist in the 4+1 model is also most dangerous at late time,
%   it has such a small coefficient that they do not washout baryon asymmetry.
% }
as interaction rates scale as $\propto T$ while the temperature 
$T$ is higher than the electroweak scale, whereas the Hubble
parameter scales as $T^2$ during radiation dominance.
Assuming that baryon/lepton asymmetry was generated when the
temperature was higher than the electroweak scale, and requiring 
baryon/lepton asymmetry not to be washed out, 
%
% constraints on bilinear R parity violation becomes % \cite{Campbell:1991at}
% \cite{washout}
one can obtain upper limits on bilinear R-parity violation.
Applying the results of~\cite{DavidsonEllis} to the case without 
trilinear R-parity violation, 
\begin{equation}
 {\rm min}_{i=1,2,3} \left(\frac{\mu_i}{\mu_0}\right) 
   \simlt 10^{-7} \times (y^d_3)^{-1} \sim 
   3 \times 10^{-6} \times \cos \beta, 
\label{eq:washout-constraint-bilin}
\end{equation} 
where $y^d_3 = y_b$ is the bottom Yukawa coupling.
The rough estimate (\ref{eq:mui2v}) of $\mu_i$ satisfies this constraint easily.
Constraints on soft bilinear R-parity violating parameters 
in~\cite{Davidson} are a little stronger: 
\begin{equation}
 \left(\frac{B_i}{B_0} - \frac{\mu_i}{\mu_0}\right) \simlt 10^{-7}. 
\end{equation}
As long as $B_i \approx {\cal O}(v \times \mu_i)$, however, 
the estimate (\ref{eq:mui2v}) also satisfies this constraint 
from washout.

Dimension-5 operators, on the other hand, are more relevant at 
higher temperature. Interaction rates of R-parity violating processes 
scale as $\Gamma \sim 10^{-2} T^3/M_i^2$ $(i=3,4,6,7,8,9,10)$, and 
hence these operators would wash out baryon/lepton asymmetry 
right after it was generated, if they ever would. 
Requiring that R-parity violating processes caused by the dimension-5 
operators are out of equilibrium at that time, 
it follows that \cite{washout} 
\begin{equation}
  M_i \simgt 10^{12} \; \GEV \times 
   \left(\frac{T_{\Delta B/\Delta L}}{10^{10} \;
    \GEV}\right)^{\frac{1}{2}},
\label{eq:washout-constraint-dim5}
\end{equation}
where $T_{\Delta B/\Delta L}$ stands for the temperature of 
baryo/lepto-genesis.\footnote{
  Limits from the absence of washout
  of baryon/lepton asymmetry are more complicated
  if $T_{\Delta B/\Delta L} \simgt 10^{12} \; \GEV$,
  because the sphaleron process is out of equilibrium
  when $T \simgt 10^{12} \; \GEV$.
} 
Note that (\ref{eq:washout-constraint-dim5}) should be satisfied
for all seven operators (\ref{eq:Op34}--\ref{eq:Op8}).
If the baryon asymmetry of the universe originates from thermal 
leptogenesis, we know that 
$T_{\Delta B/\Delta L} \simgt 2 \times 10^{9} \; \GEV$
\cite{Buchmuller:2005eh}.
In this scenario, all the effective mass scales $M_i$ in 
(\ref{eq:Op34}--\ref{eq:Op8}) have to be larger than about 
$10^{12} \; \GEV$.
We have seen in section~\ref{sec:frame} that 
% the Kaluza--Klein scale
the zeroth-order 
approximation of $1/M_{3,4,6}$ in the superpotential and 
$1/M_{7,8,9,10}$ in the K\"{a}hler potential are
$y^2/M_{\rm KK}$ and $(y^3/16\pi^2)1/M_{\rm KK}$, respectively. 
% accompanied by suppression or enhancement given by powers of 
% $(y^\nu \vev{N})/M_{\rm KK}$. 
Therefore, it is quite easy to satisfy the constraints from 
the washout of baryon/lepton asymmetry, if the Kaluza--Klein scale 
is around the GUT scale $M_{\rm GUT}$.

%%%%%%%%%%%%%%%%%%%%%%%%%%%%%%%%%%%%%%%%%%%%%%%%%%%%%%%%%%%
\subsection{LSP Decay}
\label{ssec:BBN}
%%%%%%%%%%%%%%%%%%%%%%%%%%%%%%%%%%%%%%%%%%%%%%%%%%%%%%%%%%%

Given the order-of-magnitude estimate of bilinear R-parity violation 
in (\ref{eq:mui2v}), it is very unlikely that the lightest supersymmetric
particle in the visible sector (hereafter we call it the vLSP\footnote{
  We coin a term vLSP because whether gravitino is lighter than vLSP
  is sometimes not quite important.
})
has a lifetime much longer than the age of the universe. If the temperature of the 
universe was once sufficiently high, and thermal relic of vLSP is left 
after the temperature drops below the electroweak scale, then the 
relic vLSP has to have decayed before the period of BBN. 

Constraints on R-parity violating couplings from BBN were discussed 
already in \cite{HS,Ellis85,BS87} and in more detail in other papers 
that followed. 
In this subsection, we reanalyze the BBN limits 
in the light of the latest understanding of the impact of 
hadronic energy injection or of stable charged hypothetical 
particles during the period of BBN. 
We ignore tiny couplings (\ref{eq:Op0''})
that are present in the 4+1 model,
and assume that all the renormalizable interactions violating 
R-parity originate from bilinear terms, as predicted by the 
framework in section~\ref{sec:frame}. Limits on dimension-5 
R-parity violating operators are also derived. 
The following study covers two typical possibilities:  either 
the vLSP is a bino-like neutralino or a scalar tau (stau).

The vLSP is no longer a candidate of dark matter. But 
gravitino with $m_{3/2} \approx {\cal O}(1) \; \GEV$ can be 
a good candidate of dark matter \cite{TY}. 
Axion and a strongly interacting stable particle with a mass 
of order 100 TeV (if such a particle exists) \cite{GK} 
can also be dark matter.

\paragraph{Neutralino}

Let us first begin with the bino-like neutralino vLSP.
The thermal relic density of the neutralino is very model-dependent.
Here we adopt the prediction of the so-called ``bulk'' region of 
mSUGRA parameter space:
\begin{align}
       m_{\tilde{\chi}^0} Y_{\tilde{\chi}^0}
  \sim 4 \times 10^{-10} \, \mathrm{GeV}
       \left[ \frac{m_{\tilde{\chi}^0}}{100 \, \mathrm{GeV}} \right]^{2}.
\label{eq:nlino-yield}
\end{align}
%

% If the gravitino mass is sufficiently small
% so that the vLSP can decay to gravitino before the BBN,
% the relic neutralino vLSP does not have a problem with the BBN.

Nucleons and anti-nucleons produced in jets
from the decay of semi-stable particles
would contribute to $p \leftrightarrow n$ conversion processes
and change the fraction of ${}^4 {\rm He}$.
Thus, any semi-stable hypothetical particle $X$
with baryonic branching fraction $B_h$ has to have short enough lifetime
$\tau_{X} \simlt 0.1 \; {\rm sec}$,
as long as the relic density satisfies \cite{KKM}\footnote{
  The lower bound in (\ref{eq:pn-conversion}) is sensitive
  to the estimate of the error of measurement of ${}^4$He abundance,
  and here we adopt a conservative one in \cite{Fukugita:2006xy}.
  If the error estimate gets even larger,
  it brings up the lower bound in (\ref{eq:pn-conversion}),
  but the change is rather slight.
  On the contrary, if the error estimate gets smaller,
  the constraint $\tau_{X} \simlt 0.1 \; {\rm sec}$ applies
  to a semi-stable particle $X$ with much smaller $(mY)_X \times B_h$.
}
\begin{equation}
 \left(m Y \right)_X \times B_h \simgt 10^{-10}\mbox{--}10^{-9} \; \GEV.
\label{eq:pn-conversion}
\end{equation}
If there is enough mass difference
between the vLSP neutralino and the LSP gravitino
($m_{\tilde{\chi}^0} - m_{3/2} \gg M_Z$),
two-body decay $\tilde{\chi}^0 \rightarrow \psi_{3/2} + Z$
is possible at tree level, and $B_h \sim {\cal O}(1)$.
Thus (\ref{eq:pn-conversion}) is typically satisfied in this case,
% as one can see from (\ref{eq:nlino-yield}), 
and $\tau_{\tilde{\chi}^0} \simgt 0.1 \; {\rm sec}$
would be in conflict with the primordial ${}^4 {\rm He}$ abundance. 
This sets a limit \cite{FST}
\begin{equation}
% m_{3/2} < 1.3 \, \mathrm{MeV} \times |Z_{\chi \tilde{\gamma}}|
%    \left( \frac{m_{\chi}}{100 \, \mathrm{GeV}} \right)^{5/2}, 
  m_{3/2} \simlt 0.8 \; \MEV \times |N_{1\widetilde{Z}}| 
   \times \left(\frac{m_{\tilde{\chi}^0}}{100 \; \GEV}\right)^{\frac{5}{2}}
   \times \left(\frac{\tau_{pn}}{0.1 \; {\rm sec}}\right)^{\frac{1}{2}},
    \label{BBN_nlino_2} 
\end{equation}
where $N_{1\widetilde{Z}} = - \sin \theta_W N_{1\widetilde{B}}
+ \cos \theta_W N_{1 \widetilde{W}^0}$
is the $\widetilde{Z}$ component of the neutralino vLSP. 

If the mass difference is not large enough
($m_{\tilde{\chi}^0} - m_{3/2} \simlt M_Z$), however, 
$\tilde{\chi}^0 \rightarrow \psi_{3/2} + \gamma$
is the only two-body decay at tree level,
and baryons are not contained in the decay products
of this dominant decay mode.
Baryons are produced only in three-body decay processes,
and baryonic branching fraction is of order $B_h \sim {\cal O}(10^{-3})$.
In this case, (\ref{eq:pn-conversion}) is not satisfied,
and hence the limit from $p \leftrightarrow n$ conversion
$\tau_X \simlt 0.1$ sec does not apply.
Deuteron production due to hadrodissociation of ${}^4$He
sets the most stringent limit instead; 
$\tau_X \simlt 10^{2}$ sec is required
for a hypothetical particle $X$ if \cite{KKM}
\begin{equation}
 (m Y)_X \times B_h \simgt 10^{-13} \; \GEV,  
\label{eq:d-hadrodissociation}
\end{equation}
and this condition is satisfied by the typical relic density 
of neutralino (\ref{eq:nlino-yield})
even after multiplying $B_h \sim {\cal O}(1)$. Thus,
\begin{equation}
%   m_{3/2} < 42 \, \mathrm{MeV} \times |Z_{\chi \tilde{\gamma}}|
%  \left( \frac{m_{\chi}}{100 \, \mathrm{GeV}} \right)^{5/2}.
   m_{3/2} \simlt 24 \; \MEV \times |N_{1 \widetilde{\gamma}}| 
  \times \left(\frac{m_{\tilde{\chi}^0}}{100 \; \GEV}\right)^{\frac{5}{2}}
  \times \left(\frac{\tau_{d-had}}{10^2 \; {\rm sec}}\right)^{\frac{1}{2}}, 
  \label{BBN_nlino_1}
\end{equation} 
where $N_{1\widetilde{\gamma}} = 
\cos \theta_W N_{1\widetilde{B}} + \sin \theta_W N_{1 \widetilde{W}^0}$.

If gravitino is not as light as specified above, a bino-like 
vLSP has to decay fast enough through R-parity violating operators 
before the period of BBN.
Bilinear R-parity violation induces R-parity violating 
vertices $\hat{\nu}$--$\hat{\tilde{\chi}^0}$--$Z$ and 
$\hat{e}^{\pm}$--$\hat{\tilde{\chi}^0}$--$W^\mp$ 
in mass-eigenstate basis \cite{chidecayA}
(see also (\ref{eq:nu-Z-chi}--\ref{eq:l-W-chiB}) in the appendix).
If there is enough final state phase space for the two body 
decay processes $\tilde{\chi}^0 \rightarrow \nu + Z$ and 
$\tilde{\chi}^0 \rightarrow \ell^\pm + W^\mp$ ($\ell=e,\mu,\tau$),
\begin{eqnarray}
\Gamma(\tilde{\chi}^0{}_1 \rightarrow Z + \nu/\bar{\nu}) & \sim & 
  \frac{2}{16\pi} \sum_i 
   \left|\frac{m_{\tilde{\chi}^0}}{M_Z} \frac{g_Z}{2} 
         (\xi_{\hat{\nu}_i\widetilde{B}} N_{1\widetilde{B}})^* \right|^2 
         m_{\tilde{\chi}^0}, \\
\Gamma(\tilde{\chi}^0{}_1 \rightarrow W^\pm + \ell^{\mp}) & \sim & 
  \frac{2}{16\pi} \sum_i 
   \left|\frac{m_{\tilde{\chi}^0}}{M_W} \frac{g}{\sqrt{2}} 
         (\xi_{\hat{\nu}_i\widetilde{B}} N_{1\widetilde{B}})^* \right|^2 
         m_{\tilde{\chi}^0}, 
%        \Gamma(\tilde{\chi}^0_1 \to \nu + Z)
% & % \sim \frac{1}{16 \pi}
%   %     |Z_{\chi \tilde{H}_{d}}|^{2} |Z_{\nu \tilde{H}_{d}}|^{2}
%   %     g_{Z}^{2} m_{\chi} \cr
%   \sim & \frac{g_Z^2}{??}
%         |\xi^*_{\widetilde{B}k} N_{\widetilde{B}1}|^2 m_{\chi} \\
% & % \sim (0.33 \sec)^{-1}
%   %     |Z_{\chi \tilde{H}_{d}}|^{2}
%   %     \left( \frac{|Z_{\nu \tilde{H}_{d}}|}{10^{-12}} \right)^{2}
%   %     \left( \frac{m_{\chi}}{100 \, \mathrm{GeV}} \right).
%    \sim & (??? \sec)^{-1} 
%         \left(\frac{|\epsilon'|}{10^{-?}}\right)^2
%         \left(\frac{\cos \beta}{1/10}\right)^2
%         \left( \frac{200 \;\GEV}{M_1} \right). \\
% \Gamma  (\tilde{\chi}^0_1 \to l^\pm_k + W^\mp)
% & \sim & \frac{g^2}{??}
%         |U^*_{\widetilde{H}^- k} N_{\widetilde{H}^0_d 1}|^2 m_{\chi} \\
%  &   \sim & (??? {\rm sec})^{-1} \left(|N_{\widetilde{H}^0_d1}|\right)^2
%          \left(\frac{|\mu_i/\mu_0|}{10^{-??}}\right)^2
%          \left(\frac{M_1}{200 \; \GEV}\right), 
\end{eqnarray}
where we have assumed a little hierarchy $M_{Z,W} \ll M_{\rm SUSY}$
($M_{\rm SUSY} \approx M_{1,2}, \mu_0$); 
for a bino-like neutralino vLSP, % $\hat{\tilde{\chi}^0}_1$, 
the $\xi_{\hat{\nu}_i\widetilde{B}} N_{1\widetilde{B}}$ terms
in (\ref{eq:nu-Z-chi}--\ref{eq:l-W-chiB}) are proportional to 
$(M_{Z}/M_{\rm SUSY})$, while all other terms in the vertices 
(\ref{eq:nu-Z-chi}, \ref{eq:l-W-chiB}) are\footnote{
  Note that there is a partial cancellation
  between $U_{\hat{e}_{L\;i}\widetilde{H}^-_d} N_{1\widetilde{H}^0_d}^*$
  and $(\xi_{\hat{\nu}_i\widetilde{H}^0_d} N_{1\widetilde{H}^0_d})^*$
  in (\ref{eq:l-W-chiB}); see (\ref{eq:xi-Hd}) and (\ref{eq:U's}). 
  The neutralino-$W$-charged-lepton vertex (\ref{eq:l-W-chiA})
  is negligible compared with (\ref{eq:l-W-chiB}),
  because all the coefficients in (\ref{eq:l-W-chiA}) are suppressed further
  by $m_i/\mu_0$, where $m_i$ are charged lepton masses; see (\ref{eq:V's}).
} 
to $(M_Z/M_{\rm SUSY})^3$.
Using the approximate form of $\xi_{\hat{\nu}_i \widetilde{B}}$
in (\ref{eq:xi-B}) in the appendix
and assuming $N_{1\widetilde{B}} \sim 1$, we find that
\begin{equation}
 \left(\frac{|\epsilon'|}{10^{-10.5}} \right)
   \simgt \left(\frac{200 \; \GEV}{M_1}\right)^{\frac{1}{2}}
     \times \frac{1}{10 \; \cos \beta} 
     \times \left(\frac{0.1 \; {\rm
	     sec}}{\tau_{pn}}\right)^{\frac{1}{2}}. 
\label{BBN_nlino_3}
\end{equation}
Here we used the constraint from $p \leftrightarrow n$ conversion, 
because $B_h \sim {\cal O}(1)$ in these processes, and 
(\ref{eq:nlino-yield}) satisfies (\ref{eq:pn-conversion}).
%
% previous version:
%
% This condition sets a {\em lower} bound on the bilinear R parity violation:
% for $m_{\tilde{\chi}} \gg m_{Z,W}$,
% %
% \begin{equation}
%    |Z_{\nu \tilde{H}_{d}}|
%  > 1.8 \times 10^{-12} |Z_{\chi \tilde{H}_{d}}|^{-1}
%    \left( \frac{100 \, \mathrm{GeV}}{m_{\chi}} \right)^{1/2},
%    \label{BBN_nlino_3}  
% \end{equation}
%

\begin{figure}
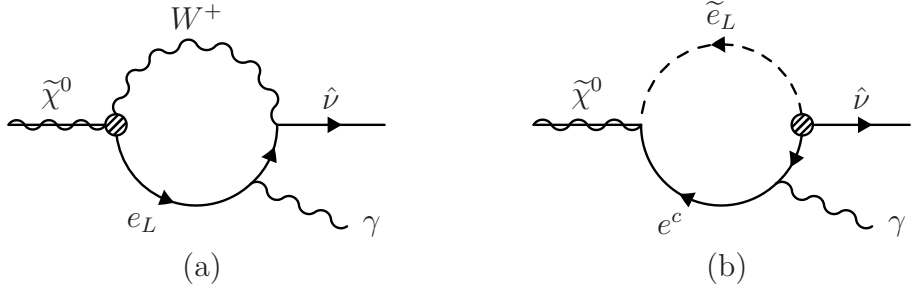

\begin{center}
\begin{tabular}{ccc}
  \input{fig_lsp_1a} & \hspace{1cm} & \input{fig_lsp_1d} \\
  (a) && (b) \\
\end{tabular}
\end{center}
\caption{
  Feynman diagrams for R-parity violating radiative decay of neutralino.
  (a) is an example of diagrams with $W^+$ and $-1$-charged fermions 
  $\psi^- = (\widetilde{W}^-,\widetilde{H}^-_d,e_{L \; k})^T$ in the loop.
  (b) is an example of those
  with charged scalar fields $\phi^\pm$ and charged fermions $\psi^\mp$,
  where $\psi^+ = (\widetilde{W}^+, \widetilde{H}^0_u,e^c_k)^T$.
}
\label{fig:neutdecay-radiative}
\end{figure}

If there is not enough final state phase space for 
$\tilde{\chi}^0 \rightarrow \nu + Z, \bar{\nu} + Z$ and $\ell^\pm + W^\mp$, 
there is no two-body decay processes at tree level. 
%%%%%%%%%%%%%%%%%%%%%%%%%%%%%%%%%%%%%%%%%%%%%%%%%%%%%%%%%%%%%%%%%%%%
%
\begin{figure}
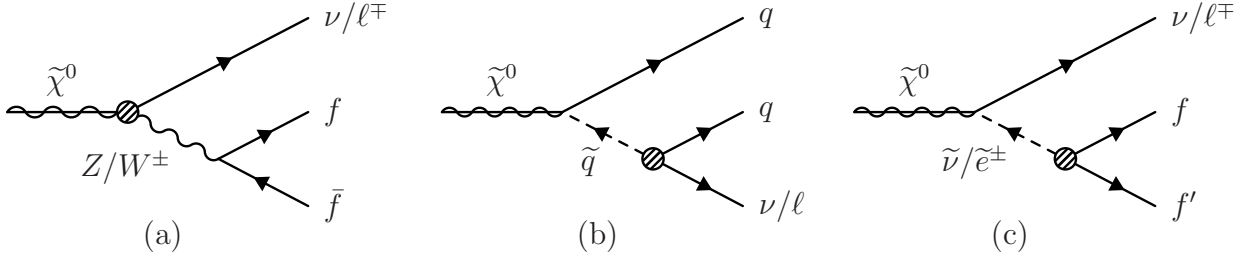

\begin{center}
\begin{tabular}{ccccc}
  \input{fig_lsp_2a} & \hspace{8mm}
& \input{fig_lsp_2d} & \hspace{5mm}
& \input{fig_lsp_2c} \\
  (a) && (b) && (c) \\
\end{tabular}
\end{center}
\caption{
  Feynman diagrams for three-body R-parity violating decay of neutralino.
  Feynman rules for the R-parity violating vertices, each shown as a blob,
  are found in \cite{chidecayA, chidecayB}
  and also in the appendix of this article. 
  (a) is with a virtual $Z/W$, (b) with a virtual squark
  and (c) with a virtual non-colored scalar. 
  (c) is only an example among several kinds of similar diagrams
  with a linear combination of
  $\phi^0 = (h^0, H^0, A^0, \tilde{\nu}_L)$,
  $\phi^+ = (H^+, \tilde{e}_k^c, \tilde{e}_{Lk}^*)$
  or its complex conjugates $\phi^-$ as the virtual particle.
}
\label{fig:neutdecay-3body}
\end{figure}
%
%%%%%%%%%%%%%%%%%%%%%%%%%%%%%%%%%%%%%%%%%%%%%%%%%%%%%%%%%%%%%%%%%%%%%%%
Examples of Feynman diagrams for three-body decay processes are shown in 
Figure~\ref{fig:neutdecay-3body}; it is easy to see that the baryonic 
branching fraction is of order unity, no matter which one of the
diagrams (a)--(c) gives the dominant contribution. 
Although a kinematically allowed two-body decay process 
$\tilde{\chi}^0 \rightarrow \nu + \gamma$ does not contain a 
baryon or even a hadron in the decay products, it is 
generated only at one-loop (Figure~\ref{fig:neutdecay-radiative}), 
%%%%%%%%%%%%%%%%%%%%%%%%%%%%%%%%%%%%%%%%%%%%%%%%%%%%%%%%%%%%%%%%%%%%
%
%%%%%%%%%%%%%%%%%%%%%%%%%%%%%%%%%%%%%%%%%%%%%%%%%%%%%%%%%%%%%%%%%%%%%%%
and is not expected to be much faster than the three-body decay processes. 
Thus, the baryonic branching fraction
of R-parity violating decay of neutralino vLSP is of order unity
even when $m_{\tilde{\chi}^0_1} \simlt m_{Z,W}$.
This is in contrast to the case of neutralino decay to {\em gravitino}.
In order not to spoil the prediction of the standard BBN, 
we need\footnote{\label{fn:BBN-pn}
  To be conservative, the constraint from $p \leftrightarrow n$ conversion 
  should be replaced by that from excessive deuteron production
  through hadrodissociation, $\tau_{\tilde{\chi}^0} \simlt 10^2$~sec,
  relaxing the lower bound on the bilinear R parity violation
  by one order of magnitude and a half.
  This is because the yield of neutralino in (\ref{eq:nlino-yield})
  can be so small for $m_{\tilde{\chi}^0} \simlt M_{Z,W}$ 
  that (\ref{eq:pn-conversion}) may not be satisfied. 
  % meaning that the constraint $\tau_{\tilde{\chi}} \simlt 0.1$~sec
  % from $p \leftrightarrow n$ conversion may not have to be imposed.
  It should also be remembered that relic density of neutralino vLSP
  can be even smaller than the estimate (\ref{eq:nlino-yield})
  in some parameter region of the SUSY breaking.
} 
$\tau_{\tilde{\chi}^0} \simlt 0.1$~sec, just like 
in the case of $m_{\tilde{\chi}^0_1} \gg M_{Z,W}$.
It is likely that the diagram Figure~\ref{fig:neutdecay-3body}~(a)
gives larger contribution to the decay rate than the others,\footnote{
  The amplitude for Figure~\ref{fig:neutdecay-3body}~(c)
  is proportional to a Yukawa coupling,
  and will remain relatively small unless $\tan \beta$ is very large.}
and 
\begin{eqnarray}
  \Gamma \left(
           \tilde{\chi}^0
           \rightarrow Z^*\nu\,[\bar{\nu}]
           \rightarrow  f \bar{f}\nu\,[\bar{\nu}]
         \right)
& \sim & \frac{2\times 3.65}{192\pi^3}
         \sum_i \left| \frac{G_F}{\sqrt{2}} C^{(Z)}_i \right|^2
         m_{\tilde{\chi}^0}^5,
  \label{eq:neut-bidecay1} \\
% \left(N^u_{gen(Z)}\left((Q^{uL}_z)^2 + (Q^{uR}_z)^2\right)
%            + N^d_{gen(Z)}\left((Q^{dL}_z)^2 + (Q^{dR}_z)^2\right)
%        \right), \\
  \Gamma \left(
           \tilde{\chi}^0
           \rightarrow \ell^\mp W^{\pm*}
           \rightarrow \ell^\mp f \bar{f}\right)
& \sim & \frac{2\times 9}{192\pi^3}
         \sum_i \left|\frac{G_F}{\sqrt{2}} C^{(W)}_i \right|^2
         m_{\tilde{\chi}^0}^5,
  \label{eq:neut-bidecay2}
%
%
% %       \Gamma (\tilde{\chi}^0_1 \to \nu Z^{*} \to \nu f \bar{f})
% %& \sim &  |Z_{\chi \tilde{H}_{d}}|^{2} |Z_{\nu \tilde{H}_{d}}|^{2}
%        %  \frac{G_{F}^{2} m_{\chi}^{5}}{192 \pi^{3}} \\
% & \approx & % (0.27 \sec)^{-1}
%      %  |Z_{\chi \tilde{H}_{d}}|^{2}
%      %  \left( \frac{|Z_{\nu \tilde{H}_{d}}|}{10^{-10}} \right)^{2}
%      %  \left( \frac{m_{\chi}}{100 \, \mathrm{GeV}} \right)^{5}.
%    (??? \sec)^{-1} \left(\frac{|\mu_i/\mu_0|}{10^{-??}}\right)^2 
%    \times \cdots ...
\end{eqnarray}
where $G_F/\sqrt{2} = g_Z^2/(8M_Z^2)=g^2/(8M_W^2)$.
%
%\begin{equation}
% N_{gen(Z) \; {\rm eff.}} = 
%   N^u_{gen(Z)}\left((Q^{uL}_z)^2 + (Q^{uR}_z)^2\right)
% + N^d_{gen(Z)}\left((Q^{dL}_z)^2 + (Q^{dR}_z)^2\right) \simeq 0.84,
%\end{equation}
%
%where 
%$N^d_{gen(Z)}=3$, $N^u_{gen(Z)}=N_{gen(W)}=2$ are the numbers
% of generations and $Q_z=T_3 - \sin^2 \theta_W Q_{\rm QED}$. 
Interference between final state neutrinos in \eqref{eq:neut-bidecay1} 
is ignored.
$C^{(Z)}_i$ and $C^{(W)}_i$ are dimensionless coefficients 
of $\hat{\nu}_i$--$\hat{\tilde{\chi}^0_1}$--$Z$ and 
$\hat{e}_k^\pm$--$\hat{\tilde{\chi}^0_1}$--$W$ vertices
in (\ref{eq:nu-Z-chi}) and (\ref{eq:l-W-chiB}), respectively,
and are of the order $\epsilon'_i$ or $\mu_i/\mu_0$.
Thus, we conclude that 
\begin{equation}
\left(\frac{\sqrt{|C^{(W)}|^2 + 0.4 |C^{(Z)}|^2}}{10^{-10}}\right)
 \simgt \left(\frac{0.1 {\rm sec}}{\tau_{pn}}\right)^{\frac{1}{2}}
  \times \left(\frac{100 \; \GEV}{m_{\tilde{\chi}^0}}\right)^{\frac{5}{2}}.
 \label{BBN_nlino_4}
\end{equation}
if $m_{\tilde{\chi}^0_1} \simlt M_{Z,W}$.

% Thus, there is an allowed range of bilinear R-parity violation 
% spanning a few orders of magnitude 
% $10^{-9} \simlt |\xi| \simlt 10^{-7}$, where the 
% upper bound comes from the washout of lepton asymmetry in 
% section~\ref{ssec:washout}.
% For a bino-like neutralino vLSP sufficiently heavier than 
% the electroweak gauge bosons, $|\mu_i/\mu_0|$ can be as small 
% as $10^{-11}$ [check these numbers].

%
%
%
%  previous version:
%%
%\begin{equation}
%   |Z_{\nu \tilde{H}_{d}}|
% > 1.6 \times 10^{-10} |Z_{\chi \tilde{H}_{d}}|^{-1}
%   \left( \frac{100 \, \mathrm{GeV}}{m_{\chi}} \right)^{5/2}.
%   \label{BBN_nlino_4}  
%\end{equation}
%%
%
%Even for $|Z_{\chi \tilde{H}_{d}}| \sim 10^{-1}$ in (\ref{BBN_nlino_4}),
%there is an allowed range of neutrino--higgsino mixing angle $\xi$ 
%spanning more than a few orders of magnitude, 
%$10^{-9} \simlt |\xi| \simlt 10^{-5}\mbox{--}10^{-6}$, where the 
%upper bound comes from neutrino-mass constraint (\ref{eq:nu-constraint}).
%For a neutralino vLSP sufficiently heavier than the electroweak 
%gauge bosons, $|\xi|$ can be as small as $10^{-11}$.
%Figure~\ref{fig:neutdecay}.
%

The bino-like neutralino vLSP can also decay through dimension-5 
operators violating R-parity before the period of BBN,
even if bilinear R-parity violation is too small to satisfy 
(\ref{BBN_nlino_3}) or (\ref{BBN_nlino_4}).
The decay modes through $\mathcal{O}_{6,8}$
are similar to those using bilinear violation, 
since lepton--Higgs mixing is induced when some of 
Higgs multiplets are replaced by their vev's.
In the case of $m_{\tilde{\chi}^0} \gg M_{W,Z}$,
two-body decay processes of a neutralino vLSP 
${\tilde{\chi}^0} \to \nu + Z$ and $\to \ell^\pm + W^\mp$.
are induced by $\mathcal{O}_{6}$.
The decay widths for these modes can be calculated
by using Goldstone equivalence theorem,
that is, treating the longitudinal component of gauge boson 
in the final state as Goldstone boson,
with mass $M_{W}$ or $M_{Z}$ and coupling equal to that of Higgs boson.
These two decay modes are comparable, and the decay width is
\begin{align}
       \Gamma (\tilde{\chi}^0 \to \nu + Z)
\simeq \frac{m_{\tilde{\chi}^0}}{16\pi}
       \left|
         4 \frac{v_d}{M_6} N_{1 \tilde{H}_u^0} \sin \beta
       + 2 \frac{v_u}{M_6} N_{1 \tilde{H}_d^0} \sin \beta
       \right|^{2}.
\end{align}
On the other hand, the only two-body decay process induced 
by $\mathcal{O}_8$ is $\tilde{\chi}^0 \to \ell^\pm + W^\mp$, 
since mixing is induced only for charged leptons.
The decay width is
\begin{align}
       \Gamma (\tilde{\chi}^0 \to \ell^\pm + W^\mp)
\simeq \frac{m_{\tilde{\chi}^0}}{16\pi}
       \left|
         \frac{m_{\tilde{\chi}^0}}{M_8} N_{1 \tilde{H}_u^0} \cos \beta
       \right|^{2}.
\end{align}
Applying the constraint $\tau_{\tilde{\chi}^0} \simlt 0.1$ sec, we have
\begin{align}
  M_6 & \simlt 10^{13} \GEV
        \times ( 10 \cos \beta )
               \left( \frac{m_{\tilde{\chi}^0}}{100 \GEV} \right)^{-1/2}, \\
  M_8 & \simlt 5 \times 10^{12} \GEV
        \times ( 10 \cos \beta )
               \left( \frac{m_{\tilde{\chi}^0}}{100 \GEV} \right)^{1/2}, 
\end{align}
where we assumed that $m_{\tilde{\chi}^0} \simeq M_{\tilde{B}}$ and
\begin{align}
       N_{1 \tilde{H}_u^0}
\simeq - \frac{M_Z}{M_{\tilde{B}}} \sin \theta_W \sin \beta,
\qquad
       N_{1 \tilde{H}_d^0}
\simeq   \frac{M_Z}{M_{\tilde{B}}} \sin \theta_W \cos \beta.
\end{align}
In the case of $m_{\tilde{\chi}^0} \simlt M_{W,Z}$,
a vLSP neutralino decays to three fermions
with a virtual gauge boson in the intermediate state.
${\cal O}_6$ with two $H_u$'s replaced by their vev's induces 
a mixing in the neutral part $H_d^0 \; L_i^0$, and consequently 
the R-parity violating couplings 
$\hat{\tilde{\chi}}{}^0$--$\hat{\nu}$--$Z$
and $\hat{\tilde{\chi}}{}^0$--$\hat{\ell}^\pm$--$W^\mp$.
% have the same form as those in (\ref{eq:nu-Z-chi}, \ref{eq:l-W-chiB}),
% with $\epsilon_i$ and $\xi_{\hat{\nu}_i \tilde{H}_u^0}$
% changed from those in (\ref{eq:xi-Hu}) to
% \begin{align}
%   \epsilon_i'
% = \frac{v_d \mu_i - v_u \kappa_i - v_i \mu_0}{v_d \mu_0},
%   \qquad
%   \xi_{\hat{\nu}_i \tilde{H}_u^0}
% = \frac{\kappa_i - m_0 \epsilon_i'}{\mu_0},
% \end{align}
% where $\kappa_i = v_u^2/M_6$ is the coefficient
% of the effective bilinear term $\kappa_i H_d^0 L_i^0$.
% Note that $\mu_i$ in $\chi^0$--$\nu$ sector is twice
% as large as that in $\chi^\pm$--$\ell^\pm$ sector,
% because of $\mathrm{SU(2)_L}$ contraction.
% Using $N_{1\alpha} \sim \mathcal{O}(1)$,
We thus find that the constraint (\ref{BBN_nlino_4}) leads to
\begin{align}
       M_6
\simlt 10^{12} \GEV
       \left(\frac{\tau_{pn}}{0.1 {\rm sec}}\right)^{\frac{1}{2}}
       \left(\frac{m_{\tilde{\chi}^0}}{100\,\GEV}\right)^{\frac{5}{2}}.
\end{align}
The dimension-5 operator $\mathcal{O}_8$ 
contributes to three-body decay of vLSP neutralino only through 
virtual-$W$ processes. 
% gives rise to
% $\hat{\tilde{\chi}}{}^0$--$\hat{\ell}^\pm$--$W^\mp$ vertices
% of the form (\ref{eq:l-W-chiA}).
% The relevant components of the mixing matrix $V$ are
% \begin{align}
%     V_{\hat{e}^c \tilde{W}^+}
%   = \frac{\sqrt{2} m_W \sin \beta}{M_{\tilde{W}}} \frac{v_d}{M_8},
%     \qquad
%     V_{\hat{e}^c \tilde{H}_u^+}
%   = - \frac{2 m_W^2 \sin \beta \cos \beta}{M_{\tilde{W}} \mu_0} \frac{v_d}{M_8}.
% \end{align}
We find, after a calculation similar to what we have had so far, that 
the upper bound of $M_8$ is
\begin{align}
       M_8
\simlt 10^{11} \GEV
\times (10 \cos \beta)
       \left(\frac{200\,\GEV}{M_2}\right)
       \left(\frac{\tau_{pn}}{0.1 {\rm sec}}\right)^{\frac{1}{2}}
       \left(\frac{m_{\tilde{\chi}^0}}{100\,\GEV}\right)^{\frac{5}{2}}.
\end{align}

\begin{figure}
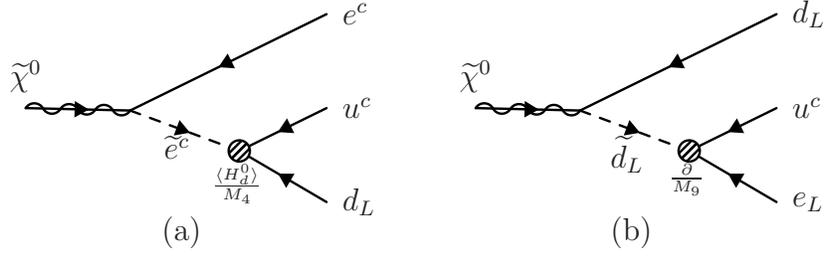

\begin{center}
\begin{tabular}{ccc}
\input{figure_neut-tri-decay2}
 & \hspace{1cm} & \input{figure_neut-tri-decay3} \\
(a) && (b) 
\end{tabular}
\end{center}
 \caption{\label{fig:neutdecayII} Examples of Feynman diagrams for 
neutralino decay that involve dimension-5 R-parity violating 
interactions. (a) uses ${\cal O}_4$, and (b) ${\cal O}_9$. }
\end{figure}

Feynman diagrams in Figure~\ref{fig:neutdecayII} show that 
there are three-body decay processes for neutralino vLSP
that involve dimension-5 R-parity violation ${\cal O}_{3,4}$ 
in (\ref{eq:Op34}) or ${\cal O}_{7,9,10}$ in (\ref{eq:Op7910}).
%
%%%%%%%%%%%%%%%%%%%%%%%%%%%%%%%%%%%%%%%%%%
%%%%%%%%%%%%%%%%%%%%%%%%%%%%%%%%%%%%%%%%%%
%
The decay width of these processes are given by 
({\it c.f.} \cite{Berezinsky})
\begin{align}
       \Gamma(\tilde{\chi}^0 \to f \tilde{f}^{*} \to f f' f'')
  \sim |\lambda_\mathrm{eff}|^{2} |N_{1 \widetilde{B}}|^{2}
       \frac{\alpha Y_{f}^{2}}{192 (2 \pi)^{2} \cos^{2} \theta_{W}}
       \frac{m_{\tilde{\chi}^0}^{5}}{m_{0}^{4}}, % \cr
% & \sim (0.53 \sec)^{-1}
%        |N_{1 \widetilde{B}}|^{2} Y_{f}^{2}
%        \left( \frac{|\lambda_\mathrm{eff}|}{10^{-8}} \right)^{2}
%        \left( \frac{m_{\tilde{\chi}^0}}{100 \, \mathrm{GeV}} \right)^{5}
%        \left( \frac{1 \, \mathrm{TeV}}{m_{0}} \right)^{4},
\label{nlino2tri}
\end{align}
where $Y_f$ is the hypercharge of a fermion $f$,
and $m_0$ the mass of a virtual sfermion $\tilde{f}$.
Effective coupling $\lambda_{\rm eff}$ is given by 
$\vev{H_d^0}/M_{3,4}$ for ${\cal O}_{3,4}$ 
and by $m_f/M_{7,9,10}$ for ${\cal O}_{7,9,10}$,
where $m_f$ is a mass of an outgoing fermion
(either $f'$ or $f''$, depending on which is the heavier one).
Combinatoric factors such as the number of colors and final states
are ignored here because it depends on which operator is dominant
and what kind of flavor structures they have.
Since a pair of quark and anti-quark is always in the final states 
in these three-body decay processes, % involving ${\cal O}_{3,4,7,9,10}$, 
baryonic branching fraction is of order unity. 
Imposing the limit $\tau_{\tilde{\chi}^0} \simlt 0.1$ sec as before,
we find that 
\begin{equation}
    \lambda_\mathrm{eff}
  > 2.3 \times 10^{-8}
    \frac{1}{|N_{1 \widetilde{B}}| Y_{f}}
    \left( \frac{100 \, \mathrm{GeV}}{m_{\tilde{\chi}^0}} \right)^{\frac{5}{2}}
    \left( \frac{m_{0}}{1 \, \mathrm{TeV}} \right)^{2} 
    \left(\frac{0.1 \; {\rm sec}}{\tau_{pn}}\right)^{\frac{1}{2}}.
    \label{BBN_nlino_5}
\end{equation}
This limit corresponds to 
\begin{eqnarray}
 M_{3,4} & \simlt & 10^9 \; \GEV \times 
   \left( 10 \; \cos \beta \right)
%  \left(\frac{101}{\tan^2 \beta + 1}\right)^{\frac{1}{2}} 
  \left(\frac{\tau_{pn}}{0.1 \; {\rm
   sec}}\right)^{\frac{1}{2}}
 \left(\frac{m_{\tilde{\chi}^0}}{100 \; \GEV}\right)^{\frac{5}{2}}
 \left(\frac{1 \; \TEV}{m_0}\right)^{2}, \label{BBN_nlino_6}\\ 
 M_{7,9,10} & \simlt & \left( 10^8 \right) \times \left(\frac{m_f}{\GEV}\right)
  \left(\frac{\tau_{pn}}{0.1 \; {\rm sec}}\right)^{\frac{1}{2}}
 \left(\frac{m_{\tilde{\chi}^0}}{100 \; \GEV}\right)^{\frac{5}{2}}
 \left(\frac{1 \; \TEV}{m_0}\right)^{2} \, \GEV. \label{BBN_nlino_7}
\end{eqnarray}
Only either one of (\ref{BBN_nlino_6}) and (\ref{BBN_nlino_7})
has to be satisfied.
%
% [the limit on $M_{1,2}$ here is obtained
%  from the corresponding equations in the appendix.
%  The limit on $M_{4,5,6}$ is obtained from that of $M_{1,2}$
%  by replacing $v_d$ by $m_f$.]

To summarize, vLSP neutralino must decay fast enough not to spoil BBN.
For $m_{\tilde{\chi}^0} \gg M_Z$ [$m_{\tilde{\chi}^0} \ll M_Z$],
neutralino decays to gravitino fast enough for gravitino mass
satisfying \eqref{BBN_nlino_2} [\eqref{BBN_nlino_1}].
For larger gravitino mass, at least either one
of R-parity violating couplings must be large enough;
either one of \eqref{BBN_nlino_3} [\eqref{BBN_nlino_4}],
(\ref{BBN_nlino_6}) and (\ref{BBN_nlino_7}) must be satisfied.

\paragraph{Stau}

Next we consider the case where the vLSP is stau.
Typical thermal relic of stau is 
\begin{equation}
 m_{\tilde{\tau}} Y_{\tilde{\tau}} \simeq 7 \times 10^{-12} \; \GEV 
 \times \left(\frac{m_{\tilde{\tau}}}{100 \; \GEV}\right)^2,
\label{eq:stau-relic}
\end{equation}
and if it decays too late, successful predictions of the BBN 
are no longer valid. 

Problems with the BBN can be avoided for sufficiently light gravitino, 
as the two-body decay $\tilde{\tau} \rightarrow \tau + \psi_{3/2}$ 
can be fast enough for stau to decay before the period of BBN.
The most stringent constraint on late-time decay 
$\tilde{\tau} \rightarrow \psi_{3/2} + \tau$ 
is to require \cite{Steffen, KKM07} that 
the primordial value of the abundance ratio
$(n_{^{6}\mathrm{Li}}/n_{^{7}\mathrm{Li}})_\mathrm{p}$ 
remains unchanged by the stau-catalyzed process 
in the presence of long-lived stau \cite{Pospelov}.
The upper bound is roughly $\tau_{\tilde{\tau}} \simlt 10^{3}$ sec,\footnote{
  The stau-catalyzed process is a problem in the range
  $m_{\tilde{\tau}} Y_{\tilde{\tau}} \simgt 10^{-13} \; \GEV$. 
  Typical thermal relic of stau vLSP (\ref{eq:stau-relic}) is within this range.
}
% \footnote{
%  Note that the constraint does not come
%  from $p \leftrightarrow n$ conversion;
%  although the 2-body decay $\tilde{\tau} \rightarrow \tau + \psi_{3/2}$
%  often ends up in energetic mesons,
%  mesons do not contribute to the $p \leftrightarrow n$ conversion process
%  very efficiently and the ${}^4$He abundance is not modified significantly
%  for a typical yield (\ref{eq:stau-relic}) 
%expected for thermal relic of stau.
%}
which corresponds to $m_{3/2} \simlt 0.1$ GeV for 
$m_{\tilde{\tau}} \sim 100$ GeV.
%
% We shall start by briefly summarizing the results of \cite{KKM07}.
% The yield of thermally produced stau is
% \begin{align}
%        m_{\tilde{\tau}} Y_{\tilde{\tau}}
% \simeq 7 \times 10^{-12} \, \mathrm{GeV}
%        \left( \frac{m_{\tilde{\tau}}}{100 \, \mathrm{GeV}} \right)^{2},
% \end{align}
% and the decay width of stau is
% \begin{align}
%          \Gamma (\tilde{\tau} \to \tau \psi_{3/2})
% & \simeq \frac{m_{\tilde{\tau}}^{5}}{48 \pi m_{3/2}^{2} M_{P}^{2}} \cr
% & \simeq (5.7 \times 10^{4} \sec)^{-1}
%          \left( \frac{100 \, \mathrm{GeV}}{m_{\tilde{\tau}}} \right)^{5}
%          \left( \frac{m_{3/2}}{1 \, \mathrm{GeV}} \right)^{2},
% \end{align}
% For a yield in the range
% $m_{\tilde{\tau}} Y_{\tilde{\tau}} \sim 10^{-(9-12)}$ GeV,
% the most stringent constraint on the lifetime is obtained
% from the primordial value of the abundance ratio
% $(n_{^{6}\mathrm{Li}}/n_{^{7}\mathrm{Li}})_\mathrm{p}$,
% which is increased by the stau-catalyzed process
% in the presence of long-lived stau.
% The upper bound is roughly $\tau_{\tilde{\tau}} \lsim 10^{3}$ sec,
% which corresponds to $m_{3/2} \lsim 0.1$ GeV
% for $m_{\tilde{\tau}} \sim 100$ GeV.

For gravitino mass $m_{3/2} \simgt 0.1 \; \GEV$, 
stau must decay through R-parity violating operators instead.
This requirement sets a lower bound on the R-parity violating couplings.
A stau decays dominantly to a pair of leptons in the presence of 
bilinear R-parity violation. Although it also decays to a pair of 
quark and anti-quark, such an amplitude is proportional to tau lepton 
mass, and the branching fraction is suppressed by a factor 
of order $(m_\tau \tan \beta/m_{\rm SUSY})^2$, which is not 
expected to be larger than about $10^{-2}$. 
See \cite{staudecay} or the appendix for more.
Since the typical yield of thermal relic of stau in
(\ref{eq:stau-relic}) with $B_h \simlt 10^{-2}$ does not satisfy 
(\ref{eq:d-hadrodissociation}), the deuteron constraint 
$\tau \simlt 10^2$~sec does not have to be imposed for stau vLSP decay.
The most stringent limit on the stau lifetime comes from 
the stau-catalyzed process,  
% also in the case stau decays through bilinear R-parity violation, 
and we need to require that $\tau_{\tilde{\tau}} \simlt 10^3$~sec.
Using (\ref{eq:stau-3pt-A}--\ref{eq:stau-3pt-C}),\footnote{
  Decay amplitudes are dominated by diagrams
  involving the vertices \eqref{eq:stau-3pt-B},
  as long as we assume large $\tan \beta$.
  But the vertex \eqref{eq:stau-3pt-A}
  can be more important for small $\tan \beta$.
} 
%%%%%%%%%%%%%%%%%%%%%%
%\begin{figure}
%\begin{center}
%\begin{tabular}{ccccc}
%\input{figure_stau-decay1} & \hspace{1cm} & % \input{figure_stau-decay2}
% & \hspace{1cm} & \input{figure_stau-decay3} \\
%(a) && % (b) 
%&& (b) 
%\end{tabular}
%\end{center}
%\caption{Feynman diagrams for R-parity violating stau decay due to  
%(a) bilinear R-parity violation and (b) dimension-5 R-parity 
%violating operator ${\cal O}_4$. }
%\label{stau_2body}
%\end{figure}
%%%%%%%%%%%%%%%%%%%%%%%
%
\begin{equation}
         \Gamma (\tilde{\tau} \to \bar{\nu}_3 \ell^{+}_k, \; 
                                  \ell_3^+ \bar{\nu}_k)
 \simeq \frac{1}{16 \pi} \, 
y_{\tau}^{2} \left|\frac{\mu_k}{\mu_0}\right|^{2} m_{\tilde{\tau}} 
 \qquad {\rm for}~k=1,2,
% & \simeq (0.33? \sec)^{-1}
%          \left( \frac{\tan \beta}{10} \right)^{2}
%          \left( \frac{|\mu_k/\mu_0|}{10^{-11}} \right)^{2}
%          \left( \frac{m_{\tilde{\tau}}}{100 \, \mathrm{GeV}} \right),
\label{BBN_stau_1}
\end{equation}
and we find that 
\begin{equation}
\sqrt{\sum_{k=1,2} \left(\frac{\mu_k}{\mu_0}\right)^2} 
 \simgt 1.8 \times 10^{-13}(10 \; \cos \beta) 
    \left(\frac{100 \; \GEV}{m_{\tilde{\tau}}}\right)
    \left(\frac{10^3 \; {\rm sec}}{\tau_{{}^6{\rm Li}\tilde{\tau}}}\right).
%   {\rm max} \left| \frac{\mu_k}{\mu_0}\right|
% \simgt 5.7 \times 10^{-13}
%   \left( \frac{100 \, \mathrm{GeV}}{m_{\tilde{\tau}}} \right)^{1/2}
%   \left( \frac{10}{\tan \beta} \right).
   \label{BBN_stau_3} 
\end{equation}

Stau may also decay fast enough to a quark--anti-quark pair 
through dimension-5 operators ${\cal O}_{4,10}$ with %(Figure~\ref{stau_2body} (b)) 
\begin{equation}
    \Gamma (\tilde{\tau} \to q \bar{q})
 \simeq \frac{3}{16 \pi} \, \lambda_\mathrm{eff}^{2} m_{\tilde{\tau}}, % \cr
% & \simeq (0.33 \sec)^{-1}
%         \left( \frac{|\lambda_\mathrm{eff}|}{10^{-12}} \right)^{2}
%         \left( \frac{m_{\tilde{\tau}}}{100 \, \mathrm{GeV}} \right).
\label{BBN_stau_2}
\end{equation}
even if bilinear R-parity violation is not large enough to satisfy (\ref{BBN_stau_3}).
The effective couplings $\lambda_{\rm eff}$ for this decay are
% $\lambda_{\rm eff} \sim (m_\tau/m_{\rm SUSY}) \theta_{03} y_b$ 
% for ${\cal O}'_0$,
$\lambda_{\rm eff.} \sim \vev{H_d^0}/M_{4}$ for ${\cal O}_{4}$ and
$\sim m_f/M_{10}$ for ${\cal O}_{10}$.
%
%As for the decay through ${\cal O}'_0$, however, $y_b \theta_{03}$ 
%is not much larger than $y_\tau |Z_{\nu \tilde{H}_{d}} + \theta_{0k}|$, and 
%an extra suppression factor $m_\tau / m_{\rm SUSY}$ is involved 
%in $\lambda_{\rm eff.}$ because of the left-right mixing. Thus, 
%it is unlikely that this decay mode is faster than 
%$\tilde{\tau}^{+} \rightarrow \bar{\nu}_3 + \ell^+, 
%\bar{\tau}^+ + \bar{\nu}$ \cite{staudecay} 
%and will not be the solution to the 
%BBN problem. Thus, if R parity violating decay of vLSP stau 
%is due to bilinear/trilinear terms rather than non-renormalizable 
%terms, then the baryoninc branching fraction remains of order 
%$B_{h} \sim 10^{-3}$ and the constraint is (\ref{BBN_stau_3}).
%
% If the $L_i$--$H_d$ mixing angle for the third generation,
% $|\theta_{03}|$, is the largest among the three $|\theta_{0k}|$, 
% then $|\theta_{03}|$ and $|U_{H\nu} + \theta_{0k}|$ are 
% of the same order of magnitude. 
% Thus the decay width (\ref{BBN_stau_2}) with $\lambda_{\rm eff.} \sim 
% \lambda'$ is comparable to (\ref{BBN_stau_1}).
% 
% In addition, two-body decay through triliner terms is dominantly through 
% operators such as ${\cal O}_{2,5,7}$, there are quarks in the final state.
% Therefore, baryonic branching fraction of stau decay can be of order 
% unity for some cases, if the BBN problem of stau is to be solved 
% by R-parity violation.
%
If the two-body decay through ${\cal O}_{4}$ or ${\cal O}_{10}$ is 
to be the solution to the BBN problem, the effective coupling 
of these operators has to be large enough,
so that $\tau_{\tilde{\tau}} \simlt 10^2$ sec;
since the baryonic branching fraction of the R-parity violating 
decay through ${\cal O}_4$ or ${\cal O}_{10}$ is of order unity, 
(\ref{eq:d-hadrodissociation}) is satisfied.
% The BBN constraint from stau-catalyzed ${}^6 {\rm Li}$ production,
% $\tau_{\tilde{\tau}} \simlt 10^3$ sec, is not the most stringent.
%
% When the hadron branching ratio is large, i.e. $B_{h} \sim 1$,
% the bound $\tau_{\tilde{\tau}} \lsim 10^{3}$ sec obtained from
% $(n_{^{6}\mathrm{Li}}/n_{^{7}\mathrm{Li}})_\mathrm{p}$
% is not the most stringent one.
%
% According to \cite{KKM}, 
%
%the limit $\tau \simlt 10^2$ sec from the deutron production 
%through hadrodissociation is applied to the range 
%$m_{\tilde{\tau}} Y_{\tilde{\tau}} \in 10^{-12}\mbox{--}10^{-9}$ GeV, 
%and the typical thermal relic density of stau (\ref{eq:stau-relic}) 
%falls in this range. 
%
% Hence the limit on $\tau_{\tilde{\tau}}$ above.  
% The typical yield (\ref{eq:stau-relic}) is so small (relatively to that
% of bino-like vLSP) that the ${}^4 {\rm He}$ 
% abundance is not modified significantly through 
% the $p \leftrightarrow n$ conversion.
% Using $\tau_{\tilde{\tau}} \simlt 10^2$ sec, 
We thus find that 
\begin{equation}
   \lambda_{\rm eff.} \simgt 3.3 \times 10^{-13} \times 
     \left(\frac{100 \; \GEV}{m_{\tilde{\tau}}}\right)^{\frac{1}{2}}
     \left(\frac{10^2 \; {\rm sec}}{\tau_{d-had}}\right)^{\frac{1}{2}}.
%     |\lambda_\mathrm{eff}|
%  > 5.7 \times 10^{-14}
%     \left( \frac{100 \, \mathrm{GeV}}{m_{\tilde{\tau}}} \right)^{1/2}.
\label{BBN_stau_4} 
\end{equation}
%
% As we have explained before, (\ref{BBN_stau_4}) is almost the same as 
% (\ref{BBN_stau_3}) when $\lambda_{\rm eff.} \sim \lambda'_{33*}$ 
% (where we used $y_b/y_\tau \sim {\cal O}(1)$).
This lower bound on $\lambda_{\rm eff.}$ is equivalent to either 
\begin{equation}
   M_4 \simlt 5 \times 10^{13} \; \GEV \times 
      (10 \cos \beta) 
     \times \left(\frac{m_{\tilde{\tau}}}{100 \; \GEV}\right)^{\frac{1}{2}}
     \left(\frac{\tau_{d-had}}{10^2 \; {\rm sec}}\right)^{\frac{1}{2}}, 
%  M_{4} \simlt 3 \times 10^{15} \; \GEV \times \cos \beta \times 
%    \left( \frac{m_{\tilde{\tau}}}{100 \; \GEV}\right)^{\frac{1}{2}}, 
\label{BBN_stau_5}
\end{equation}
or 
\begin{equation}
  M_{10} \simlt 3 \times 10^{12} \times
% 2 \times 10^{13} \times
  m_f \times 
  \left(\frac{m_{\tilde{\tau}}}{100 \; \GEV}\right)^{\frac{1}{2}}
  \left(\frac{\tau_{d-had}}{10^2 \; {\rm sec}}\right)^{\frac{1}{2}}.
\label{BBN_stau_6}
\end{equation}
It is sufficient if either one of these is satisfied.

The dimension-5 operator ${\cal O}_6$ induces stau decay through effective bilinear R-parity violation.\footnote{
  Although ${\cal O}_6$ also provides effective trilinear R-parity violating vertices
  with only one of Higgs fields replaced by its vev,
  their contribution to stau-vLSP decay is negligible in the limit where left-right mixing is ignored,
  since we assume that the vLSP $\tilde{\tau}$ is the scalar partner of $\tau_R^c$.
}
On the condition that $v_u \gg v_d$, dominant contribution comes from
$W \ni \left( v_u{}^2/M_6\right) L^0 H_d^0$,
which causes lepton--Higgs mixing only in the neutral part.
A decay mode using bino--neutrino mixing gives dominant contribution\footnote{
  Another decay mode using down-type Higgsino--neutrino mixing 
  can give comparable contribution only if $\tan\beta$ is very large.
}
\begin{equation}
 \Gamma (\tilde{\tau} \to \tau \bar{\nu}_k)
 \simeq \frac{1}{16 \pi} \, 
\left(\sqrt{2}g Y\right)^2 \left|\frac{m_Z}{m_{\tilde{\chi}^0}}\frac{v_u{}^2}{M_6 \mu_0}\,\sin\beta\sin\theta_W\right|^2 m_{\tilde{\tau}} 
\end{equation}
and the constraint from BBN is given by 
\begin{equation}
  M_6 \simlt 4 \times 10^{13} \times
  \left(\frac{m_{\tilde{\tau}}}{100 \; \GEV}\right)^{\frac{1}{2}}
  \left(\frac{\mu_0}{200 \; \GEV}\right)^{-1}
  \left(\frac{m_{\tilde{\chi}}}{200 \; \GEV}\right)^{-1}
  \left(\frac{\tau_{{}^6{\rm Li}\tilde{\tau}}}{10^3 \; {\rm sec}}\right)^{\frac{1}{2}}.
\end{equation}

The dimension-5 operator ${\cal O}_8$ also contributes to stau decay
through effective bilinear R-parity violation, with one of Higgs fields 
replaced by its vev. In contrast to the case with ${\cal O}_6$,
lepton--Higgs mixing occurs only in the charged part.
The decay width of stau vLSP depends on whether 
gravitino mass is large or not. 
In addition, if $m_{\tilde{\tau}} \gg M_Z + M_W$, a two-body decay 
process $\tilde{\tau}^\pm \rightarrow Z + W^\pm$ is also possible, 
without any of Higgs fields replaced by their vev's.\footnote{
\label{fn:stauDecay}
It is interesting to note that the branching fraction of various 
decay modes stau vLSP vary so much, depending on which R-parity
violating operator is responsible primarily for the vLSP decay.
${\cal O}_6$ predicts that stau-vLSP decays dominantly into final 
states including $\tau$, whereas the branching fractions of 
decay modes to $\tau$ and $\mu$ or $e$ have fixed ratio if 
the conventional bilinear R-parity violation dominates \cite{staudecay}. 
A branching fraction to $\mu$ or $e$ can be larger that 
that to $\tau$, if ${\cal O}_8$ dominates. If R-parity violating 
decay of stau vLSP is observed at the LHC, branching fractions 
of various decay modes provide valuable information on physics 
behind R-parity violation.}

To summarize, if the vLSP is stau, and 
if $m_{3/2} \simgt 100 \; \MEV$, one of the R-parity violating couplings
should be sufficiently large, so that stau can decay fast enough
and the standard predictions of the BBN are not spoiled.
It is sufficient if either one of R-parity violating operators allow
stau-vLSP to decay.

The neutrino mass bound (\ref{eq:nu-constraint}) and the absence 
of washout of baryon/lepton asymmetry 
(\ref{eq:washout-constraint-bilin}, \ref{eq:washout-constraint-dim5}) 
require that none of R-parity violating couplings are too large. 
On the other hand, some of R-parity violating couplings should be 
sufficiently large so that the vLSP can decay before the period of 
the BBN, unless gravitino mass is sufficiently small.
Figure~\ref{fig:para-space}~(a)--(d) summarize the relation 
among these phenomenological constraints in simplified two-dimensional 
parameter space characterizing the order of magnitude of 
bilinear and dimension-5 R parity violation. 

Flavor dependence of R-parity violating couplings has been ignored 
so far. Let the effective operator ${\cal O}_4$ in (\ref{eq:Op34}) 
has a flavor dependent coefficient 
$W \ni C^{(4)}_{ijk} Q_i \bar{U}_j \bar{E}_k H_d / M_4$, for example. 
It is $\sum_{i, j} |C^{(4)}_{ij3}|^2$ that determines the decay rate 
of stau vLSP, while what matters to the washout of lepton asymmetry 
is the smallest among $\sum_{i, j} |C^{(4)}_{ijk}|^2$ $(k=1,2,3)$; 
baryon/lepton asymmetry is washed out completely, only when all three 
$B/3 - L_k$ ($k=1,2,3$) symmetries are broken by some interactions in 
the thermal equilibrium.
Therefore, with flavor structure of R-parity violation, 
the true allowed region in the parameter space tends to be  
wider than it appears in Figure~\ref{fig:para-space}.

%%%%%%%%%%%%%%%%%%%%%%%%%%%%%%%%%%%%%%%%%%%%%%%%%%%%%%%%%%%%%%
\subsection{Nucleon Decay}
\label{ssec:proton}
%%%%%%%%%%%%%%%%%%%%%%%%%%%%%%%%%%%%%%%%%%%%%%%%%%%%%%%%%%%%%%

Dimension-5 R-parity violating operators ${\cal O}_3$ in (\ref{eq:Op34})
and ${\cal O}_7$ in (\ref{eq:Op7910}) break baryon number symmetry, 
and hence proton may decay. In the 4+1 model, baryon number is 
broken also in the trilinear R-parity violation ${\cal O}''_0$ 
in (\ref{eq:Op0''}).
In section \ref{sssec:p-lifetime}, we derive limits
on R-parity violating couplings from proton lifetime.
We further discuss in section \ref{sssec:N-decay}
how to probe R-parity violation through nucleon decay experiments.

%%%%%%%%%%%%%%%%%%%%%%%%%%%%%%%%%%%%%%%%%%%%%%%%%%%%%%%%%%%%%%
\subsubsection{Limits on R-Parity Violating Couplings from Proton Lifetime}
\label{sssec:p-lifetime}
%%%%%%%%%%%%%%%%%%%%%%%%%%%%%%%%%%%%%%%%%%%%%%%%%%%%%%%%%%%%%%

Decay products of a proton contain at least one fermion, 
and it must be one of $e, \mu, \nu$ or their anti-particles 
if it is a particle in the Standard Model.
Thus, proton decay is induced by squark exchange diagrams 
combining two R-parity violating operators;
one breaks baryon number and the other lepton number.
${\cal O}_3$, ${\cal O}_7$ and ${\cal O}''_0$ are candidates for the former,
while renormalizable interactions 
$\tilde{u}^*_{L/R} \bar{e} P_{L/R} d$ in (\ref{eq:ubardebar}), 
$\tilde{d}^*_{L/R} \overline{e^c} P_{L/R} u$ in (\ref{eq:udbare}) 
and $\tilde{q}^* \bar{\nu}q + {\rm h.c.}$ in (\ref{eq:uLunu}--\ref{eq:dRdnu}) 
for the latter.
The interactions (\ref{eq:ubardebar}--\ref{eq:dRdnu})
originate from bilinear R-parity violation.
R-parity violating dimension-5 operators
${\cal O}_{4}$ in (\ref{eq:Op34}) and ${\cal O}_{9,10}$ in (\ref{eq:Op7910})
can also play the same role for necessary lepton number violation\footnote{
  Dimension-5 operators $\mathcal{O}_{6,8}$
  also violate R-parity and lepton number.
  They induce effective bilinear R-parity violating operators 
  when some of Higgs fields are replaced with its vev.
  The consequence can be discussed by borrowing the constraint on
  bilinear R-parity violation \eqref{eq:pdecay-bi-limit},
  but it turns out that such constraint is looser if $M_{6,8}$
  are assumed to be in the same order as $M_{4,9,10}$,
  and hence we do not discuss them here.
}.

\begin{figure}
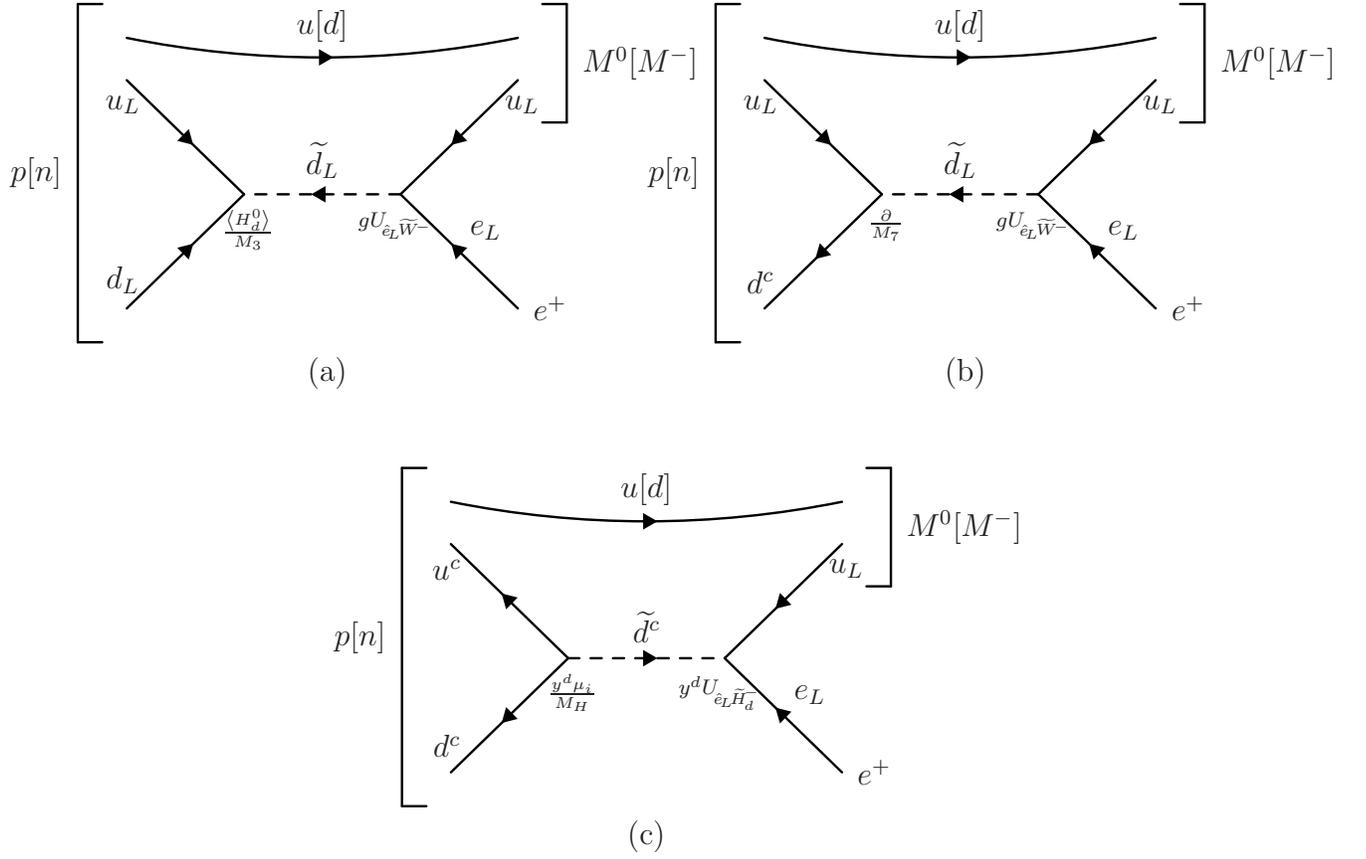

\begin{center}
\begin{tabular}{ccc}
\input{fig_pdecay_bi1} & \hspace{1cm} & \input{fig_pdecay_bi2} \\
(a) && (b)\\
\vspace{5mm} &&\\
\multicolumn{3}{c}{\input{fig_pdecay_bi3}} \\
\multicolumn{3}{c}{(c)}
\end{tabular}
\end{center}
\caption{\label{fig:Op3vsOp7vsOp0''} Feynman diagrams leading to proton decay 
$p \rightarrow M^0 + \ell^+$ (and neutron decay $n \rightarrow M^- + \ell^+$)
that originate from the combination of 
two operators ${\cal O}_3$--(\ref{eq:udbare}) in (a),  
${\cal O}_7$--(\ref{eq:udbare}) in (b) and 
${\cal O}''_0$--(\ref{eq:udbare}) in (c).}
\end{figure}

Feynman diagrams in Figure~\ref{fig:Op3vsOp7vsOp0''}
show squark-exchange diagrams combining
baryon-number violating ${\cal O}_3$, ${\cal O}_7$ and ${\cal O}''_0$
and a lepton-number violating (\ref{eq:udbare}). 
Diagrams using (\ref{eq:udbare}) contribute to nucleon decay processes 
such as $p \rightarrow \pi^0 + e^+$ 
where a positively charged lepton is in the decay products.
%%%%%%%%%%%%%%%%%%%%%%
%%%%%%%%%%%%%%%%%%%%%%
After integrating out SUSY particles, effective operators 
become 
\begin{eqnarray}
 {\cal O}_3\mbox{--}(\ref{eq:udbare}) & & {\cal L}_{\rm eff.} 
    = \frac{1}{m^2_{\tilde{d}_L}} \; 
      \frac{v_d}{M_3} \; \left(g U_{\hat{e}_L\widetilde{W}^-}\right) \; 
      (u_L d_L)(u_L e_L) , 
\label{eq:Op3+ren} \\
 {\cal O}_7 \mbox{--}(\ref{eq:udbare}) & & {\cal L}_{\rm eff.}
    = \frac{1}{m^2_{\tilde{d}_L}} \;
      \frac{1}{M_7} \; \left(g U_{\hat{e}_L\widetilde{W}^-} \right) \; 
             \partial_\mu (d_R \bar{\sigma}^\mu u_L) (u_L e_L), 
\label{eq:Op7+ren} \\
 {\cal O}''_0 \mbox{--}(\ref{eq:udbare}) & & {\cal L}_{\rm eff.} 
    = \frac{1}{m^2_{\tilde{d}^c}} \; 
        \left(- y^d \frac{\mu_i}{M_H}\right) \; 
        \left(y^d U_{\hat{e}_L\widetilde{H}^-_d}\right) \;
            (u_R d_R) (u_L e_L).
 \label{eq:Op0''+ren}
\end{eqnarray}
Generation indices are dropped,
and the CKM matrix elements that appear in (\ref{eq:udbare}) are ignored.
$d_{L/R}$ [resp.\!\! $e_{L/R}$] in this subsection
just means down-type quarks [resp.\!\! charged leptons], 
not just the down quark [resp.\!\! electron] in the first generation.
$U_{\hat{e}_L\widetilde{W}^-}$ and $U_{\hat{e}_L\widetilde{H}^-_d}$ are matrix elements 
appearing in diagonalization of charged fermion mass matrix, and are 
proportional to bilinear R-parity violation. 
For their definition, see \cite{chidecayA, chidecayB} 
or the appendix of this article.
Hadronic matrix elements 
\begin{eqnarray}
      \bra{M^0(\vec{p})}u_L (u_L d_L)\ket{p(\vec{k})}
& = & W(q^2) u_L(\vec{k}), \\
      \bra{M^0(\vec{p})}u_L (u_R d_R)\ket{p(\vec{k})}
& = & W'(q^2) u_L(\vec{k})
\end{eqnarray} 
have been obtained by lattice simulation \cite{JLQCD, Aoki:2006ib} 
for neutral pseudoscalars $M^0$ such as $\pi^0$ and $K^0$,
and the form factors $W$ and $W'$ are of order 
$\Lambda^2_{\rm QCD}$, 
where $\Lambda_{\rm QCD} \sim 300 \; \MEV$ is the QCD scale.
%
% where $\vec{k}$ and $\vec{p}$ are momentum of a proton and 
% a meson in the decay products in $p \rightarrow M^+ + \nu$, 
% and $M^+$ is a positively charged meson such as 
% $\pi^+$ and $K^+$. 
% $u_R(\vec{k})$ on the right-hand side 
% is a polarization spinor of an initial proton.
%
Although the matrix element 
\begin{equation}
  \bra{M^0(\vec{p})} u_L \; 
     \partial_\mu (d_R \bar{\sigma}^\mu u_L) \ket{p(\vec{k})}  =  
  W''(q^2) u_L(\vec{k}), \label{eq:MEw1D}
\end{equation}
is not known to our knowledge, we would expect that the form factor 
$W''$ is of order $\Lambda^3_{\rm QCD}$.

We have discussed in the previous section how the effective mass 
scales of dimension-5 operators depend on parameters of microscopic 
descriptions such as the Kaluza--Klein scale $M_{\rm KK}$ and a
suppression (sometimes enhancement) factor associated with a U(1) 
symmetry breaking $(y \vev{N}/M_{\rm KK})$. 
Writing down the three contributions to the proton decay amplitudes 
in terms of those parameters, we find that they are proportional to 
\begin{eqnarray}
 \frac{v_d}{M_3} (g U_{\hat{e}_L\widetilde{W}^-})& \approx & 
   y^2 \; \; \frac{y\vev{N}}{M_{\rm KK}} \frac{v_d}{M_{\rm KK}} 
   (g U_{\hat{e}_L\widetilde{W}^-}), 
\label{eq:compare-Op3}\\
 \frac{\Lambda_{\rm QCD}}{M_7}(g U_{\hat{e}_L\widetilde{W}^-})& \approx & 
   \frac{y^2}{16\pi^2} y
   \frac{y\vev{N}}{M_{\rm KK}} \frac{\Lambda_{\rm QCD}}{M_{\rm KK}} 
   (g U_{\hat{e}_L\widetilde{W}^-}),
\label{eq:compare-Op7}\\
 y^d \frac{\mu_i}{M_H} (y^d U_{\hat{e}_L\widetilde{H}^-_d}) & \approx &
   \frac{y^2}{16\pi^2} y
   \frac{y\vev{N}}{M_{\rm KK}} \frac{m_{3/2}}{M_{\rm KK}}
   (y^d U_{\hat{e}_L\widetilde{H}^-_d}), 
 \label{eq:compare-Op0''}
\end{eqnarray} 
where we ignored a possibility that some of the dimension-5 operators 
are enhanced when there are vector-like pair of particles whose 
masses are of order $y \vev{N}$. 
Rough ratio between these three operators is\footnote{
See (\ref{eq:2summary_1}, \ref{eq:2summary_2}) for the meaning of variations of Yukawa coupling constants
$y_{\rm 00H}$, $y_{\rm 0HH}$ and $y_{\rm HHH}$.
}
\begin{align}
  &\text{\eqref{eq:compare-Op3}}
: \text{\eqref{eq:compare-Op7}}
: \text{\eqref{eq:compare-Op0''}}
 \nonumber\\ &\sim
  \left[ (y_{\rm 00H})^2 \left( \frac{\cos \beta}{0.1} \right)^2\right]
: \left[ 10^{-4} \times (y_{\rm 0HH})^3 \left( \frac{\cos \beta}{0.1} \right) \right]
: \left[ 10^{-2.5} \times (y_{\rm 00H})(y_{\rm 0HH})(y_{\rm HHH}) \left( \frac{0.1}{\cos \beta} \right) \right],
\end{align}
where we used expressions for $U_{\hat{e}_L\widetilde{W}^-}$
and $U_{\hat{e}_L\widetilde{H}^-_d}$ in (\ref{eq:U's}),
and assumed that $y^d \sim m_s / v_d$ ($m_s$ is the $s$ quark mass).
Thus the combination ${\cal O}_3$--(\ref{eq:udbare})
is likely to contribute the most.
It should be remembered, however, that the order-of-magnitude estimates
in section~\ref{sec:frame} have very large uncertainties.
% Assuming that the Kaluza--Klein scale is around the GUT scale,
% The last contribution is just as important as the second one
% when $m_{3/2} \sim 10^{2.5} \cos^2 \beta \; \GEV$,
% since $y^d$ in (\ref{eq:compare-Op0''})
% which gives rise to the largest possible amplitude
% should be $y_s \sim 10^{-3}/ \cos \beta$,
% however it is unlikely for (\ref{eq:compare-Op0''})
% to surpass (\ref{eq:compare-Op3})
% unless at least $m_{3/2} \simgt 10^7 \cos^3 \beta \; \GEV$[check].
% It will be subdominant whenever $m_{3/2} \ll 1 \; \GEV$.

Experimental limits on proton lifetime are roughly around 
$\tau(p \rightarrow M^0 + \ell^+) \simgt 10^{32}\mbox{--}10^{33} \; {\rm yrs.}$ 
for major decay modes. 
This limit can be used to set an upper bound 
on the bilinear--dimension-5 R-parity violation. 
Using the partial amplitude from the combination ${\cal O}_3$--(\ref{eq:udbare}), 
we find\footnote{
  For consistent comparison with constraints from the vLSP decay and others, 
  we evaluate effective mass scales like $M_3$ renormalized at weak scale here.
} 
\begin{equation}
 \left(\frac{10^{15} \; \GEV}{M_3}\right)
 \left(\frac{\epsilon'_i}{10^{-9}}\right)
%  \left(\frac{gU_{\widetilde{W}^-}}{10^{-10}}\right)
  \simlt \frac{m^2_{\tilde{d}_L}}{(1 \; \TEV)^2} 
         \left(\frac{10 \; \GEV}{v_d}\right)
         \left(\frac{g\sqrt{2}M_W/M_2 \; \cos \beta}{10^{-1.5}}\right)^{-1}.
  \label{eq:pdecay-bi-limit}
\end{equation}
A similar constraint is obtained for the partial amplitude 
from ${\cal O}_7$--(\ref{eq:udbare}) when $M_3$ is replaced 
by $M_7$ and $v_d$ by $\Lambda_{\rm QCD}$. Unless different 
partial amplitudes cancel one another, each of these constraints 
have to be satisfied, and the one with ${\cal O}_3$ gives stronger 
constraint if $M_3$ and $M_7$ is in the similar order. 

\begin{figure}
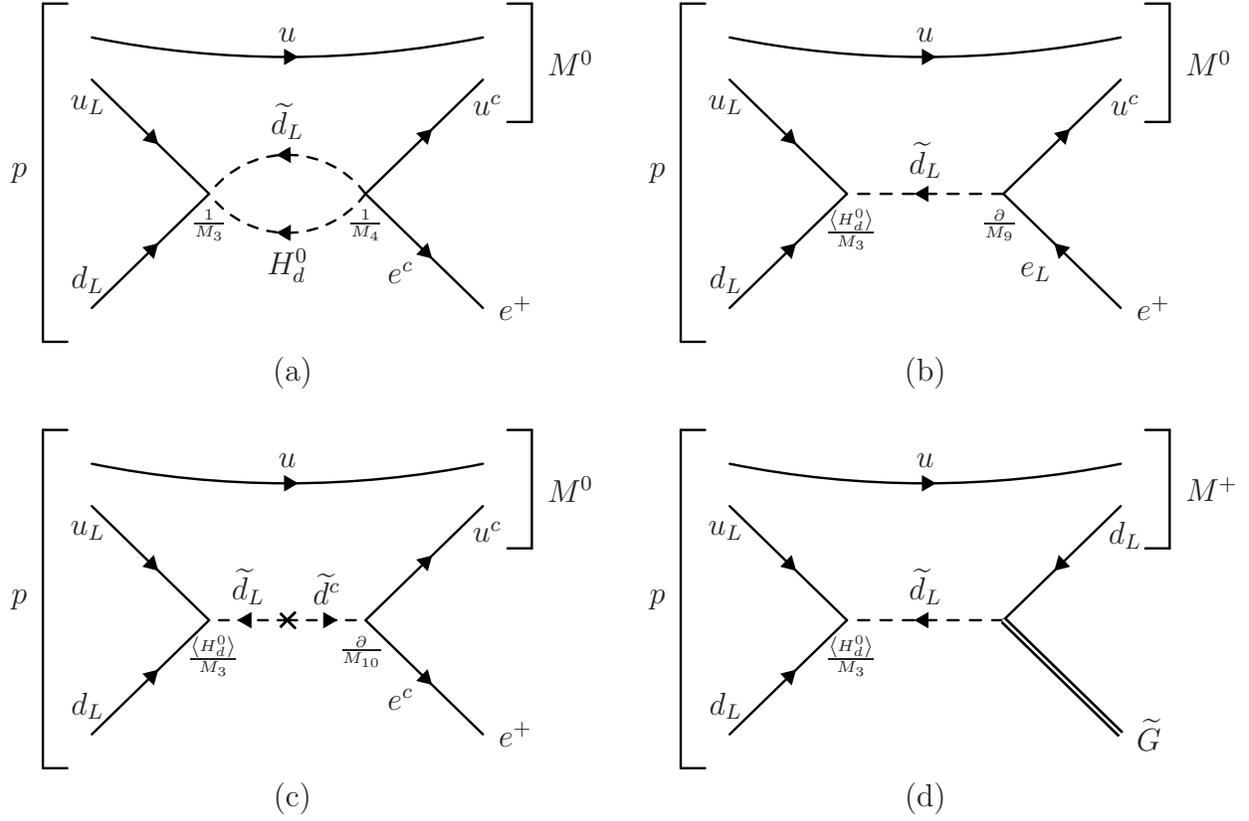

\begin{center}
\begin{tabular}{ccc}
\input{figure_pndecay3} & \hspace{1cm} & \input{figure_pndecay4}\\
(a) && (b)\\
&&\\
\input{figure_pndecay6} & \hspace{1cm} & \input{figure_pndecay5}\\
(c) && (d) \\
\end{tabular}
\end{center}
\caption{\label{fig:Op4910G} Feynman diagrams for proton decay 
involving (a): ${\cal O}_3$--${\cal O}_4$, 
(b): ${\cal O}_3$--${\cal O}_{9}$, 
(c): ${\cal O}_3$--${\cal O}_{10}$ with left-right mixing, 
(d): ${\cal O}_3$--${\cal O}_{\tilde{G}}$. }
\end{figure}

For even smaller bilinear R-parity violation, proton decay amplitudes 
are dominated by squark-exchange diagrams combining two R-parity
violating dimension-5 operators. Typical Feynman diagrams\footnote{
The 1-loop amplitude of Figure~\ref{fig:Op4910G}~(a) has a larger 
contribution than a tree-level one with a propagating Higgs boson 
replaced by its vev. The 1-loop diagram is logarithmically divergent, 
and the divergence is cut off at the energy scale of masses of 
heavy particles that are already integrated out in section~\ref{ssec:dim5}.
1-loop numerical factor including logarithm and $(1/16\pi^2)$ 
is about $0.4$, thanks to the large logarithm.} are found 
in Figure~\ref{fig:Op4910G}~(a)--(c).
%%%%%%%%%%%%%%%%%%%%%%%%%%%%%%%%%%%%%%%%
%%%%%%%%%%%%%%%%%%%%%%%%%%%%%%%%%%%%%%%%
Requiring that each partial amplitude is small enough, we find that 
\begin{eqnarray}
 \left(\frac{10^{15} \; \GEV}{M_3}\right)
 \left(\frac{10^{15} \; \GEV}{M_4}\right) & \simlt & 1, \\
 \left(\frac{10^{15} \; \GEV}{M_3}\right)
 \left(\frac{10^{15} \; \GEV}{M_{9}}\right) & \simlt & 
   \frac{m^2_{\tilde{d}_L}}{(1 \; \TEV)^2}
   \left(\frac{10 \; \GEV}{v_d}\right)
   \left(\frac{10^{-0.5} \; \GEV}{\Lambda_{\rm QCD}}\right)
   \times 10^{5}, \\
%    \left(\frac{m^2_{\tilde{d}_L}}{v_d \Lambda_{\rm QCD}}\right), \\
 \left(\frac{10^{15} \; \GEV}{M_3}\right)
 \left(\frac{10^{15} \; \GEV}{M_{10}}\right) & \simlt & 
  \frac{m^2_{\tilde{d}}}{(1 \; \TEV)^2}
   \left(\frac{10 \; \GEV}{v_d}\right)
   \left(\frac{10^{-0.5} \; \GEV}{\Lambda_{\rm QCD}}\right)
   \frac{10^{-2}}{\theta_{LR}}
   \times 10^{7}, 
%     \left(\frac{m^2_{\tilde{d}_L}}{v_d \; \Lambda_{\rm QCD} \; 
%            \theta_{LR}}\right), 
\end{eqnarray}
where $\theta_{LR}$ is a mixing angle between left-handed and
right-handed squarks.

If gravitino is lighter than a proton, then gravitino can be the fermion 
in the decay products of a proton. A squark exchange diagram in 
Figure~\ref{fig:Op4910G}~(d) induces proton decay, by combining 
baryon number violating ${\cal O}_3$ and 
\begin{equation}
 {\cal O}_{\tilde{G}} = \frac{i}{\sqrt{3}} \frac{1}{m_{3/2} M_G} 
    D_\nu \tilde{q}^\dagger 
    \psi_q \sigma^\mu \bar{\sigma}^\nu \partial_\mu \tilde{G},
\label{eq:def-OpG}
\end{equation}
where $\tilde{G}$ is the Goldstino field
and $(\tilde{q}, \psi_q)$ is a pair of complex scalar and chiral fermion
of any one of chiral multiplets of the MSSM.
The effective operator corresponding to the diagram is 
\begin{equation}
  {\cal O}_3\mbox{--}{\cal O}_{\tilde{G}} \qquad  {\cal L}_{\rm eff.} 
    = \frac{1}{m^2_{\tilde{d}_L}}
      \frac{v_d}{M_3}
      \frac{1}{\sqrt{3} m_{3/2} M_G}
      (\partial_\mu \tilde{G}_L \sigma^\nu \bar{\sigma}^\mu d_L)
      \partial_\nu (u_L d_L).
\label{eq:Op1+OpG} 
\end{equation}
Relevant hadronic matrix element is 
of the form \cite{Choi}
\begin{equation}
 \bra{M^+(\vec{p})} \sigma^\nu \bar{\sigma}^\mu d_L 
                     \partial_\nu (u_L d_L)
 \ket{p(\vec{k})} = p^\mu W''' u_L(\vec{k}),  
\label{eq:MEw2D}
\end{equation}
and the form factor $W'''$ will be of the order of 
$\Lambda^2_{\rm QCD}$. Thus, we should require that 
\begin{equation}
 \left(\frac{10^{15} \; \GEV}{M_3}\right)
 \left(\frac{3 \; \EV}{m_{3/2}}\right) \simlt 
 \frac{m^2_{d_L}}{(1 \; \TEV)^2}
 \left(\frac{10 \; \GEV}{v_d}\right),
\end{equation}
or otherwise the partial amplitude from 
${\cal O}_3$--${\cal O}_{\tilde{G}}$ would predict proton decay 
faster than the experimental bound.

%%%%%%%%%%%%%%%%%%%%%%%%%%%%%%%%%%%%%%%%%%%%%%
\begin{figure}
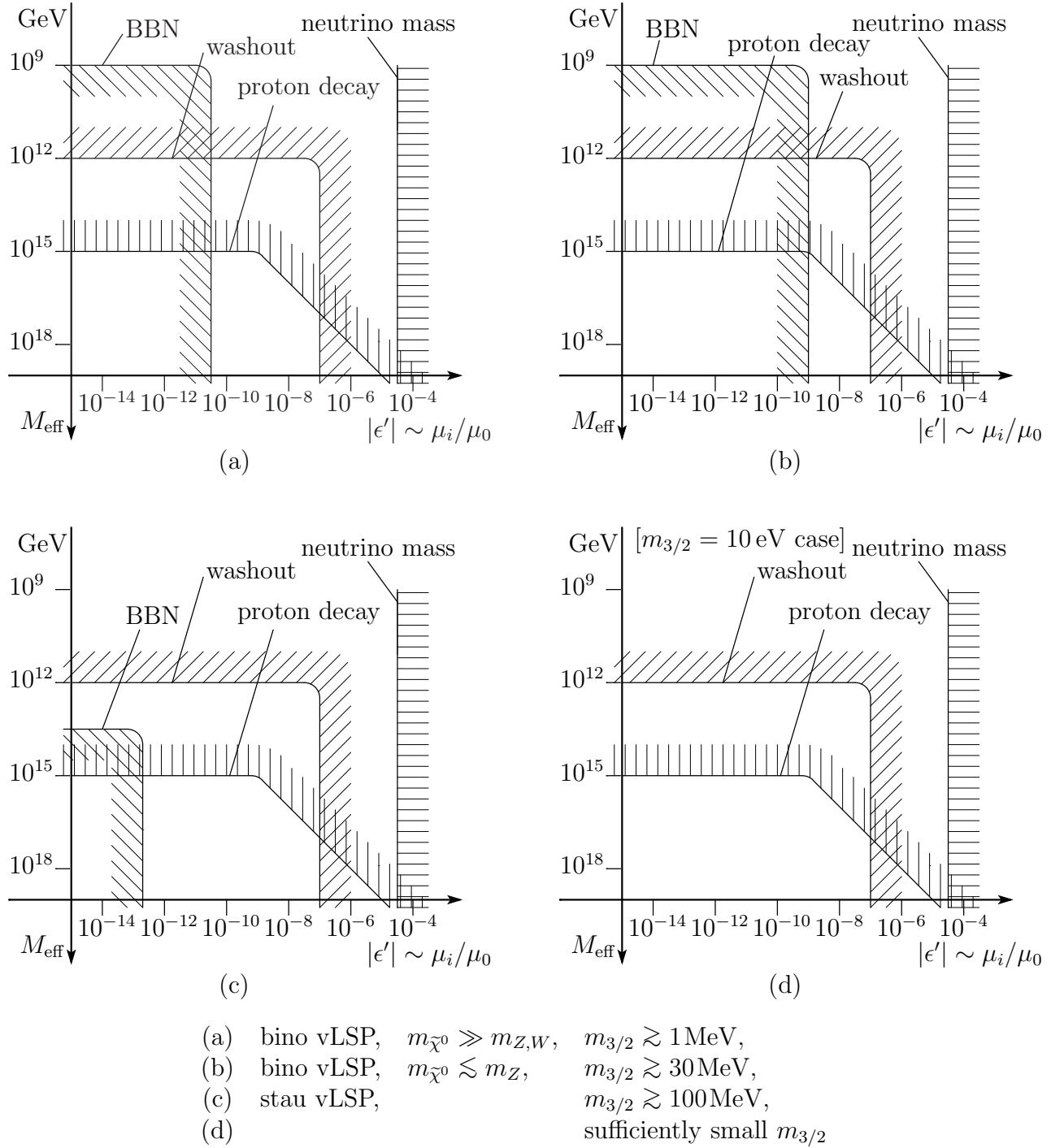

\begin{center}
\begin{tabular}{ccc}
\input{figure_region_a} & \hspace{8mm} & \input{figure_region_b} \\
(a) && (b) \\
\vspace{3mm} && \\
\input{figure_region_c} & \hspace{8mm} & \input{figure_region_d} \\
(c) && (d)
\end{tabular}
\vspace{3mm} \\
\begin{tabular}{clll}
  (a) & bino vLSP, & $m_{\tilde{\chi}^0} \gg m_{Z,W}$,
      & $m_{3/2} \simgt 1 \; \MEV$,\\
  (b) & bino vLSP, & $m_{\tilde{\chi}^0} \simlt m_{Z}$,
      & $m_{3/2} \simgt 30 \; \MEV$, \\
  (c) & stau vLSP, & & $m_{3/2} \simgt 100 \; \MEV$, \\
  (d) & & & sufficiently small $m_{3/2}$
\end{tabular}
\end{center}
\vspace{-5mm}
\caption{\label{fig:para-space}
Simplified picture of allowed parameter space of R-parity violation.
$|\epsilon'|$ parametrizes bilinear R-parity violation,
and $M_{\rm eff.} \approx M_{3,4,6,7,8,9,10}$ sets the scale
of dimension-5 R-parity violating operators.
Hatched areas are excluded.
We assumed that $T_{\Delta B/\Delta L} \sim 10^{10}$ GeV
and vLSP's are thermally produced.
}
\end{figure}
%%%%%%%%%%%%%%%%%%%%%%%%%%%%%%%%%%%%%%%%%%%%%%

Upper bounds on dimension-5 and bilinear R-parity violation
derived from proton decay are shown in Figure~\ref{fig:para-space},
along with constraints from neutrino mass (section \ref{ssec:nu}),
washout of baryon/lepton asymmetry (section~\ref{ssec:washout})
and BBN (section~\ref{ssec:BBN}).
The four parameter-space plots in Figure~\ref{fig:para-space}
are intended just to provide a big picture of the allowed parameter space
of bilinear--dimension-5 R-parity violation.
As the effective mass scale $M_{\rm eff.}$ of dimension-5 R-parity  
violation,
we adopt $M_{3,4}$ of the operators ${\cal O}_{3,4}$ for demonstration.
If $M_{3,4,6,7,8,9,10}$ are assumed to be in the same order of magnitude,
the operators ${\cal O}_{3,4}$ are the most effective pair for proton  
decay.
For washout constraint, there are no significant dependence on which  
operators are used.
On the other hand, some of dimension-5 R-parity violating operators among  
${\cal O}_{6,7,8,9,10}$
let the vLSP decay faster than through ${\cal O}_{3,4}$.
In such cases, the constraint from BBN on dimension-5 R-parity violation  
is looser
than shown in Figure~\ref{fig:para-space},
but general perspective of allowed parameter region is not largely altered.

If baryon asymmetry was generated
when the temperature is below the electroweak scale,
then the upper bound on the bilinear R-parity violation
is replaced by the neutrino mass bound from cosmology.
The allowed parameter space is extended to larger $|\epsilon'|$
by about two orders of magnitude.

Flavor structures are not taken into account in  
Figure~\ref{fig:para-space}.
For the case of stau vLSP, allowed parameter
space may extend to arbitrary small $|\epsilon'|$ with
$M_{\rm eff.} \approx 10^{15} \; \GEV$, if flavor structures of
dimension-5 operators are taken into account; see the comments
at the end of section~\ref{ssec:BBN}.

The theoretical framework in section~\ref{sec:frame} provides rough
estimates of bilinear and dimension-5 R-parity violation,
$|\epsilon'| \sim {\cal O}(10^{-8})$ and $M_{\rm eff.} \sim {\cal  
O}(M_{\rm KK})$.
Here $M_{\rm KK} \sim M_{\rm GUT}$ in a scenario
where ${\rm SU(5)_{GUT}}$ symmetry breaking
is associated with the compactification of spacetime.
It roughly corresponds to the upper-right corner of the parameter
space that survives all the phenomenological constraints
discussed in sections~\ref{ssec:nu}--\ref{ssec:proton}
in the big-picture parameter space in Figure~\ref{fig:para-space}.

%%%%%%%%%%%%%%%%%%%%%%%%%%%%%%%%%%%%%%%%%%%%%%%%%%%%%%%
\subsubsection{Probing R-Parity Violation with Nucleon Decay}
\label{sssec:N-decay}
%%%%%%%%%%%%%%%%%%%%%%%%%%%%%%%%%%%%%%%%%%%%%%%%%%%%%%% 

No evidence for proton decay has been found so far. 
However, once nucleon decay is discovered by experiments, 
then data of branching fractions of various decay modes 
can be used to study physics behind the nucleon decay.

Nucleon decay has been discussed in the literature mainly 
in the context of unified theories.
GUT gauge boson exchange and colored Higgsino exchange
predict dimension-6 and -5 operators
in the K\"{a}hler potential and superpotential:
%which become dimension-6 operators of the standard model fields:
%
\begin{eqnarray}
 K & \ni & \bar{E}^\dagger Q \bar{U}^\dagger Q  + 
           \bar{D}^\dagger L \bar{U}^\dagger Q, \label{eq:4fermi6} \\
%   & & \longrightarrow (u_L d_L) (u_R e_R) +
%    (u_R d_R) \left[(u_L e_L) - (d_L \nu_L) \right], \\
 W & \ni & Q Q Q L + \bar{U} \bar{U} \bar{E} \bar{D}. \label{eq:4fermi5}
%   & & \longrightarrow  (u_L d_L)\left[(u_L e_L) - (d_L \nu_L) \right]
%        + (u_R d_R) (u_R e_R). 
\end{eqnarray}
All these operators happen to preserve $B-L$. 
Therefore, $B-L$ number is preserved in all the nucleon decay processes 
$p [n] \rightarrow M^0 [M^-] + \ell^+$ and
$p [n] \rightarrow M^+ [M^0] + \bar{\nu}$ 
which are predicted by conventional (SUSY) GUT's. 
It is also known \cite{WWZ} that the $B-L$ number is 
preserved in all the baryon-number violating dimension-6 
operators of the Standard Model that preserve the gauge group 
$\SU(3)_C \times \SU(2)_L \times \U(1)_Y$. 
%In fact, there are no dimension-6 baryon-number violating operators 
%of the Standard-Model fields that preserve 
%the $\SU(3)_C \times \SU(2)_L \times \U(1)_Y$ gauge group
%other than the four operators above \cite{WWZ}. 
 
\begin{figure}
\begin{center}
\input{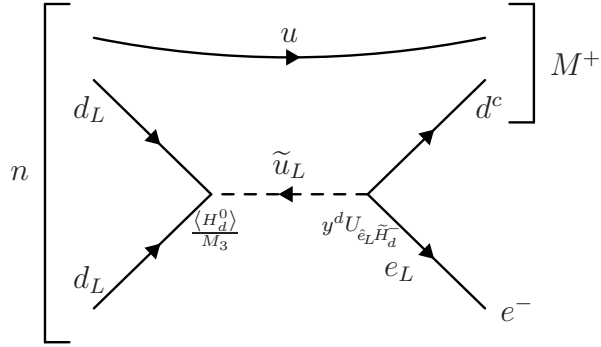} 
\end{center}
\caption{\label{fig:B-Lviolation} A Feynman diagram for $B-L$ violating 
neutron decay.  In this example the combination of 
${\cal O}_3$--(\ref{eq:ubardebar}) is used. }
\end{figure}

A Feynman diagram in Figure~\ref{fig:B-Lviolation} shows that 
a $B-L$ breaking nucleon decay $n \to M^{+} + \ell^{-}$ is possible 
%%%%%%%%%%%%%%%%%%%%%%%%%%%%%%%%%%%%
%%%%%%%%%%%%%%%%%%%%%%%%%%%%%%%%%%%%
in the presence of R-parity violation. Squark-exchange diagrams
involving $\tilde{u}^*\bar{e}d$ vertex (\ref{eq:ubardebar}) induce 
neutron decay with a negatively charged lepton in the final state.
This $B-L$ breaking neutron decay, coming from a partial amplitude 
${\cal O}_3$--(\ref{eq:ubardebar}), is rather a robust prediction 
of R-parity violation.  As one can see in Table~\ref{tab:pdecay}, 
all but one combination of two operators for nucleon decay allow 
the $B-L$ breaking neutron decay in the third row.\footnote{Although 
proton decay processes $p \rightarrow M^+ + \nu$ also break the 
$B-L$ symmetry, there is no way to confirm in experiments whether 
the missing particle is $\nu$ or something else.}
Even in gauge mediation scenario with very light gravitino of $m_{3/2} \sim {\cal O}(1\mbox{--}10\,\EV)$,
the amplitudes involving ${\cal O}_{\tilde{G}}$ is at most comparable with other decay modes, so
%Except in gauge mediation scenario with a very light gravitino, 
%when the amplitudes involving ${\cal O}_{\tilde{G}}$ 
%dominate in the nucleon decay processes,  
$n \to M^{+} + \ell^{-}$ may be observed with a considerable fraction.
%%%%%%%%%%%%%%%%%%%%%%%%%%%%%%%%%%%%%%%%%%%%%%%%
\begin{table}[tb]
\begin{center}
\begin{tabular}{c||c|c||c||c||c|c||c}
  & ${\cal O}_3$-(ren.)
  & ${\cal O}_7$-(ren.)
  & ${\cal O}_{3,7}$-${\cal O}_4$ 
  & ${\cal O}_{3,7}$-${\cal O}_{9}$ 
  & ${\cal O}_3$-${\cal O}_{10}$ 
  & ${\cal O}_7$-${\cal O}_{10}$ 
  & ${\cal O}_{3,7}$-${\cal O}_{\tilde{G}}$ \\ 
\hline
   $p[n] \rightarrow M^0[M^-] + \ell^+$
 & $\surd$ & $\surd$ & $\surd$ & $\surd$ & (LR) & $\surd$ & \\
\hline 
   $p[n] \rightarrow M^+[M^0] + f^0$
 & $\surd$ & $\surd$ & & $\surd$ & & & $\surd$ \\
\hline
   $n \rightarrow M^+ + \ell^-$ % $l^-$
 & $\surd$ & $\surd$ & (LR) & (LR) & (LR) & (LR) &
\end{tabular} 
\caption{\label{tab:pdecay}
  Check marks in this table show that
  a combination of operators in a given column 
  contributes to nucleon decay processes specified in the three rows. % :
  % $p [n] \rightarrow M^0 [M^-] + \ell^+$, 
  % $p [n] \rightarrow M^+ [M^0] + f^0$
  % and $n \rightarrow M^+ + \ell^-$.
  $M^{\pm, 0}$ are mesons with given electric charges,
  $\ell^{\pm} = e^{\pm}, \mu^{\pm}$,
  and $f^0$ a neutral fermion such as $\bar{\nu}, \nu$ and $\tilde{G}$.
  % ``Dim.-5'' and ``Dim.-6'' 
  % in the last two columns correspond to conventional dimension-5 
  % nucleon decay operators 
  % $W \ni Q Q Q L + \bar{U} \bar{U} \bar{E} \bar{D}$, and gauge-boson 
  % exchange dimension-6 operators, 
  % $K \ni \bar{E}^\dagger Q \bar{U}^\dagger Q +
  % \bar{D}^\dagger L \bar{U}^\dagger Q$. 
  %
  % Three rows are for 
  % two-body nucleon decay with a postive-charged lepton, neutral fermion 
  % and negative-charged lepton, respectively, in the final state.
  % Check marks in the first row, therefore, mean that nucleon decay 
  % $p\rightarrow M^0 + l^+$ and $n \rightarrow M^- + l^+$ can be induced 
  % with $M^0$ and $M^\pm$ a neutral and charged meson and 
  % $l^+=e^+, \mu^+$. 
  % Those in the second row are for the process 
  % $p \rightarrow M^+ + f$ and $n \rightarrow M^0 + f$ with 
  % $f = \bar{\nu}, \nu, \tilde{G}$. Corresponding operators induce 
  % a $B-L$ violating decay $n\rightarrow M^+ + l^-$ if check marks 
  % are in the third row; $l^-=e^-, \mu^-$ here. 
  ``(ren.)'' in the first two columns mean renormalizable interactions 
  listed in (\ref{eq:ubardebar}--\ref{eq:dRdnu}).
  Some nucleon decay partial amplitudes are generated 
  with an insertion of left-right mixing in the virtual squark propagator 
  (see Figure~\ref{fig:Op4910G}~(c)).
  Such cases are indicated in the table by ``(LR)''. 
}
\end{center}
\end{table}
%%%%%%%%%%%%%%%%%%%%%%%%%%%%%%%%%%%%%%%%%%%%%%%%%%%%

The $B-L$ violating nucleon decay is possible 
because all the effective operators that we obtain 
after integrating out squarks are dimension-7 or higher. 
Either Higgs boson (vev) or derivatives are involved in those operators.
% \footnote{Some of those 
% effective operators become dimension-6 4-fermion operators 
% by replacing the Higgs doublet field by its vev, but 
% such dimension-6 operators are not $\SU(2)_L \times \U(1)_Y$ invariant.}
For example,
\begin{equation}
 {\cal O}_3\mbox{--}(\ref{eq:ubardebar}) \qquad 
 {\cal L}_{\rm eff.} \sim \frac{1}{m^2_{\tilde{q}}} 
   \frac{y^d (\mu_i/\mu_0)}{M_3} 
   \overline{d^c} (l q) (q H_d),
\end{equation}
is a dimension-7 operator in the Standard-Model fields, 
with the coefficient $1/m_{\tilde{q}}^2$ coming from squark exchange.
After a Higgs doublet is replaced by its vev, 
this operator effectively becomes a 4-fermion dimension-6 operator 
with a coefficient suppressed only by a single power of a large energy scale $M_3$. 
Thus, such an effective operator can be more important 
than the nucleon decay operators (\ref{eq:4fermi6}, \ref{eq:4fermi5}) 
of conventional unified theories.
%
%%%%%%%%%%%%%%%%%%%%%%%%%%%%%%%%%%%%%%%%%%%%%%%%%%%%
\begin{figure}
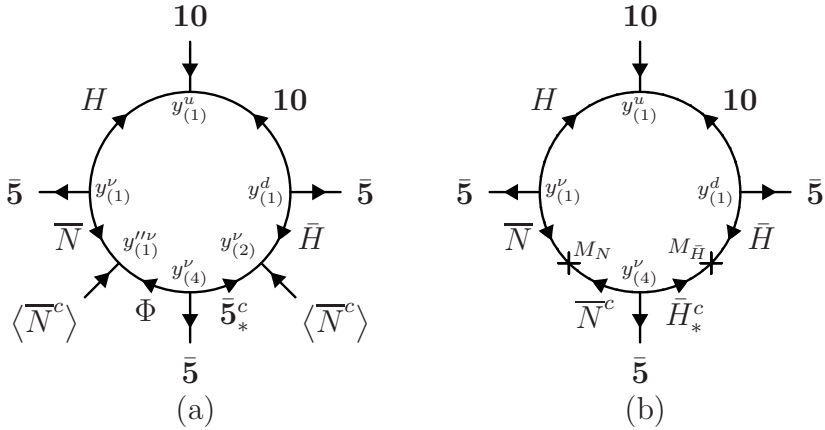

\begin{center}
\begin{tabular}{ccc}
\input{figure_10555_1} & \hspace{1cm} &\input{figure_10555_2} \\
&&\\
 %& \hspace{1cm} & \input{figure_stau-decay3} \\
(a) && (b)
\end{tabular}
\end{center}
\caption{\label{fig:kahler-10555}
Feynman diagrams for 
$K \ni \bar{\bf 5}^\dag\;\bar{\bf 5}^\dag\;\bar{\bf 5}^\dag\;{\bf 10}$,
(a) for the 4+1 model and (b) for the 3+2 model.}
\label{generating_55510}
\end{figure}
%%%%%%%%%%%%%%%%%%%%%%%%%%%%%%%%%%%%%%%%%%%%%%%%%%%%
%

\begin{figure}
\begin{center}
\input{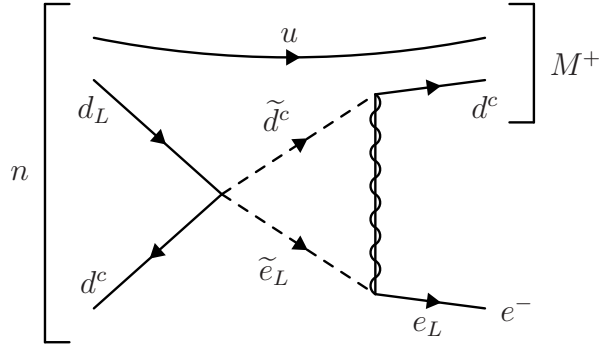}
\end{center}
\caption{\label{fig:pdecay-10555}
A Feynman diagram for $B-L$ violating neutron decay through
$K \ni \bar{\bf 5}^\dag\;\bar{\bf 5}^\dag\;\bar{\bf 5}^\dag\;{\bf 10}$.}
\label{ndecay_55510}
\end{figure}

The $B-L$ violating neutron decay is regarded as a signal different 
from prediction of conventional unified theories, and certainly 
R-parity violation is one of possible explanations. One should 
keep in mind, however, that an effective dimension-6 operator 
\begin{equation}
 K \ni \bar{\bf 5}^\dagger \; \bar{\bf 5}^\dagger \; 
       \bar{\bf 5}^\dagger \; {\bf 10} = 
      L^\dagger \; \bar{D}^\dagger \; \bar{D}^\dagger \; Q 
   + \bar{D}^\dagger \; \bar{D}^\dagger \; \bar{D}^\dagger \bar{E}
   + L^\dagger \; L^\dagger \; \bar{D}^\dagger \; \bar{U}
\label{eq:55510}
\end{equation}
also breaks $B-L$ symmetry, while R-parity is conserved.
It is true that this operator cannot be generated
by gauge-boson exchange processes
in any unified theories on 3+1 dimensions,
as such processes would result
in two anti-chiral multiplets and two chiral multiplets.
But, it will be possible to come up with a model of SUSY GUT
with some new matter multiplets and their interactions in the 
superpotential, so that this operator is generated.\footnote{
This operator exists even in the 4+1 model and 3+2 model
that we explained in section~\ref{ssec:4132}. 
See Figure~\ref{generating_55510}.}
The first two operators of the MSSM in (\ref{eq:55510}) generate 
dimension-7 effective operators in the standard model, after 
sfermions and gauginos are integrated out (Figure~\ref{ndecay_55510}).
Therefore, just an observation of $B-L$ violating neutron decay
is not enough to conclude that R-parity is not preserved.
It is an interesting question whether it is possible
to distinguish nucleon decay due to R-parity violation
from one through (\ref{eq:55510}),
using data on branching fraction of various decay modes.
But this question is beyond the scope of this paper.
%
%%%%%%%%%%%%%%%%%%%%%%%%%%%%%%%%%%%%%%%%%%%%%%%%%%%%
%%%%%%%%%%%%%%%%%%%%%%%%%%%%%%%%%%%%%%%%%%%%%%%%%%%%
%

It should be possible to derive predictions on flavor dependent 
branching fractions by exploiting flavor structure of R-parity 
violating operators. All three chiral multiplets $Q$ 
in ${\cal O}_3$ cannot be in the same generation; one chiral 
multiplet $Q$ in the second generation is necessarily involved.
On the other hand, operators (\ref{eq:ubardebar}--\ref{eq:dRdnu}) and 
${\cal O}_{\tilde{G}}$ does not introduce a generation mixing 
(except for predictable small flavor mixing from CKM matrix elements 
in (\ref{eq:ubardebar}, \ref{eq:udbare})). Thus, dominant decay modes will 
include a $K$ meson in the final state, if the nucleon decay is 
dominated by partial amplitudes 
${\cal O}_3$--(\ref{eq:ubardebar}--\ref{eq:dRdnu}) or 
${\cal O}_3$--${\cal O}_{\tilde{G}}$.
Quantitative predictions, however, are not covered in this article. 

%If nucleon decay amplitude is dominated by a combination of operators 
%${\cal O}_{3,7}$--${\cal O}_{\tilde{G}}$, then all the decay processes 
%do not apparently break $B-L$ symmetry; it will be impossible to 
%find out experimentally that the outgoing missing particle is not 
%an anti-neutrino but a gravitino.
%In this case in the gauge mediation scenario, however, we have 
%a prediction that the branching fraction of nucleon decay 
%with a charged lepton in the final state vanishes,\footnote{
%The dimension-5 proton decay operator in the minimal 
%SU(5)$_{\rm GUT}$ model predicts that the dominant decay mode of proton 
%is $p \rightarrow K^+ + \bar{\nu}$. But it also predicts non-vanishing 
%branching fraction of proton decay that goes to $\mu^+$ or $e^+$ \cite{HMY}.} 
%unless other contributions to nucleon decay are comparable. 
%Therefore, setting a stronger upper limit on the branching fraction 
%of $p\rightarrow M^0 + l^+$ from experiments helps strengthen 
%the case for nucleon decay dominated by 
%${\cal O}_{3,7}$--${\cal O}_{\tilde{G}}$.

%%%%%%%%%%%%%%%%%%%%%%%%%%%%%%%%%%%%%%%%%%%%%%%%%%%%%%%
\subsection{Brief Summary}
\label{ssec:summary-phen}
%%%%%%%%%%%%%%%%%%%%%%%%%%%%%%%%%%%%%%%%%%%%%%%%%%%%%%%

In section~\ref{sec:phen}, we studied phenomenological constraints 
on R-parity violation, specialized to cases when there are only 
bilinear and dimension-5 R-parity violation. The (virtual) absence of 
trilinear R-parity violation is a prediction of the theoretical 
framework in section~\ref{sec:frame}. BBN constraints were reanalyzed 
in section~\ref{ssec:BBN}, where we exploited the latest understanding 
of the impact of hadronic energy injection and presence of stable 
charged particle during the period of BBN. Section~\ref{ssec:proton}
placed a limit on bilinear--dimension-5 R-parity violation through 
proton decay. An allowed parameter space of bilinear--dimension-5 
R parity violation is presented in Figure~\ref{fig:para-space}, 
where constraints from BBN and proton decay are shown along with 
those from cosmological neutrino mass bound and absence of washout 
of baryon/lepton asymmetry at high temperature. 
Theoretical framework in section~\ref{sec:frame} predicts a 
theoretically likely region in the parameter space (with very 
large uncertainties). The region is roughly around the upper 
right corner of the space that is still allowed by all the 
phenomenological constraints discussed above.

Because the likely region is around the upper right corner, 
the framework in the section \ref{sec:frame}
prefer stronger R-parity violation within the allowed parameter region.
This means that there is a chance that R-parity violation is confirmed by experiments. 
R-parity violating decay of the LSP in the visible sector (vLSP) may be 
observed inside the detectors of the LHC for large bilinear R-parity 
violation. Studies of this signal of R-parity violation are found 
in the literature, and we have nothing to add in this article.
The R-parity violating decay of vLSP may not 
be observed in accelerators, however, for smaller bilinear R-parity violation and/or
light gravitino in the gauge mediation scenario. 

As we discussed in 
section~\ref{ssec:proton}, nucleon decay can be an alternative 
way to probe R-parity violation.
Especially, non-vanishing branching fraction of $B-L$ breaking 
neutron decay $n \rightarrow M^+ + \ell^-$ is a robust prediction 
of bilinear--dimension-5 R-parity violation.
$n \rightarrow M^+ + \ell^-$ is always predicted except in
the decay mode into gravitino, and the decay ratio into gravitino 
can be at most comparable with other decay mode, 
even if gravitino mass is very small as ${\cal O}(1\mbox{--}10\,\EV)$.
This is a notable feature of bilinear--dimension-5 R-parity violation
because the observation of $n \rightarrow M^+ + \ell^-$ enables us
to distinguish it from conventional (SUSY) GUT's, 
which predict only $B-L$ preserving nucleon decay processes.
%First, decay rates of nucleons are enhanced by bilinear R-parity 
%violation with $\xi \simgt 10^{-9}$ or light gravitino 
%($m_{3/2} \simlt 1 \; \KEV$). 
%Second, non-vanishing branching fraction of $B-L$ breaking 
%neutron decay $n \rightarrow M^+ + l^-$ is a robust prediction 
%of bilinear--dimension-5 R-parity violation, as long as 
%$m_{3/2} \simgt 1 \; \KEV$. Conventional (SUSY) GUT's predict only 
%$B-L$ preserving nucleon decay processes.
%For $m_{3/2} \simlt 1 \; \KEV$ (nature of dark matter suggests that 
%$m_{3/2} \simlt {\cal O}(10 \; \EV)$ in this case), nucleon decay 
%process $p[n] \rightarrow M^+[M^-] + \psi_{3/2}$ is the dominant 
%mode, and it is predicted that the branching fraction is negligible 
%for the decay processes with charged leptons in the final states. 
Therefore, nucleon decay experiments can be complementary to 
the R-parity violating decay of the vLSP in the detectors of 
the LHC in probing R-parity violation.

%%%%%%%%%%%%%%%%%%%%%%%%%%%%%%%%%%%%%%%%%%%%%%%
%\section*{Acknowledgements} % preferred spelling in BrE
\section*{Acknowledgments}  % more usual in AmE

We thank K. Hamaguchi, T. Kawano, K. Kohri, and T.T. Yanagida 
for useful comments and fruitful discussion. 

%%%%%%%%%%%%%%%%%%%%%%%%%%%%%%%%%%%%%%%%%%%%%%%%%%%
\appendix

%%%%%%%%%%%%%%%%%%%%%%%%%%%%%%%%%%%%%%%%%%%%%%%%%%%%%
\section{Notes on Bilinear R-parity Violation}
%%%%%%%%%%%%%%%%%%%%%%%%%%%%%%%%%%%%%%%%%%%%%%%%%%%%%

In the MSSM with bilinear R-parity violation, superpotential 
is given by 
\begin{equation}
 W = y^u_{ij} \; \bar{U}_i \; Q_j \; H_u
   - y^d_{kj} \; \bar{D}_k \; Q_j \; H_d 
   - y^e_{kj} \; L_k \; \bar{E}_j \; H_d 
   + \mu_0 \; H_u \; H_d 
   + \mu_i \; H_u \; L_i, 
\end{equation}
and soft SUSY breaking potential by 
\begin{eqnarray}
 V_{\rm soft} & = & m^2_{Q \; ij} \; \tilde{q}^\dagger_i \tilde{q}_j 
   + m^2_{D \; ij} \; \tilde{d}^{c \dagger}_i \tilde{d}^c_j 
   + m^2_{U \; ij} \; \tilde{u}^{c \dagger}_i \tilde{u}^c_j
   + m^2_{E \; ij} \; \tilde{e}^{c \dagger}_i \tilde{e}^c_j 
   + m^2_{Hu} |h_u|^2 \\
  & + & m^2_{L \; 00} \; |h_d|^2
      + m^2_{L \; ij} \; \tilde{l}^\dagger_i \tilde{l}_j
      + m^2_{L \; 0i} \; h_d^\dagger \tilde{l}_i + {\rm h.c.} \\
  & + & B_0 h_u h_d + B_i h_u \tilde{l}_i \\
  & + & A^u y^u \tilde{u}^c \; \tilde{q} \; h_u 
    - A^d y^d \tilde{d}^c \; \tilde{q} \; h_d 
    - A^e y^e \tilde{l} \; \tilde{e}^c \; h_d.
\end{eqnarray}
In bilinear R-parity violating scenario,
there exists a basis of chiral multiplets $(H_d, L_{1,2,3})$ 
such that the R-parity violation appears only in bilinear terms 
as above. This basis is called bilinear basis.
We basically follow the convention of \cite{Martin}.

We will assume that all the R-parity violating parameters, 
$\mu_i$, $B_i$ and $m^2_{L \; 0i}$, are small compared with 
$\mu_0$, $B_0$ and $m^2_{L \; 00}$ and $m^2_{L \; ij}$, 
respectively. Anything that are quadratic in the R-parity violating 
parameters are ignored in this article. 
Sneutrino vev's in the bilinear basis 
are given approximately by \cite{Hempfling}
%
% Nakajima's version
%\begin{equation}
%       v_i
%\equiv \vev{\tilde{\nu}_i}
%\simeq \frac{B_i v_u - 2v_d(\mu_0 \mu_i + m^2_{L \; 0i})}
%            {2m^2_{L \; ii} + M_Z^2 \cos(2 \beta)},
%\end{equation}
%
%
\begin{equation}
       v_i \simeq \frac{B_i v_u - v_d(\mu_0 \mu_i + m^2_{L \; 0i})}
            {m^2_{L \; ii} + \frac{M_Z^2}{2} \cos(2 \beta)},
\end{equation}
where $v_i \equiv \vev{\tilde{\nu}_i} = \vev{L^0_i}$, 
$v_d \equiv \vev{H^0_d}$, $v_u \equiv \vev{H_u^0}$, and 
$\tan \beta \equiv v_u/v_d$, so $v_i \ll v_d$. 
We do not distinguish $\sqrt{v_d^2 + \sum_i v_i^2}$ from $v_d$, 
as the difference between them is quadratic in the small 
R-parity violating parameters.
%
% HN's version 
%
%\begin{equation}
%       v_d
%\equiv \vev{\tilde{h}_d^0} \simeq 
%       \frac{B_0 v_u}
%            {2(\mu_0^2 + m^2_{L \; 00}) - M_Z^2 \sin^2 \beta},
%\end{equation}
%
Using the leading-order part of the $H_d^0$ minimization condition
\begin{equation}
 (m^2_{L \; 00} + \mu_0^2) + \frac{M_Z^2}{2} \cos (2\beta) = 
  B_0 \tan \beta,
\end{equation}
the expression above can be rewritten in a useful form \cite{NP}:
%
% HN's version
%
%\begin{equation}
%       \frac{v_i}{v_d}
%\simeq \frac{B_i - 2\cot\beta (\mu_0\mu_i + m^2_{L \; 0i})}
%            {B_0 - 2\cot\beta (\mu_0^2 + m^2_{L \; 00} - m^2_{L \; ii})}.
%\label{eq:vivd}
%\end{equation}
%
%
\begin{equation}
       \frac{v_i}{v_d} \simeq
       \frac{B_i - \cot\beta (\mu_0\mu_i + m^2_{L \; 0i})}
            {B_0 - \cot\beta (\mu_0^2 + m^2_{L \; 00} - m^2_{L \; ii})}.
\label{eq:vivd}
\end{equation}

Misalignment parameters $\epsilon'_i$ are defined by 
\cite{Banks:1995by}\footnote{The misalignment parameter $\xi$ 
in \cite{Banks:1995by, Barbieri} corresponds to $|\epsilon'|$, 
norm of $\epsilon'_i$. Many references introduced different notations 
for these misalignment parameters. To name a few, 
Ref. \cite{NP} introduces $\alpha_i \equiv v_i/v_d$, 
$\gamma_i \equiv \mu_i/\mu_0$, $\delta_i \equiv B_i/B_0$, and 
difference between arbitrary two out of $\alpha, \gamma$ and $\delta$ 
is basis independent. For example, $(\gamma_i - \alpha_i) = \epsilon'_i$. 
Reference~\cite{chidecayA} uses $\Lambda_i = \epsilon'_i \mu_0 v_d$.
$\zeta$ in \cite{Barbieri} is norm of $(\delta_i-\alpha_i)$.}
\begin{equation}
 \epsilon'_i  \equiv  \left(\frac{\mu_i}{\mu_0} - \frac{v_i}{v_d}\right),
  \label{eq:epsi-def}
\end{equation}
and its $\cot \beta$ expansion is given by
\begin{eqnarray}
 \epsilon'_i & = &  
   \left(\frac{\mu_i}{\mu_0} - \frac{B_i}{B_0}\right)
   \left(1 + \cot \beta \frac{\mu_0^2}{B_0}\right)
 - \frac{B_i}{B_0} \cot \beta \frac{m^2_{L \; 00}- m^2_{L \; ii}}{B_0}
 + \cot \beta \frac{m^2_{L \; 0i}}{B_0} \nonumber \\
&  + &  {\cal O}\left(\frac{\mu_i}{\mu_0}\cot^2 \beta,\,\frac{B_i}{B_0}\cot^2 \beta\right).
\label{eq:epsi-cotB}
\end{eqnarray}
In minimal SUGRA mediation scenario, the first three terms 
in (\ref{eq:epsi-cotB}) vanish in the UV initial condition, 
and the remaining terms are suppressed by $\tan^{-2} \beta$.  
Thus, $\epsilon'_i$ can be much smaller than $\mu_i/\mu_0$ or 
$B_i/B_0$. For large $\tan \beta$, however, the first two 
terms become larger when renormalized at the electroweak scale,  
as they are generated through radiative corrections involving 
bottom Yukawa couplings \cite{Hempfling, NP}; 
there is still a cancellation of about $10^{-4} \tan^2 \beta$ in 
$(\mu_i/\mu_0 - B_i/B_0)$ and about $10^{-3} \tan^2 \beta$ in 
$(m^2_{L \; 00} - m^2_{L \; ii})$.  
In the end, $\epsilon'_i$ can be as small as 
$10^{-2} \times {\cal O}(\mu_i / \mu_0)
 \approx 10^{-2} \times {\cal O}(B_i/B_0)$, but 
it is unlikely that $\epsilon'_i$'s are even smaller than that.

%%%%%%%%%%%%%%%%%%%%%%%%%%%%%%%%%%%%%%%%%%%%%%%%%%%%%%%%%%%%%%%%
\subsection{Mass Matrices}
%%%%%%%%%%%%%%%%%%%%%%%%%%%%%%%%%%%%%%%%%%%%%%%%%%%%%%%%%%%%%%%%

\paragraph{Neutralino--Neutrino Mixing}
Because of $v_i \neq 0$ and $\mu_i \neq 0$, mass matrices of neutralinos and neutrinos 
are mixed up. In the gauge-eigenstate basis 
$\psi^0 = (\widetilde{B}, \widetilde{W}^0, \widetilde{H}^0_u,
\widetilde{H}^0_d,\nu_i)^T$, the neutralino--neutrino mass matrix 
${\cal L} \ni - \psi^{0 \; T} \; M_{\tilde{N}; \; 7 \times 7} \;
\psi^0/2$ becomes
\begin{equation}
  M_{\tilde{N} ; \; 7 \times 7}
= \left(
    \begin{array}{cc}
      M_{\tilde{N} ; \; 4 \times 4} & m_{\tilde{N}}^T \\ 
% m_{\tilde{N} ; \; 3 \times 4}^T \\
      m_{\tilde{N} ; \; 3 \times 4} & m_{\rm ss}
    \end{array}
   \right) ,
\label{eq:MN-7x7}
\end{equation}
where $m_{\rm ss}$ is the contribution from the see-saw mechanism 
involving right-handed neutrinos, and 
\begin{eqnarray}
 M_{\tilde{N} ; \; 4 \times 4} & = & \left( 
 \begin{array}{cccc}
  M_1 & & - g' v_d/\sqrt{2} & g' v_u /\sqrt{2} \\
  & M_2 &  g v_d/\sqrt{2} & - g v_u / \sqrt{2} \\
  - g' v_d/\sqrt{2} & g v_d/\sqrt{2} & 0 & -\mu_0 \\
  g' v_u/\sqrt{2} & - g v_u/\sqrt{2} & -\mu_0 & 0
 \end{array}
  \right), \\
 m_{\tilde{N} ; \; 3 \times 4} & = & \left(
\begin{array}{cccc}
 - g' v_i/\sqrt{2}, &   gv_i/\sqrt{2}, & 0,  & - \mu_i \\
\end{array}
\right).
\end{eqnarray}
Gauge-eigenstate basis $\psi^0$ and mass-eigenstate basis $\hat{\psi}^0$
are related by $\hat{\psi}^0 = N_{7 \times 7} \cdot \psi^0$, where 
a unitary matrix $N_{7 \times 7}$ makes 
$N_{7 \times 7}^* \cdot M_{\tilde{N}; 7 \times 7} \cdot 
N^{-1}_{7 \times 7}$ diagonal.
Four mass eigenvalues are supposed to be around the electroweak scale or SUSY scale, 
and three others are very small. 
We will write these mass eigenstates as $\hat{\tilde{\chi}^0}_{1,2,3,4}$ and 
$\hat{\nu}_{1,2,3}$. Ignoring $m_{\rm ss}$ and anything that 
comes at the second order of bilinear R-parity violation, the
diagonalization matrix is expressed in a form 
\begin{equation}
       N_{7 \times 7}
\simeq \left(
         \begin{array}{cc}
           N_{4 \times 4} & \\
           & 1_{3 \times 3}
         \end{array}
       \right) \;
       \left(
         \begin{array}{cc}
           1_{4 \times 4} & \xi^T \\
                    - \xi^* & 1_{3 \times 3}
         \end{array}
       \right),
\end{equation}
where 
% HN's version
% $\xi^* M_{\tilde{N}; \; 4 \times 4} = m_{\tilde{N}; \; 3 \times 4}$
$\xi  \cdot M_{\tilde{N}; \; 4 \times 4} = m_{\tilde{N}; \; 3 \times 4}$
at this order, and
$N_{4 \times 4}^* M_{\tilde{N}; \; 4 \times 4} N_{4 \times 4}^{-1}$
is diagonal.
The suffixes of $\xi_{\alpha\beta}$ run over $\alpha \in \{ \hat{\nu}_1,\hat{\nu}_2,\hat{\nu}_3\}$ and
$\beta \in \{ \widetilde{B}, \widetilde{W}^0,
\widetilde{H}^0_d, \widetilde{H}^0_u\}$.
% Mass-eigenstate basis
% $\hat{\psi}^0 = (\hat{\chi}^0_{1,2,3,4}, \; \hat{\nu}_{1,2,3})$
% and gauge-eigenstate basis $\psi^0$
% are related by $\hat{\psi}^0 = N_{7 \times 7} \cdot \psi^0$.
We will keep the lower-right $3 \times 3$ part of $N_{7 \times 7}$
as $1_{3 \times 3}$, so that
the three neutrino-like ``massless'' eigenstates
$\hat{\nu}_{1,2,3}$ become $\nu_{e,\mu, \tau}$ approximately.

Assuming a little hierarchy between the electroweak symmetry breaking 
scale and the SUSY-breaking scale, i.e., $M_{Z,W} \ll M_{1,2}, \mu_0$, 
simple expressions for $\xi_{\alpha\beta}$ are obtained.
%
% HN's version
%
%\begin{eqnarray}
%           \xi_{i \widetilde{B}}
%& \simeq & \frac{M_Z}{M_1} \sin \theta_W \cos \beta \; \epsilon'_i, \\
%           \xi_{i \widetilde{W}^0}
%& \simeq & - \frac{M_Z}{M_2} \cos \theta_W \cos \beta \; \epsilon'_i, \\
%           \xi_{i \widetilde{H}^0_u}
%& \simeq & - \frac{m_0}{\mu_0} \; \epsilon'_i, \\
%           \xi_{i \widetilde{H}^0_d}
%& \simeq & \frac{\mu_i}{\mu_0}
%           - \frac{m_0}{\mu_0} \tan \beta
%           \left(
%             \frac{\mu_i}{\mu_0} + \frac{v_i}{v_d}
%           \right)
%\end{eqnarray}
%
%
\begin{eqnarray}
           \xi_{\hat{\nu}_i \widetilde{B}}
& \simeq & \frac{M_Z}{M_1} \sin \theta_W \cos \beta \; \epsilon'_i, 
         \label{eq:xi-B}\\
           \xi_{\hat{\nu}_i \widetilde{W}^0}
& \simeq & - \frac{M_Z}{M_2} \cos \theta_W \cos \beta \; \epsilon'_i, 
         \label{eq:xi-W}\\
           \xi_{\hat{\nu}_i \widetilde{H}^0_d}
& \simeq & \frac{\mu_i + m_0 \tan \beta \epsilon'_i}{\mu_0}, 
         \label{eq:xi-Hd}\\
           \xi_{\hat{\nu}_i \widetilde{H}^0_u}
& \simeq & - \frac{m_0}{\mu_0} \; \epsilon'_i, 
         \label{eq:xi-Hu}
\end{eqnarray}
where 
\begin{equation}
           m_0
 \equiv  \frac{(M_Z \cos \beta)^2 
          (M_1 \cos^2 \theta_W + M_2 \sin^2 \theta_W)}{M_1 M_2}. % \\
%           \epsilon'_i
% & \equiv & \frac{\mu_i v_d - \mu_0 v_i}{\mu_0 v_d}
%       =   \frac{\mu_i}{\mu_0} - \frac{v_i}{v_d}. 
%  \label{eq:epsi-def}
\end{equation}
The tree-level contribution to the low-energy neutrino mass from 
the higgsino--neutrino mixing is given by $m_0 \epsilon'_i \epsilon'_j$
\cite{Banks:1995by}.

\paragraph{Chargino--Charged-Lepton Mixing}

Charginos and charged leptons also have a mixed mass matrix. 
In the gauge-eigenstate basis
$\psi^+ = (\widetilde{W}^+, \widetilde{H}^+_u, e^c_i)^T$ and
$\psi^- = (\widetilde{W}^-, \widetilde{H}^-_d, e_{L \; i})^T$,
the mass matrix $M_{\tilde{C} ; \; 5 \times 5}$ in
${\cal L} \ni - \psi^{- T} \; M_{\tilde{C}; \; 5 \times 5} \; \psi^+ + {\rm h.c.}$
is given by
\begin{equation}
  M_{\tilde{C}; \; 5 \times 5}
= \left(
    \begin{array}{ccc}
      M_2   & g v_u &   0         \\
      g v_d & \mu_0 & - y^e_i v_i \\
      g v_i & \mu_i &   y^e_i v_d \\
    \end{array}
  \right).
\end{equation}
Mass eigenstates 
$\hat{\psi}^+ = (\hat{\tilde{\chi}}^+_{1,2}, \; \hat{e}^c_{1,2,3})$
and $\hat{\psi}^- = (\hat{\tilde{\chi}}^-_{1,2}, \; \hat{e}_{L \; 1,2,3})$
are unitary transforms of the original gauge eigenstates;
\begin{equation}
 \hat{\psi}^- = U \cdot \psi^-, \qquad 
 \hat{\psi}^+ = V \cdot \psi^+,
\end{equation}
where unitary matrices $U$ and $V$ make 
$U^* (M_{\tilde{C}} M^\dagger_{\tilde{C}}) U^T$ and 
$V (M^\dagger_{\tilde{C}} M_{\tilde{C}}) V^{-1}$ diagonal.

Under the same approximation as above, eigenvectors for the 
$\{ e_L, \mu_L, \tau_L \}$-like mass eigenstates 
$\hat{e}_{L \; 1,2,3}$ are
\begin{equation}
       U_{\hat{e}_{L\; i} \widetilde{W}^-}
\simeq \frac{\sqrt{2} M_W \cos \beta}{M_2} \epsilon'_i,
       \qquad
       U_{\hat{e}_{L\; i} \widetilde{H}_d^-} 
\simeq - \frac{\mu_i}{\mu_0}.
       \label{eq:U's}
\end{equation}
Eigenvectors for the $\{ e^c, \mu^c, \tau^c\}$-like mass eigenstates 
$\hat{e}^c_{1,2,3}$  are 
\begin{equation}
       V_{\hat{e}^c_i  \widetilde{W}^+}
\simeq \frac{\sqrt{2}M_W (M_2 \sin \beta + \mu_0 \cos \beta)}{M_2^2}
       \frac{m^e_i}{\mu_0} \; \epsilon'_i,
       \qquad
       V_{\hat{e}^c_i  \widetilde{H}_u^+}
\simeq - \frac{m^e_i}{\mu_0} \; \epsilon'_i.
\label{eq:V's}
\end{equation}

\paragraph{Charged Scalar Mixing}

There are eight complex scalar fields with $+1$ electric charge in 
the MSSM, namely, $H_u^+, H_d^{-*}, \tilde{e}^c_i, \tilde{e}_{L \; i}^*$, one of which 
is the longitudinal component of $W^+$. It is convenient to 
adopt the $v_i = 0$  basis, ($H'_d$, $L'_i$). 
The $v_i=0$ basis and the bilinear basis are related by 
\begin{equation}
  \left(
    \begin{array}{c}
      H_d' \\
      L_i'
    \end{array}
  \right)
\simeq 
  \left(
    \begin{array}{cc}
      1 & v_i/v_d \\
    - v_i/v_d & 1_{3 \times 3}
    \end{array}
  \right)
  \left(
    \begin{array}{c}
      H_d \\
      L_i
    \end{array}
  \right).
\end{equation}
In the new basis $\phi^+ = (H^+, H^{- \prime *}, \tilde{e}^c_i,\tilde{e}'_{L \; i})^T$, 
mass matrix in ${\cal L} \ni -\phi^{+\dagger} \; M_{C; \; 8\times 8} \; \phi^+$ 
is given by 
\begin{equation}
  M_{C; \; 8 \times 8}
= \left(
  \begin{array}{cc|cc}
    M_u^2 & B_0' & m_q^e \mu_q' & B'_j \\
    B_0' & M_{d}^2 & m_q^e \mu_q' \tan \beta & M_{L \; 0j}^2 \\
    \hline
    m_p^e \mu_p' & m_p^e \mu_p' \tan \beta & M_{E \; pq}^2 & M_{{\rm mix} \; pj}^2 \\
    B_i' & M_{L \; i0}^2 & M_{{\rm mix} \; i q}^2 & M_{L \; ij}^2 
  \end{array}
  \right),
\end{equation}
where
\begin{align}
B_0'   & \simeq B_0 + \frac{1}{2} M_W^2 \sin (2 \beta), \\
B_i'   & \simeq B_i - (v_i/v_d)B_0, \\
\mu_i' & \simeq \mu_i - (v_i/v_d)\mu_0 = \epsilon'_i \mu_0
\end{align}
(i.e., $\{ B_i', \mu_i' \}$ are $\{ B_i, \mu_i \}$ in the $v_i=0$ basis), and
\begin{eqnarray}
 M_u^2 & \equiv & m_{H_u}^2
                  - \frac{1}{2} M_Z^2 \cos (2 \beta)
                  +  M_W^2 \cos^2 \beta
                  + \mu_0^2 + \mu_i^{\prime \; 2}, \\
 M_d^2 & \equiv & m_{L \; 00}^{\prime \; 2}
                  + \frac{1}{2} M_Z^2 \cos (2 \beta)
                  +  M_W^2 \sin^2 \beta
                  + \mu_0^2, \\
 M_{L \; ij}^2
       & \equiv & m_{L \; ij}^{\prime \; 2}
                  + \left( \frac{1}{2} M_Z^2 - M_W^2 \right)
                    \cos (2 \beta) \delta_{ij}
                  + \mu_i' \mu_j'
                  + (m_i^e)^2 \delta_{ij}, \\
 M_{E \; pq}^2
       & \equiv & m_{E \; pq}^2
                  + (M_Z^2 - M_W^2) \cos (2 \beta) \delta_{pq}
                  + (m_p^e)^2 \delta_{pq}, \\
 M_{L \; 0j}^2
       & \equiv & m_{L \; 0j}^{\prime \; 2} + \mu_0 \mu_j', \\
 M_{{\rm mix} \; i q}^2
       & \equiv & m_{iq}^e (A - \mu_0 \tan \beta).
\end{eqnarray}
Here
$m_{L \; 00}^{\prime \; 2}$,
$m_{L \; i0}^{\prime \; 2}$,
$m_{L \; 0j}^{\prime \; 2}$ and
$m_{L \; ij}^{\prime \; 2}$ are non-holomorphic SUSY-breaking mass-square 
parameters in the $v_i=0$ basis. The eigenvector corresponding to 
the would-be Goldstone boson is $\propto (v_u , - v_d, 0,0,0,0,0,0)$.
After removing this mode, the mass matrix becomes 
\begin{equation}
   M_{C; \; 7 \times 7}
= \left(
    \begin{array}{c|cc}
      M_{H^\pm}^2 & m_q^e \mu_q'/\cos\beta  & B'_i/\cos\beta \\
      \hline
      m_p^e \mu_p'/\cos\beta & M_{E \; pq}^2 & M_{{\rm mix} \; pj}^2 \\
      B'_i/\cos\beta & M_{{\rm mix} \; iq}^2 & M_{L \; ij}^2
    \end{array}
  \right).
\end{equation}
In the limit that charged lepton masses $m^e$ are ignored,
both $M_{{\rm mix} \; i \ell}^2$ and $m^e_\ell\mu_\ell'$ vanish,
and hence $\tilde{e}^c$'s do not mix with the charged Higgs boson
and left-handed charged sleptons.

% Assuming that $M_E^2$ and $M_L^2$ are diagonal,
% eigenvectors for these $\tilde{e}^c_i$-like mass eigenstates
% can be obtained up to the leading order
% of the left-right mixing and bilinear R-parity violation:
%
% \begin{equation}
%  \phi^+ \simeq \left( \frac{B'_i}{M^2_{H^\pm} - M^2_{e \; ii}} 
%                       \frac{m_iA}{M^2_{l \; ii} - M^2_{e \; ii}}, 
%                       \frac{\tan \beta \; B'_i}{M^2_{H^\pm} - M^2_{e \; ii}} 
%                       \frac{m_iA}{M^2_{l \; ii} - M^2_{e \; ii}},
%                       \delta_{ij}, 
%                       - \frac{m_iA}{M^2_{l \; ii} - M^2_{e \; ii}} 
%                       \delta_{ij} 
%                 \right)^T \; \hat{\tilde{e}^c}_i. 
% \end{equation}
%

%%%%%%%%%%%%%%%%%%%%%%%%%%%%%%%%%%%%%%%%%%%%%%%%%%%%%%%%%%%%%%%
\subsection{Three Point Vertices}

We present some of three point interactions of the MSSM with bilinear 
R-parity violation. Three point interactions involving 
one SUSY-particle-like mass eigenstate and two Standard-Model-particle 
like mass eigenstates are relevant to two-body decay of the
visible-sector LSP, and to nucleon decay amplitudes.

\paragraph{Squark Yukawa Couplings}
Here is a list of all the three point couplings 
that involve a squark and two Standard-Model fermions. 
They are found in \cite{chidecayB}.
These interactions are combined with ${\cal O}_3$ \eqref{eq:Op34}, 
${\cal O}_7$ \eqref{eq:Op7910}, ${\cal O}''_0$ \eqref{eq:Op0-def}
to generate nucleon decay amplitudes.

The three point interactions involving charged leptons are
\begin{eqnarray}
 \Delta {\cal L} & = & 
   - (g V_{\hat{e}^c_k \widetilde{W}^+}V^{CKM}_{ij}) \; 
     \tilde{u}^*_{L \; i} (\bar{e}_k P_L d_j) \nonumber \\ & & 
   + (y^d_j U_{\hat{e}_{L\;k} \widetilde{H}^-_d})^* V^{CKM}_{ij} \; 
     \tilde{u}^*_{L \; i} (\bar{e}_k P_R d_j) 
   + (y^u_i V_{\hat{e}^c_k \widetilde{H}^+_u}V^{CKM}_{ij}) \;
     \tilde{u}^*_{R \; i} (\bar{e}_k P_L d_j) + {\rm h.c.}, 
  \label{eq:ubardebar}\\
 \Delta {\cal L} & = & 
   - (g U_{\hat{e}_{L\;k} \widetilde{W}^-}) V^{CKM *}_{ji} \; 
     \tilde{d}^*_{L\; i} (\overline{e^c}_k P_L u_j) \nonumber \\ & &
   + (y^u_j V_{\hat{e}^c_k \widetilde{H}^+_u} V^{CKM}_{ji})^* \; 
     \tilde{d}^*_{L\; i} (\overline{e^c}_k P_R u_j) 
   + (y^d_i U_{\hat{e}_{L\;k} \widetilde{H}^-_d}) V^{CKM *}_{ji} \;
     \tilde{d}^*_{R\; i} (\overline{e^c}_k P_L u_j) + {\rm h.c.}
  \label{eq:udbare}
\end{eqnarray}
Here, $\tilde{u}_{L \; i}$ and $V^{CKM}_{ij} \tilde{d}_{L \; j}$ 
are complex scalar fields in the chiral multiplets 
$(u_{L \; i}, V^{CKM}_{ij} d_{L \; j})$, and $\tilde{u}_R$ and $\tilde{d}_R$ 
[resp. $\tilde{u}^*_R = \tilde{u}^c$ and $\tilde{d}^*_R=\tilde{d}^c$] 
in the anti-chiral multiplets $\bar{U}^\dagger$ and $\bar{D}^\dagger$ 
[resp. in the chiral multiplets $\bar{U}$ and $\bar{D}$]. 
Four component notations $\bar{e} P_L d$ and $\bar{e} P_R d$ mean 
$\hat{e}^c d_L$ and $\overline{\hat{e}_L} \overline{d^c}$, respectively, and 
$\overline{e^c} P_{L} u = \overline{u^c} P_{L} e$ and 
$\overline{e^c} P_{R} u = \overline{u^c} P_{R} e$ correspond to 
$\hat{e}_L u_L$ and $\overline{\hat{e}^c} \overline{u^c}$, respectively.

Those involving neutrinos are 
\begin{eqnarray}
 \Delta {\cal L} & = & 
   - \sqrt{2} \left(\frac{g'}{6}\xi_{\hat{\nu}_k \widetilde{B}}
                   +\frac{g}{2}\xi_{\hat{\nu}_k \widetilde{W}^0}\right)
      \tilde{u}^*_{L \; i} (\bar{\nu}_k P_L u_i) 
   - (y^u_i \xi_{\hat{\nu}_k \widetilde{H}^0_u})^* 
      \tilde{u}^*_{L \; i} (\bar{\nu}_k P_R u_i) + {\rm h.c.}, 
      \label{eq:uLunu} \\
  &  &
   - \sqrt{2} \left(\frac{g'}{6}\xi_{\hat{\nu}_k \widetilde{B}}
                   -\frac{g}{2}\xi_{\hat{\nu}_k \widetilde{W}^0}\right)   
     \tilde{d}^*_{L \; i} (\bar{\nu}_k P_L d_i) 
   - (y^d_i \xi_{\hat{\nu}_k \widetilde{H}^0_d})^*
      \tilde{d}^*_{L \; i} (\bar{\nu}_k P_R d_i) + {\rm h.c.}, 
     \label{eq:dLdnu}\\
  &  &
   - \sqrt{2}\left(- \frac{2}{3}g' \xi_{\hat{\nu}_k \widetilde{B}}\right)^*
      \tilde{u}^*_{R \; i} (\bar{\nu}_k P_R u_i)
   - (y^u_i \xi_{\hat{\nu}_k \widetilde{H}^0_u})
      \tilde{u}^*_{R \; i} (\bar{\nu}_k P_L u_i) + {\rm h.c.}, 
     \label{eq:uRunu}\\
  &  & 
   - \sqrt{2} \left(+\frac{1}{3}g' \xi_{\hat{\nu}_k \widetilde{B}}\right)^*
      \tilde{d}^*_{R \; i} (\bar{\nu}_k P_R d_i) 
   - (y^d_i \xi_{\hat{\nu}_k \widetilde{H}^0_d})
      \tilde{d}^*_{R \; i} (\bar{\nu}_k P_L d_i) + {\rm h.c.}
 \label{eq:dRdnu}
\end{eqnarray}
The four component spinor $\bar{\nu}_k \equiv 
(\hat{\nu}_k , \overline{\hat{\nu}}_k)$ is Majorana.

\paragraph{Neutralino Three-Point Vertices}

Here, we list three-point vertices involving the neutralino-like 
mass eigenstate $\hat{\tilde{\chi}^0}_1$ that allow its 
two-body decay to two Standard-Model fields.
\begin{eqnarray}
 \Delta {\cal L} & = & -
   \overline{\hat{\nu}}_k \bar{\sigma}^\mu Z_\mu \hat{\tilde{\chi}^0}_1 
     \times \frac{g_Z}{2} \left(
      \xi_{\hat{\nu}_k\widetilde{B}} N_{4 \times 4; \; 1\widetilde{B}}
    + \xi_{\hat{\nu}_k\widetilde{W}^0} N_{4 \times 4; \; 1\widetilde{W}^0}
    + 2 \xi_{\hat{\nu}_k\widetilde{H}^0_u} N_{4 \times 4; \; 1\widetilde{H}^0_u}
\right)^* + {\rm h.c.} 
  \label{eq:nu-Z-chi}\\
 &  & - \overline{\hat{e}^c}_k \bar{\sigma}^\mu W_\mu^+ \hat{\tilde{\chi}^0}_1 
    \times \frac{g}{\sqrt{2}} \left( 
      -  \sqrt{2} V_{\hat{e}^c_k\widetilde{W}^+} N_{1\widetilde{W}^0}^*
      +  V_{\hat{e}^c_k\widetilde{H}^+_u} N_{1\widetilde{H}^0_u}^* 
                       \right)+ {\rm h.c.}, \label{eq:l-W-chiA}\\
 &  & - \overline{\hat{e}_{L \; k}} \bar{\sigma}^\mu W^-_\mu
  \hat{\tilde{\chi}^0}_1
     \times \frac{g}{\sqrt{2}} \left( 
         \sqrt{2} U_{\hat{e}_{L\;k} \widetilde{W}^-} N_{1\widetilde{W}^0}^*
       + U_{\hat{e}_{L\;k} \widetilde{H}^-_d} N_{1\widetilde{H}^0_d}^* 
       + \sum_\alpha \xi^*_{\hat{\nu}_k\alpha} N_{1\alpha}^*\right)
    + {\rm h.c.} \label{eq:l-W-chiB}
\end{eqnarray}
In the last line, $\alpha$ runs over 
$\{ \widetilde{B}, \widetilde{W}^0, \widetilde{H}^0_d, \widetilde{H}^0_u\}$.
Higgs--neutralino--neutrino three point vertices are omitted. 

% Three point vertices relevant to 
% $\tilde{\chi}^0 \rightarrow H^0 + \nu$ and 
% $\tilde{\chi}^0 \rightarrow H^\pm + l^\mp$ are omitted. 
% if necessary, see \cite{chidecayA, chidecayB}.

\paragraph{Stau Three Point Vertices}

When the left-right mixing is ignored, $\tilde{\tau}^c$ is a
mass-eigenstate $\hat{\tilde{\tau}}^c$ itself.
In this limit, R-parity violating vertices including stau,
which can be used for the calculation of stau-vLSP decay,
are given as
\begin{eqnarray}
 \Delta {\cal L} & = & (\sqrt{2} g' \xi_{\hat{\nu}_k \widetilde{B}}^*) \; 
   \tilde{\tau}^c (\bar{\nu}_k P_R \tau) \qquad (k= e,\mu,\tau) 
  \label{eq:stau-3pt-A}\\ & &
   + (y^e_3 \xi_{\hat{\nu}_k \widetilde{H}^0_d }) \;
      \tilde{\tau}^c (\bar{\nu}_k P_L \tau)
   + (y^e_3 U_{\hat{e}_{L\;k} \widetilde{H}^-_d}^* ) \; 
      \tilde{\tau}^c (\bar{\nu}_\tau P_L e_k) \qquad (k \neq \tau)
   \label{eq:stau-3pt-B} \\&&
   + (y^e_3 (U_{\hat{\tau}_L \widetilde{H}^-_d }^* + \xi_{\hat{\nu}_\tau\widetilde{H}^0_d }))
      \; \tilde{\tau}^c (\bar{\nu}_\tau P_L \tau) + {\rm h.c.}
   \label{eq:stau-3pt-C}
\end{eqnarray}
%

%%%%%%%%%%%%%%%%%%%%%%%%%%%%%%%%%%

\end{fmffile}

\begin{thebibliography}{99}

%
\bibitem{Barbieri}
%
  R.~Barbier {\it et al.},
  ``R-parity violating supersymmetry,''
  Phys.\ Rept.\  {\bf 420}, 1 (2005)
  [arXiv:hep-ph/0406039].
  %%CITATION = PRPLC,420,1;%%
%
\bibitem{Allanach}
%
  B.~C.~Allanach, A.~Dedes and H.~K.~Dreiner,
  ``The R parity violating minimal supergravity model,''
  Phys.\ Rev.\  D {\bf 69}, 115002 (2004)
  [Erratum-ibid.\  D {\bf 72}, 079902 (2005)]
  [arXiv:hep-ph/0309196].
  %%CITATION = PHRVA,D69,115002;%%
%
\bibitem{WeinbergDim5}
%
  S.~Weinberg,
  ``Supersymmetry At Ordinary Energies. 1. Masses And Conservation Laws,''
  Phys.\ Rev.\  D {\bf 26}, 287 (1982).
  %%CITATION = PHRVA,D26,287;%%%
%
\bibitem{FaraggiA}
%
  A.~E.~Faraggi,
  ``Proton stability in superstring derived models,''
  Nucl.\ Phys.\  B {\bf 428}, 111 (1994)
  [arXiv:hep-ph/9403312].
  %%CITATION = NUPHA,B428,111;%%
%
\bibitem{SUSY-0}
%
  Y.~Nir and N.~Seiberg,
  ``Should squarks be degenerate?,''
  Phys.\ Lett.\  B {\bf 309}, 337 (1993)
  [arXiv:hep-ph/9304307].
  %%CITATION = PHLTA,B309,337;%%
%
\bibitem{GR}
%
  G.~F.~Giudice and R.~Rattazzi,
  ``R-parity violation and unification,''
  Phys.\ Lett.\  B {\bf 406}, 321 (1997)
  [arXiv:hep-ph/9704339].
  %%CITATION = PHLTA,B406,321;%%
%
\bibitem{TW1}
%
  R.~Tatar and T.~Watari,
  ``Proton decay, Yukawa couplings and underlying gauge symmetry in string
  theory,''
  Nucl.\ Phys.\  B {\bf 747}, 212 (2006)
  [arXiv:hep-th/0602238].
  %%CITATION = NUPHA,B747,212;%%
%
\bibitem{TW2}
%
  R.~Tatar and T.~Watari,
  ``A stable proton without R-parity: Implications for the LSP,''
  Phys.\ Lett.\  B {\bf 646}, 258 (2007)
  [arXiv:hep-ph/0605315].
  %%CITATION = PHLTA,B646,258;%%
%
\bibitem{DESY-Rparity}
%
  W.~Buchmuller, L.~Covi, K.~Hamaguchi, A.~Ibarra and T.~Yanagida,
  ``Gravitino dark matter in R-parity breaking vacua,''
  JHEP {\bf 0703}, 037 (2007)
  [arXiv:hep-ph/0702184].
  %%CITATION = JHEPA,0703,037;%%
%
\bibitem{HS}
%
  L.~J.~Hall and M.~Suzuki,
  ``Explicit R-Parity Breaking In Supersymmetric Models,''
  Nucl.\ Phys.\  B {\bf 231}, 419 (1984).
  %%CITATION = NUPHA,B231,419;%%
%
\bibitem{Penn5-2}
%
  R.~Donagi, Y.~H.~He, B.~A.~Ovrut and R.~Reinbacher,
  ``Moduli dependent spectra of heterotic compactifications,''
  Phys.\ Lett.\  B {\bf 598}, 279 (2004)
  [arXiv:hep-th/0403291].
  %%CITATION = PHLTA,B598,279;%%
%
\bibitem{WittenWilson}
%
  E.~Witten,
  ``Symmetry Breaking Patterns In Superstring Models,''
  Nucl.\ Phys.\  B {\bf 258}, 75 (1985).
  %%CITATION = NUPHA,B258,75;%%
%
\bibitem{Munich}
%
  R.~Blumenhagen, G.~Honecker and T.~Weigand,
  ``Loop-corrected compactifications of the heterotic string with line
  bundles,''
  JHEP {\bf 0506}, 020 (2005)
  [arXiv:hep-th/0504232].
  %%CITATION = JHEPA,0506,020;%%
%
\bibitem{Het-super}
%
  E.~Witten,
  ``Dimensional Reduction Of Superstring Models,''
  Phys.\ Lett.\  B {\bf 155}, 151 (1985); \\
  %%CITATION = PHLTA,B155,151;%%
%
  J.~P.~Derendinger, L.~E.~Ibanez and H.~P.~Nilles,
  ``On The Low-Energy D = 4, N=1 Supergravity Theory Extracted From The D = 10,
  N=1 Superstring,''
  Phys.\ Lett.\  B {\bf 155}, 65 (1985); \\
  %%CITATION = PHLTA,B155,65;%%
%
  A.~Strominger and E.~Witten,
  ``New Manifolds For Superstring Compactification,''
  Commun.\ Math.\ Phys.\  {\bf 101}, 341 (1985); \\
  %%CITATION = CMPHA,101,341;%%
%
  J.~P.~Derendinger, L.~E.~Ibanez and H.~P.~Nilles,
  ``On The Low-Energy Limit Of Superstring Theories,''
  Nucl.\ Phys.\  B {\bf 267}, 365 (1986); \\
  %%CITATION = NUPHA,B267,365;%%
%
  A.~Strominger,
  ``Yukawa Couplings In Superstring Compactification,''
  Phys.\ Rev.\ Lett.\  {\bf 55}, 2547 (1985).
  %%CITATION = PRLTA,55,2547;%%
%
\bibitem{AGW}
%
  N.~Arkani-Hamed, T.~Gregoire and J.~G.~Wacker,
  ``Higher dimensional supersymmetry in 4D superspace,''
  JHEP {\bf 0203}, 055 (2002)
  [arXiv:hep-th/0101233].
  %%CITATION = JHEPA,0203,055;%%
%
\bibitem{WittenSU(3)}
%
  E.~Witten,
  ``New Issues In Manifolds Of SU(3) Holonomy,''
  Nucl.\ Phys.\  B {\bf 268}, 79 (1986).
  %%CITATION = NUPHA,B268,79;%%
%
\bibitem{FaraggiB}
%
  C.~Coriano, A.~E.~Faraggi and M.~Guzzi,
  ``A Novel String Derived Z' With Stable Proton, Light-Neutrinos and
  R-parity violation,''
  arXiv:0704.1256 [hep-ph].
  %%CITATION = ARXIV:0704.1256;%%
%
\bibitem{ws-instanton}
%
  M.~Dine, N.~Seiberg, X.~G.~Wen and E.~Witten,
  ``Nonperturbative Effects on the String World Sheet,''
  Nucl.\ Phys.\  B {\bf 278}, 769 (1986); \\
  %%CITATION = NUPHA,B278,769;%%
%
  ``Nonperturbative Effects on the String World Sheet. 2,''
  Nucl.\ Phys.\  B {\bf 289}, 319 (1987).
  %%CITATION = NUPHA,B289,319;%%
%
\bibitem{DIN2}
%
 The fourth paper in \cite{Het-super} by Derendinger, Ibanez 
 and Nilles.
%
\bibitem{Donoghue}
%
  V.~Agrawal, S.~M.~Barr, J.~F.~Donoghue and D.~Seckel,
  ``Anthropic considerations in multiple-domain theories and the scale of
  electroweak symmetry breaking,''
  Phys.\ Rev.\ Lett.\  {\bf 80}, 1822 (1998)
  [arXiv:hep-ph/9801253]; \\
  %%CITATION = PRLTA,80,1822;%%
%
  ``The anthropic principle and the mass scale of the standard model,''
  Phys.\ Rev.\  D {\bf 57}, 5480 (1998)
  [arXiv:hep-ph/9707380].
  %%CITATION = PHRVA,D57,5480;%%
%
\bibitem{Hall-mu}
%
  L. J. Hall, in Proceedings of the Winter School
  ``Supersymmetry and Supergravity, Nonperturbative QCD,''
  Mahabaleshwar, India, Jan 5-19, 1984,
  ed. P. Roy and V. Singh, Springer, 1984.
%
\bibitem{DGP}
%
  G.~R.~Dvali, G.~F.~Giudice and A.~Pomarol,
  ``The $\mu$-Problem in Theories with Gauge-Mediated Supersymmetry Breaking,''
  Nucl.\ Phys.\  B {\bf 478}, 31 (1996)
  [arXiv:hep-ph/ 9603238].
  %%CITATION = NUPHA,B478,31;%%
%
\bibitem{SKnucleondecay}
%
  K.~Kobayashi {\it et al.}  [Super-Kamiokande Collaboration],
  ``Search for nucleon decay via modes favored by supersymmetric grand
  unification models in Super-Kamiokande-I,''
  Phys.\ Rev.\  D {\bf 72}, 052007 (2005)
  [arXiv:hep-ex/0502026].
  %%CITATION = PHRVA,D72,052007;%%
%
\bibitem{MurayamaPierce}
%
  H.~Murayama and A.~Pierce,
  ``Not even decoupling can save minimal supersymmetric SU(5),''
  Phys.\ Rev.\  D {\bf 65}, 055009 (2002)
  [arXiv:hep-ph/0108104].
  %%CITATION = PHRVA,D65,055009;%%
%
\bibitem{HarnikLarsonMurayamaThormeier}
%
  R.~Harnik, D.~T.~Larson, H.~Murayama and M.~Thormeier,
  ``Probing the Planck scale with proton decay,''
  Nucl.\ Phys.\  B {\bf 706}, 372 (2005)
  [arXiv:hep-ph/0404260].
  %%CITATION = NUPHA,B706,372;%%
%
\bibitem{DIN1}
%
 The second paper in \cite{Het-super} by Derendinger, Ibanez 
 and Nilles.
%
\bibitem{DKL}
%
% \bibitem{Dixon:1990pc}
  L.~J.~Dixon, V.~Kaplunovsky and J.~Louis,
  ``Moduli dependence of string loop corrections to gauge coupling constants,''
  Nucl.\ Phys.\  B {\bf 355}, 649 (1991).
  %%CITATION = NUPHA,B355,649;%%
%
\bibitem{GHMR}
%
%\bibitem{Gross:1985rr}
  D.~J.~Gross, J.~A.~Harvey, E.~J.~Martinec and R.~Rohm,
  ``Heterotic String Theory. 2. The Interacting Heterotic String,''
  Nucl.\ Phys.\  B {\bf 267}, 75 (1986).
  %%CITATION = NUPHA,B267,75;%%
%
\bibitem{Yahikozawa}
%
  S.~Yahikozawa,
  ``Evaluation Of One Loop Amplitude In Heterotic String Theory,''
  Phys.\ Lett.\  B {\bf 166}, 135 (1986).
  %%CITATION = PHLTA,B166,135;%%
%
\bibitem{EJM}
%
%\bibitem{Ellis:1987dc}
  J.~R.~Ellis, P.~Jetzer and L.~Mizrachi,
  ``ONE LOOP STRING CORRECTIONS TO THE EFFECTIVE FIELD THEORY,''
  Nucl.\ Phys.\  B {\bf 303}, 1 (1988).
  %%CITATION = NUPHA,B303,1;%%
%
\bibitem{Nakajima}
%
% \bibitem{Nakajima:2007hu}
  H.~Nakajima,
  ``Testing the seesaw mechanism at collider energies in the Randall-Sundrum
  model,''
  arXiv:0705.3527 [hep-ph].
  %%CITATION = ARXIV:0705.3527;%%
%
\bibitem{GrSl}
%
% \bibitem{Gross:1986mw}
  D.~J.~Gross and J.~H.~Sloan,
  ``The Quartic Effective Action for the Heterotic String,''
  Nucl.\ Phys.\  B {\bf 291}, 41 (1987).
  %%CITATION = NUPHA,B291,41;%%
%
\bibitem{Banks:1995by}
%
  T.~Banks, Y.~Grossman, E.~Nardi and Y.~Nir,
  ``Supersymmetry without R-parity and without lepton number,''
  Phys.\ Rev.\  D {\bf 52}, 5319 (1995)
  [arXiv:hep-ph/9505248].
  %%CITATION = PHRVA,D52,5319;%%
%
\bibitem{Hempfling}
%
  R.~Hempfling,
  ``Neutrino Masses and Mixing Angles in SUSY-GUT Theories with explicit
  R-Parity Breaking,''
  Nucl.\ Phys.\  B {\bf 478}, 3 (1996)
  [arXiv:hep-ph/9511288].
  %%CITATION = NUPHA,B478,3;%%
%
\bibitem{NP}
%
  H.~P.~Nilles and N.~Polonsky,
  ``Supersymmetric neutrino masses, R-symmetries, and the generalized mu
  problem,''
  Nucl.\ Phys.\  B {\bf 484}, 33 (1997)
  [arXiv:hep-ph/9606388].
  %%CITATION = NUPHA,B484,33;%%
%
\bibitem{washout}
%
  B.~A.~Campbell, S.~Davidson, J.~R.~Ellis and K.~A.~Olive,
  ``Cosmological baryon asymmetry constraints on extensions of the standard
  model,''
  Phys.\ Lett.\  B {\bf 256}, 484 (1991); \\
  %%CITATION = PHLTA,B256,484;%%
% B.~A.~Campbell, S.~Davidson, J.~R.~Ellis and K.~A.~Olive,
  ``On B+L Violation In The Laboratory In The Light Of Cosmological And
  Astrophysical Constraints,''
  Astropart.\ Phys.\  {\bf 1}, 77 (1992); \\
  %%CITATION = APHYE,1,77;%%
%
  W.~Fischler, G.~F.~Giudice, R.~G.~Leigh and S.~Paban,
  ``Constraints On The Baryogenesis Scale From Neutrino Masses,''
  Phys.\ Lett.\  B {\bf 258}, 45 (1991); \\
  %%CITATION = PHLTA,B258,45;%%
%
  H.~K.~Dreiner and G.~G.~Ross,
  ``Sphaleron Erasure Of Primordial Baryogenesis,''
  Nucl.\ Phys.\  B {\bf 410}, 188 (1993)
  [arXiv:hep-ph/9207221].
  %%CITATION = NUPHA,B410,188;%%
%
\bibitem{DavidsonEllis}
%
  S.~Davidson and J.~R.~Ellis,
  ``Flavour-dependent and basis-independent measures of R violation,''
  Phys.\ Rev.\  D {\bf 56}, 4182 (1997)
  [arXiv:hep-ph/9702247].
  %%CITATION = PHRVA,D56,4182;%%
%
\bibitem{Davidson}
%
  S.~Davidson,
  ``Basis independent parametrisations of R parity violation in the soft SUSY
  breaking sector,''
  Phys.\ Lett.\  B {\bf 439}, 63 (1998)
  [arXiv:hep-ph/9808425].
  %%CITATION = PHLTA,B439,63;%%
%
\bibitem{Buchmuller:2005eh}
%
  W.~Buchmuller, R.~D.~Peccei and T.~Yanagida,
  ``Leptogenesis as the origin of matter,''
  Ann.\ Rev.\ Nucl.\ Part.\ Sci.\  {\bf 55}, 311 (2005)
  [arXiv:hep-ph/0502169].
  %%CITATION = ARNUA,55,311;%%
%
\bibitem{Ellis85}
%
  J.~R.~Ellis, G.~Gelmini, C.~Jarlskog, G.~G.~Ross and J.~W.~F.~Valle,
  ``Phenomenology Of Supersymmetry With Broken R-Parity,''
  Phys.\ Lett.\  B {\bf 150}, 142 (1985).
  %%CITATION = PHLTA,B150,142;%%
%
\bibitem{BS87}
%
  A.~Bouquet and P.~Salati,
  ``R PARITY BREAKING AND COSMOLOGICAL CONSEQUENCES,''
  Nucl.\ Phys.\  B {\bf 284}, 557 (1987).
  %%CITATION = NUPHA,B284,557;%%
%
\bibitem{TY}
%
  F.~Takayama and M.~Yamaguchi,
  ``Gravitino dark matter without R-parity,''
  Phys.\ Lett.\  B {\bf 485}, 388 (2000)
  [arXiv:hep-ph/0005214].
  %%CITATION = PHLTA,B485,388;%%
%
\bibitem{GK}
%
  K.~Griest and M.~Kamionkowski,
  ``Unitarity Limits on the Mass and Radius of Dark Matter Particles,''
  Phys.\ Rev.\ Lett.\  {\bf 64}, 615 (1990).
  %%CITATION = PRLTA,64,615;%%
%
\bibitem{KKM}
%
  M.~Kawasaki, K.~Kohri and T.~Moroi,
  ``Hadronic decay of late-decaying particles and big-bang nucleosynthesis,''
  Phys.\ Lett.\  B {\bf 625}, 7 (2005)
  [arXiv:astro-ph/0402490]; \\
  %%CITATION = PHLTA,B625,7;%%
%
% M.~Kawasaki, K.~Kohri and T.~Moroi,
  ``Big-bang nucleosynthesis and hadronic decay of long-lived massive
  particles,''
  Phys.\ Rev.\  D {\bf 71}, 083502 (2005)
  [arXiv:astro-ph/0408426].
  %%CITATION = PHRVA,D71,083502;%%
%
\bibitem{FST}
%
  J.~L.~Feng, S.~Su and F.~Takayama,
  ``Supergravity with a gravitino LSP,''
  Phys.\ Rev.\  D {\bf 70}, 075019 (2004)
  [arXiv:hep-ph/0404231].
  %%CITATION = PHRVA,D70,075019;%%
%
\bibitem{chidecayA}
%
  M.~Hirsch, M.~A.~Diaz, W.~Porod, J.~C.~Romao and J.~W.~F.~Valle,
  ``Neutrino masses and mixings from supersymmetry with bilinear R-parity
  violation: A theory for solar and atmospheric neutrino oscillations,''
  Phys.\ Rev.\  D {\bf 62}, 113008 (2000)
  [Erratum-ibid.\  D {\bf 65}, 119901 (2002)]
  [arXiv:hep-ph/0004115].
  %%CITATION = PHRVA,D62,113008;%%
%
\bibitem{chidecayB}
%
  W.~Porod, M.~Hirsch, J.~Romao and J.~W.~F.~Valle,
  ``Testing neutrino mixing at future collider experiments,''
  Phys.\ Rev.\  D {\bf 63}, 115004 (2001)
  [arXiv:hep-ph/0011248].
  %%CITATION = PHRVA,D63,115004;%%
%
\bibitem{Fukugita:2006xy}
%
  M.~Fukugita and M.~Kawasaki,
  ``Primordial Helium Abundance: A Reanalysis of the Izotov-Thuan Spectroscopic
  Sample,''
  Astrophys.\ J.\  {\bf 646}, 691 (2006)
  [arXiv:astro-ph/0603334].
  %%CITATION = ASJOA,646,691;%%
%
\bibitem{Berezinsky}
%
  V.~Berezinsky, A.~S.~Joshipura and J.~W.~F.~Valle,
  ``Gravitational violation of R parity and its cosmological signatures,''
  Phys.\ Rev.\  D {\bf 57}, 147 (1998)
  [arXiv:hep-ph/9608307].
  %%CITATION = PHRVA,D57,147;%%
%
\bibitem{Steffen}
%
  F.~D.~Steffen,
  ``Constraints on gravitino dark matter scenarios with long-lived charged
  sleptons,''
  AIP Conf.\ Proc.\  {\bf 903}, 595 (2007)
  [arXiv:hep-ph/0611027].
  %%CITATION = APCPC,903,595;%%
%
\bibitem{KKM07}
%
  M.~Kawasaki, K.~Kohri and T.~Moroi,
  ``Big-Bang Nucleosynthesis with Long-Lived Charged Slepton,''
  Phys.\ Lett.\  B {\bf 649}, 436 (2007)
  [arXiv:hep-ph/0703122].
  %%CITATION = PHLTA,B649,436;%%
%
\bibitem{Pospelov}
%
  M.~Pospelov,
  ``Particle physics catalysis of thermal big bang nucleosynthesis,''
  Phys.\ Rev.\ Lett.\  {\bf 98}, 231301 (2007)
  [arXiv:hep-ph/0605215].
  %%CITATION = PRLTA,98,231301;%%
%
\bibitem{staudecay}
%
  M.~Hirsch, W.~Porod, J.~C.~Romao and J.~W.~F.~Valle,
  ``Probing neutrino properties with charged scalar lepton decays,''
  Phys.\ Rev.\  D {\bf 66}, 095006 (2002)
  [arXiv:hep-ph/0207334].
  %%CITATION = PHRVA,D66,095006;%%
%
\bibitem{JLQCD}
%
  S.~Aoki {\it et al.}  [JLQCD Collaboration],
  ``Nucleon decay matrix elements from lattice QCD,''
  Phys.\ Rev.\  D {\bf 62}, 014506 (2000)
  [arXiv:hep-lat/9911026].
  %%CITATION = PHRVA,D62,014506;%%
%
\bibitem{Aoki:2006ib}
%
  Y.~Aoki, C.~Dawson, J.~Noaki and A.~Soni,
  ``Proton decay matrix elements with domain-wall fermions,''
  Phys.\ Rev.\  D {\bf 75}, 014507 (2007)
  [arXiv:hep-lat/0607002].
  %%CITATION = PHRVA,D75,014507;%%
%
\bibitem{Choi}
%
  K.~Choi, E.~J.~Chun and J.~S.~Lee,
  ``Proton decay with a light gravitino or axino,''
  Phys.\ Rev.\  D {\bf 55}, 3924 (1997)
  [arXiv:hep-ph/9611285].
  %%CITATION = PHRVA,D55,3924;%%
%
\bibitem{WWZ}
%
  S.~Weinberg,
  ``Baryon And Lepton Nonconserving Processes,''
  Phys.\ Rev.\ Lett.\  {\bf 43}, 1566 (1979); \\
  %%CITATION = PRLTA,43,1566;%%
%
  F.~Wilczek and A.~Zee,
  ``Operator Analysis Of Nucleon Decay,''
  Phys.\ Rev.\ Lett.\  {\bf 43}, 1571 (1979).
  %%CITATION = PRLTA,43,1571;%%
%
\bibitem{Martin} 
%
  S.~P.~Martin,
  ``A supersymmetry primer,''
  arXiv:hep-ph/9709356.
  %%CITATION = HEP-PH/9709356;%%
%

\end{thebibliography}
\end{document}